\documentclass[useAMS,usenatbib,useverbatim]{mn2e}

\usepackage{verbatim,epsfig}  

\def\nms{\mathsurround=0pt}
\def\oversim#1#2{\lower 2pt\vbox{\baselineskip 0pt \lineskip 1pt
    \ialign{$\nms#1\hfil##\hfil$\crcr#2\crcr\sim\crcr}}}
\def\ltsim{\mathrel{\mathpalette\oversim<}} 
\def\gtsim{\mathrel{\mathpalette\oversim>}} 

\title[Giants in $\omega$ Centauri]{Giants in the globular cluster $\omega$ Centauri: dust production, mass loss and distance}
\author[McDonald et al.]{Iain McDonald$^{1}$, Jacco Th. van Loon$^{1}$, Leen Decin$^{2}$, Martha L. Boyer$^{3}$, \newauthor Andrea K. Dupree$^{4}$, Aneurin Evans$^{1}$, Robert D. Gehrz$^{3}$, Charles E. Woodward$^{3}$\\
$^1$Astrophysics Group, School of Physical \& Geographical Sciences, Keele University, Staffordshire ST5 5BG, UK\\
$^2$Department of Physics and Astronomy, Institute for Astronomy, K.\ U.\ Leuven, Celestijnenlaan 200B, 3001 Leuven, Belgium\\
$^3$Department of Astronomy, 116 Church Street, S.E., University of Minnesota, Minneapolis, MN 55455, USA\\
$^4$Harvard-Smithsonian Center for Astrophysics, 60 Garden Street, Cambridge, MA 02138, USA\\
}

\begin{document}

\date{Received date; accepted date}

\pagerange{\pageref{firstpage}--\pageref{lastpage}} \pubyear{2002}

\maketitle

\label{firstpage}

\begin{abstract}

We present spectral energy distribution modelling of 6875 stars in $\omega$ Centauri, obtaining stellar luminosities and temperatures by fitting literature photometry to state-of-the-art {\sc marcs} stellar models. By comparison to four different sets of isochrones, we provide a new distance estimate to the cluster of $4850 \pm 200$ (random error) $\pm 120$ (systematic error) pc, a reddening of $E(B-V) = 0.08 \pm 0.02$ (random) $\pm 0.02$ (systematic) mag and a \emph{differential} reddening of $\Delta E(B-V) < 0.02$ mag for an age of 12 Gyr. Several new post-early-AGB candidates are also found. Infra-red excesses of stars were used to measure total mass-loss rates for individual stars down to $\sim 7 \times 10^{-8}$ M$_\odot$ yr$^{-1}$. We find a total dust mass-loss rate from the cluster of $1.3 \pm ^{0.8}_{0.5} \times 10^{-9}$ M$_\odot$ yr$^{-1}$, with the total gas mass-loss rate being $> 1.2 \pm ^{0.6}_{0.5} \times 10^{-6}$ M$_\odot$ yr$^{-1}$. Half of the cluster's dust production and 30\% of its gas production comes from the two most extreme stars -- V6 and V42 -- for which we present new Gemini/T-ReCS mid-infrared spectroscopy, possibly showing that V42 has carbon-rich dust. The cluster's dust temperatures are found to be typically $\gtsim$ 550 K. Mass loss apparently does not vary significantly with metallicity within the cluster, but shows some correlation with barium enhancement, which appears to occur in cooler stars, and especially on the anomalous RGB. Limits to outflow velocities, dust-to-gas ratios for the dusty objects and the possibility of short-timescale mass-loss variability are also discussed in the context of mass loss from low-metallicity stars. The ubiquity of dust around stars near the RGB-tip suggests significant dusty mass loss on the RGB; we estimate that typically 0.20--0.25 M$_\odot$ of mass loss occurs on the RGB. From observational limits on intra-cluster material, we suggest the dust is being cleared on a timescale of $\ltsim 10^5$ years.
\end{abstract}


\section{Introduction}

\subsection{Mass loss and globular clusters}

Galactic globular clusters (GCs) are unique stellar laboratories, containing roughly co-eval populations of stars at known distances, covering the range [Fe/H] $\approx$ --2.3 to solar \citep{Harris96}. They are ideal locations in which to probe the later stages of development of low-mass ($\sim$0.8 M$_\odot$) stars, and their large ($\approx 10^{5}-10^{6}$) populations mean they can harbour objects in very short phases of evolution.

Stellar mass loss, and its implications for both stellar evolution and the cluster's fate are of particular importance. All stars over $\sim$0.8 M$_\odot$ are thought to lose $\gtsim$30\% of their mass on the Red and Asymptotic Giant Branches (RGB/AGB) (e.g.\ \citealt{Rood73,LDZ94,CDA08}). Warm giants may drive an outflow of $\sim 10^{-9}$ M$_\odot$ yr$^{-1}$ via magnetic or hydrodynamic processes, (e.g.~\citealt{DHA84,MCP06}), while the coolest, most luminous giants are thought to be able to sustain a wind of at least $\sim 10^{-6}$ M$_\odot$ yr$^{-1}$ through a combination of pulsations and radiation pressure on circumstellar dust (e.g.~\citealt{GW71,BW91}). Mid-infrared (IR) spectra can be used to identify the species, temperature, mass and nature of these circumstellar dust grains and improve estimates of the total mass-loss rate from the star \citep{LPH+06,vLMO+06}.

Mass loss has important consequences for the remainder of the star's evolution and the chemical enrichment of interstellar material. Significant mass loss can decrease the number of thermal pulses such a star undergoes, thus inhibiting the growth of the stellar core, and reducing dredge-up, which affects the chemistry and characteristics of the post-AGB and planetary nebula phase \citep{VW93}. Mass-loss evolutionary processes also determine whether the star remains an oxygen-rich AGB star or becomes a carbon star \citep{vLZW+98,FCL+98}. Ultimately, mass loss will limit the white dwarf mass. Many aspects of the mass-loss process remain poorly understood. There is no clear consensus even on the primary driving mechanism behind mass loss, likely pulsation or continuum-driving of dust. By examining mass loss along the giant branches of populations of different metallicities, we can gain insight into the poorly-determined relationships between metallicity and the mass-loss rate, the wind speed and gas-to-dust ratio of the stars; and the correlation between mass loss and gravity (e.g.~\citealt{MvL07}). The clearing of lost mass from within the cluster will also liberate mass from the cluster and could exacerbate its evaporation. Study of the clearing can also allow us to infer conditions in the cluster's local environment \citep{OKYM07}.

The \emph{Spitzer Space Telescope} \citep{WRL+04,GRW+07} has, for the first time, allowed a complete census of dust production within GCs, due to its unprecedented sensitivity and angular resolution. A comprehensive analysis of mass loss from stars in 47 Tuc was recently performed by \cite{ORFF+07} (hereafter O+07). In this study, mid-IR flux excesses from \emph{Spitzer} data were used to construct the mass-loss rate from individual stars in the cluster, attaining an empirical relation for mass loss from metal-poor giant stars. This relation suggests that significant dust production occurs along a considerable part of the giant branch. We compare our results to this analysis in Section \ref{MdotRGBSect}.

\subsection{Omega Centauri}

On the basis of our own \emph{Spitzer} data, published in \cite{BMvL+08} (hereafter B+08), and new mid-IR spectra of the two stars (V6 \& V42) with the strongest IR excess, we here present an estimate of the mass-loss evolution in the most massive Galactic globular cluster: $\omega$ Centauri.

$\omega$ Cen is unique in the wealth of information available -- it is comparatively nearby at a relatively well-known distance of $\sim$5 kpc \citep{Harris96,vLlPB+00,vdVvdBVdZ06,dPPS+06}; its high radial velocity of $v_{\rm LSR} = +232$ km s$^{-1}$ \citep{MM86,vdVvdBVdZ06} allows easy spectroscopic membership and radial velocity determination (\citealt{vLvLS+07}; hereafter vL+07), and it also has independent membership determinations from proper motion measurements (\citealt{vLlPB+00}; hereafter vL+00). The importance of this is highlighted in B+08, which also contains an assessment of the effects of stellar blending on our longer-wavelength (lower-resolution) data. The cluster contains a statistically-significant sample of stars in most advanced stages of evolution (vL+07). Coupling this with a large existing photometric dataset, it is possible to positively identify cluster members and probe the stellar population well down the giant branch at all wavelengths relevant to spectral energy distribution (SED) modelling.

The bulk of the stars in the cluster have a relatively low metallicity -- [Fe/H] $\approx$ --1.7 to --1.6 \citep{Norris96,SSC+00}. However, a helium-enriched, metal-intermediate population at [Fe/H] $\approx$ --1.2 \citep{Norris96,Norris04} also exists; along with a metal-rich component, the `anomalous' RGB (RGB-a), with metallicities up to [Fe/H] $\approx$ --0.7, which together comprise perhaps 10\% of the cluster \citep{LJS+99,PFB+00,PPH+02}. Variations in surface abundances are also present, including oxygen-rich, M-type stars with titanium oxides; stars enhanced in CH and/or CN; and genuine carbon stars with molecular carbon (C$_2$; vL+07).

The origin of these sub-populations is undecided \citep{SPF+05,SdCNC06a,VPK+07}, though there is evidence that $\omega$ Cen may be the remnant nucleus of a tidally disrupted dwarf spheroidal galaxy (dSph) \citep{ZKD+88,Freeman93}. Understanding $\omega$ Cen may therefore assist our understanding of mass loss and chemical enrichment in other nucleated dSphs, as well as in the earlier, more metal-poor Universe.

\subsection{Individual stars}

Within this paper, we also present the first mid-infrared spectra of the two brightest, most IR-excessive stars in the cluster, V6 (LEID 33062, ROA 162) and V42 (LEID 44262, ROA 90). \cite{GF73,GF77} measured the IR colours of V6, confirming it as a very bright IR source and showing it to be variable at near-IR wavelengths, with a possible $L$-band excess. Its period is uncertain, being listed as 73.513 days in \cite{SawyerHogg73} and 100--120 days in \cite{DFLE72}. Classified as an M4--5 emission line variable, it has a possible radial velocity variation of up to 40 km s$^{-1}$ with a mean velocity of +213.6$\pm$3.7 km s$^{-1}$ (Dickens et al.~1972; \citealt{Webbink81}; quoted uncertainty $\sim$13 km s$^{-1}$). Its temperature has been estimated at 3300--3600 K with log($g$) of 0.0 (assuming $M = 0.8$ M$_\odot$; \citealt{PCM+80,Frogel83}). It contains relatively large amounts of H$_2$O compared to the other cluster long-period variables (LPVs) and is CN and NH enhanced \citep{CB86}. It is also known to be a TiO variable and shows variable hydrogen emission (\citealt{LloydEvans83b,LloydEvans83c,LloydEvans86}). V6 appears to belong to the metal-intermediate sub-population, with [Fe/H] $\sim$ --1.19 \citep{ZW84,NdC95b,VWS02}.

\cite{Feast65} implies V42 should be classified as a semi-regular variable of type SRd, but its spectrum is somewhat later than the F--K-type this implies. An M1--2.5 emission line variable, it may also exhibit radial velocity variations between about +253 and +272 km s$^{-1}$ (Dickens et al.~1972), though this variation may be largely attributable to insufficient signal-to-noise. V42 may represent a star bridging the gap between SRd stars and emission-line LPVs. As an apparent fundamental-mode pulsator, its 148.64$\pm$0.03 day period (vL+00) is the longest in $\omega$ Cen and among the longest in globular cluster variables as a group \citep{Clement97}. Dickens et al.\ also report that the TiO bands weaken and hydrogen emission lines are present only near optical photometric maximum (see also \citealt{LloydEvans83c,LloydEvans83d}), suggesting an additional opacity source. It is also CN and NH enhanced \citep{CB86}. \cite{CF83} estimate log($g$) = 0.5 and $T$ = 3950 K, whereas \cite{MW85} calculate a much lower temperature of 2818 K, based on IR data, and show V42 undergoes substantial variability -- some $\sim$60\% of its mean luminosity -- even in the $L$-band. No reliable information on the star's metallicity is available, although vL+07 place it at [Fe/H] $\sim -1.25$; a high-resolution optical spectrum shows significant H-$\alpha$ emission, attributable to shocks propagating in the stellar wind (\citealt{MvL07}, hereafter MvL07).

The cluster also contains a number of other interesting objects. Five carbon stars have now been reported (vL+07). Several post-AGB stars are also known. Most famously, Fehrenbach's Star (LEID 16018, HD 116745) appears to have already undergone thorough mixing and mass loss \citep{FD62,DP73,GW92}. Another, V1 (LEID 32029), is an irregularly-pulsating star with multiple periods that is thought to have undergone gas-dust separation, but not thought to have undergone third dredge-up (surface enrichment during thermal pulses) and may thus be a post-early-AGB star (\citealt{Gonzalez94,MHLN98}; \citealt{TKD+06,TKD+07}; vL+07). V29 (LEID 43105), V43 (LEID 39156), V48 (LEID 46162) and V92 (LEID 26026) have also been suggested to be post-AGB stars \citep{Gonzalez94,GW94}.

The remainder of the paper is organised as follows: Section 2 describes the SED input data and models; Section 3 details the corrections to the photometry and the process of creation of the SEDs; Section 4 compares our observed temperatures and luminosities with those predicted by a variety of stellar evolution models, providing a new estimate of the distance and reddening to the cluster; Section 5 presents new mid-infrared spectra of V6 and V42, and discusses subsequent estimation of mass loss from individual stars; Section 6 discusses the implications of the dataset, including calculating the total mass-loss rate of the cluster, correlations with various stellar parameters, and the evolutionary nature of mass loss; finally, Section 7 presents our conclusions.

\section{The Input Datasets}

\subsection{Literature photometry and variability data}

Our input list of objects was derived from the optical photometry of vL+00. We have selected from this catalogue those stars that have at least one detection in the mid-IR \emph{Spitzer} IRAC (3.6, 4.5, 5.8 and 8 $\mu$m) and MIPS (24 $\mu$m) imaging published in B+08. We have combined these data with photometry from 2MASS \citep{SCS+06}. Of the resultant list of 6875 stars, 6018 are proper motion members, 1145 are radial velocity members in vL+07, and 1701 have photometry at both 8 and 24 $\mu$m.

\subsection{The model spectra}

As many of our stars -- particularly the cooler objects -- depart significantly from a blackbody spectrum, we have used model spectra to derive the stellar parameters. A grid of spectra were created using the Model Atmosphere in Radiative and Convective Scheme ({\sc marcs}) code \citep{GBEN75,GEE+08} from 4000 K to 6500 K in steps of 250 K, with an additional dataset at 3500 K; gravities were sampled from $\log(g) = 0.0$ to $3.5$ in steps of 0.5. At each grid point, a further dimension in metallicity was sampled at [Z/H] = 0.0 with solar abundances; at [Z/H] = --1.0 with [$\alpha$/Fe] = +0.3; and at [Z/H] = --1.5 and --2.0, with [$\alpha$/Fe] = +0.4. For each grid point, a synthetic spectrum was created with $R =$ 20\,000 over the range 0.13 to 20 $\mu$m. Some models did not converge; these were replaced with a neighbouring convergent model that had the same temperature and the closest available combination of metallicity and gravity.

The spectra were extrapolated beyond 20 $\mu$m using a Rayleigh-Jeans tail fit to the range 18--20 $\mu$m. This was used in preference to the normally more-appropriate Engelke function \citep{Engelke92,DE07} as the Engelke function over-estimated the flux we observe by up to 20--40\%, while the over-estimation from a Rayleigh-Jeans tail is only around 10--20\%. Both a Rayleigh-Jeans tail or Engelke function are expected to \emph{under}-estimate the flux we observe at 24 $\mu$m for our cooler stars, though the reverse may be true in warmer giants. The immediate reasons behind the apparent \emph{over}-estimation these functions provide when compared to our data are not clear and we make an empirical correction in Section \ref{SystematicSection}.

\section{Spectral Energy Distributions}
\label{SEDSect}

\begin{figure}
\resizebox{\hsize}{!}{\includegraphics[angle=270]{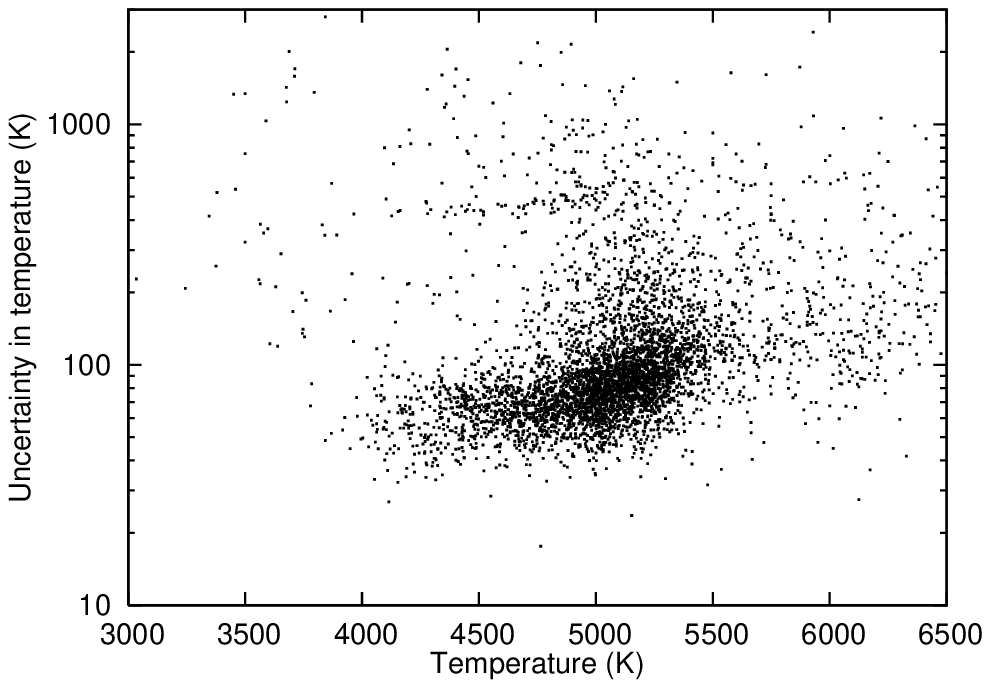}}
\caption[]{Distribution of internal errors in temperature over the giant branch stars. Errors increase to typically $\pm\sim$1000 K for stars not covered by our hottest models ($>$ 6500 K).}
\label{TempErrorFig}
\end{figure}

\begin{figure}
\resizebox{\hsize}{!}{\includegraphics[angle=270]{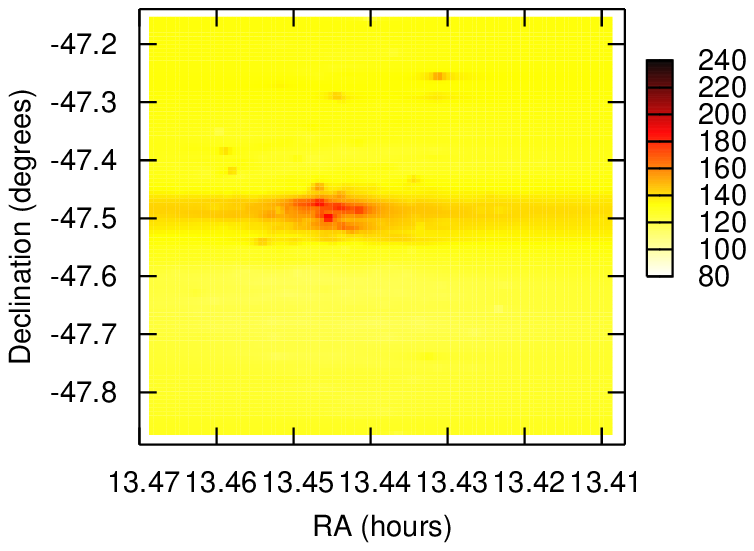}}
\caption[]{Distribution of internal errors in temperature over the cluster, containing only stars on the giant branches (here defined at $T < 5500$ K). Errors increase towards the cluster centre due to blending. The horizontal smearing is an interpolation artefact.}
\label{TempErrorMap}
\end{figure}

Firstly, the literature photometry and model spectra were converted into $F_\nu$ (Janskys). The models were then degraded in resolution by a factor of 100 to $R = 200$ to save on computing time and, using a cubic spline, interpolated in wavelength onto filter transmission data in the relevant filters. Comparing the photometric flux from the original spectrum and the interpolated spectrum, we find the differences to be $\ll$ 1\% in all cases except $B$-band, where the lower-resolution sampling of strong spectral lines lead to the systematic over-estimation of the flux by up to $\sim$3\% for the cooler stars. These are comparable to the errors present in the photometry, and much less than the observed photometric variation of our mass-losing candidates, thus we do not consider it to be a major source of error in our analysis. An additional source of systematic error comes from uncertainties in the filter and imaging device responses, and instrumental photometric zero points. Though difficult to quantify, we expect these effects to affect our photometry at the level of 3\% or less and correspondingly systematically alter our temperature and luminosities by $\ltsim$1\%, as the errors will tend to average out among the different filters.

A blackbody was fitted to the observed photometry to provide a first-order estimate of temperature and luminosity. From this we derive an initial value for log($g$), assuming a mass of 0.8 M$_\odot$.

An initial grid of {\sc marcs} models was set up to coarsely define the temperature. For computing purposes, the grid was chosen to be between 4096 and 6144 K, in steps of 512 (= 2$^8$) K. For stars falling outside this range (as determined by the fitting procedure), the grid was extended to 3072 K in the case of cooler stars, and to 18\,442 K for warmer stars. The step size was chosen to provide the computationally-fastest determination of temperature, while ensuring the true best-fit temperature is reached. For each of these temperatures, a spectrum was produced from the closest two model grid temperatures by linearly interpolating in [Fe/H] and log($g$). These two spectra ($F_1$ and $F_2$) were then combined into a single spectrum ($F_{\rm final}$) by dividing the spectra by blackbodies of the corresponding grid point temperatures, averaging the spectra, then multiplying by a blackbody of the final required temperature.

By comparing the interpolated spectrum with the original, we calculate that our interpolation method can accurately reproduce a model value for a photometric band at any temperature above 4000 K, with any gravity and metallicity within our model grid to within 1\%. For stars below 4000 K, a lack of converging models at 3750 K and $<$ 3500 K means that the models will tend to over-estimate the photometric flux at short wavelengths ($B$- and $V$-band) by perhaps 3--5\% for the very coolest stars we sample, which is still (much) less than the expected stellar variability, differential interstellar reddening or (in some cases) photometric uncertainty in this regime.

Since $\omega$ Cen exhibits moderate but noticeable interstellar reddening, it was necessary to redden the model spectrum appropriately. Estimates of the reddening towards the cluster vary considerably from $E(B-V)$ = 0.09 to 0.15 mag \citep{Djorgovski93,Harris96,Kovacs02}. A differential reddening across the cluster of $\Delta E(B-V) \ltsim 0.04$ mag has also been inferred (\citealp{NB75,CSB+05}; vL+07). For our analysis, we adopted a reddening of $E(B-V) = 0.08$ mag across the entire cluster to better fit the evolutionary isochrones (see Section \ref{IsocSect}). Assuming $A_{\rm V} / E(B-V) \approx 3.05$ \citep{Whittet92}, this yields an extinction of $A_{\rm V} \approx 0.24$ mag.

To compute the reddening as a function of wavelength, we have used the relationship from \cite{Draine89}, with an empirical correction for wavelengths of $< 1$ $\mu$m based on Draine's Figure 1. Using $E(J-K) = 0.52 E(B-V)$ (after \citealt{Whittet92}), this becomes:
\begin{equation}
	\frac{A_\lambda}{E(B-V)} = \left\{ \begin{array}{ll}
			1.248 \lambda^{-1.75 - (\lambda - 1) / 1.3} &\mbox{  for $\lambda \leq$ 1\ $\mu$m} \\
			1.248 \lambda^{-1.75} &\mbox{  for $\lambda \geq$ 1\ $\mu$m} \\
			         \end{array} \right. .
\end{equation}

By examining the effective temperature and luminosity at different reddening factors on a subset of stars, we can deduce the size of this correction. We find that the corresponding temperature and luminosity, $T^\prime$ and $L^\prime$ (in Kelvin and solar luminosities), are well approximated (for an arbitrary $E(B-V)$, denoted $e$) by:
\begin{equation}
	T^\prime = T + \frac{e-0.08}{0.13} \left(\left(\frac{T-3000}{45}\right)^{1.5}-50\right) ,
	\label{TCorrEqn}
\end{equation}
\begin{equation}
	L^\prime = L \left(1 + \frac{e-0.08}{0.13} \left(\frac{L}{13200} -0.155\right)\right) \left(\frac{d}{5300}\right)^2 , 
\end{equation}
where $L$ and $T$ are the luminosity and temperature for $E(B-V) = 0.08$ mag and $d$ is the distance in parsecs.

Draine estimates a 10\% error in his relationship, which would alter our corresponding corrections by a similar amount. Our composite relation described above differs from that of \cite{Whittet92} by $10-15\%$ over most of the spectrum, suggesting a similar inherent uncertainty in our reddening correction. The uncertainty in $E(B-V)$ affects the shorter wavelengths the most, and is typically greater than that produced by our filter convolution and interpolation above. Differential reddening \emph{across} the cluster thus leads to a scatter of up to $\pm 50-100$ K in temperature and $\pm 1-4\%$ in luminosity. Uncertainty in the \emph{average} extinction to the cluster, which is of the same order, is the largest systematic error in our temperature and luminosity estimates.

We do not consider circumstellar reddening here, as it only becomes significant in individual stars for the most extreme cases, where the effect is dwarfed (by a factor of $\sim$10) by the variability due to pulsation at any particular wavelength. This will not significantly affect our determination of distance or extinction to the cluster, but may lead to a slight under-estimation of the temperature and luminosity of the most enshrouded stars.

Once appropriately reddened, the spectrum was convolved with the filter responses to obtain the expected flux density received in that filter. A true $\chi^2$ statistic was impossible due to incomplete error information; a pseudo-$\chi^2$ statistic was calculated between the observed and expected flux densities, setting all observed photometric errors to have equal weight. The 24-$\mu$m data were not included in this calculation as we aim to fit the underlying photosphere, not the photosphere plus the wind's dust component. 

The process was repeated for each of the points in the one-dimensional temperature grid and the $\chi^2$ minimum identified. The value of log($g$) was re-determined for the temperature of the $\chi^2$ minimum ($T_{\rm min}$). The $\chi^2$ was re-determined at half the grid spacing, $T_{\rm min} \pm 256$ K, and a $\chi^2$ minimum re-defined. This process was iteratively carried out to refine the temperature to a \emph{precision} of 1 K. The resulting $\chi^2$ distribution suggests that the internal \emph{accuracy} is typically $\pm$70 K, but increases towards lower luminosities and higher temperatures. Note that while colour-$T_{\rm eff}$ transformations often give more precise values for $T_{\rm eff}$, our procedure yields the most consistent values for $T_{\rm eff}$ \emph{and} luminosity, taking into account the full observed SED. We plot the temperature errors in Fig.\ \ref{TempErrorFig}.

In Table \ref{OffsetTable}, we take the average and standard deviation of the difference between the {\sc marcs} modelled flux and the observed flux in each photometric band. We have limited this to stars with $T_{\rm eff} < 6500$ K, due to our lack of models at higher temperatures. Note also that the accuracy of our temperatures is mostly determined by the accuracy of our $B$ and $V$-band flux.

We include an interpolated map of the temperature errors in Fig.\ \ref{TempErrorMap}, showing they increase toward the cluster centre. The giant-branch stars with larger temperature errors ($\gtsim$200 K) are caused by errant photometry, particularly in the $B$- and $V$-bands, due to source blending in the dense cluster core. As illustrated in Fig.\ \ref{TempErrorFig}, there is only a very weak dependence between the magnitude of the error and the stellar temperature (and thus luminosity if the star is on the giant branch). The frequency of errors due to blending does increase as one heads to lower temperatures and luminosities. Crucially, however, the very presence of these errors allows us to determine which stars suffer from blending. We find that, on the upper giant branch, where mass loss is taking place, the only stars to suffer from substantial errors (130--330 K) are known variables (specifically V42, V6, V152 and V148), whose temperatures are expected to be uncertain due to their inherent variability.

\begin{figure*}
\resizebox{\hsize}{!}{\includegraphics[angle=270]{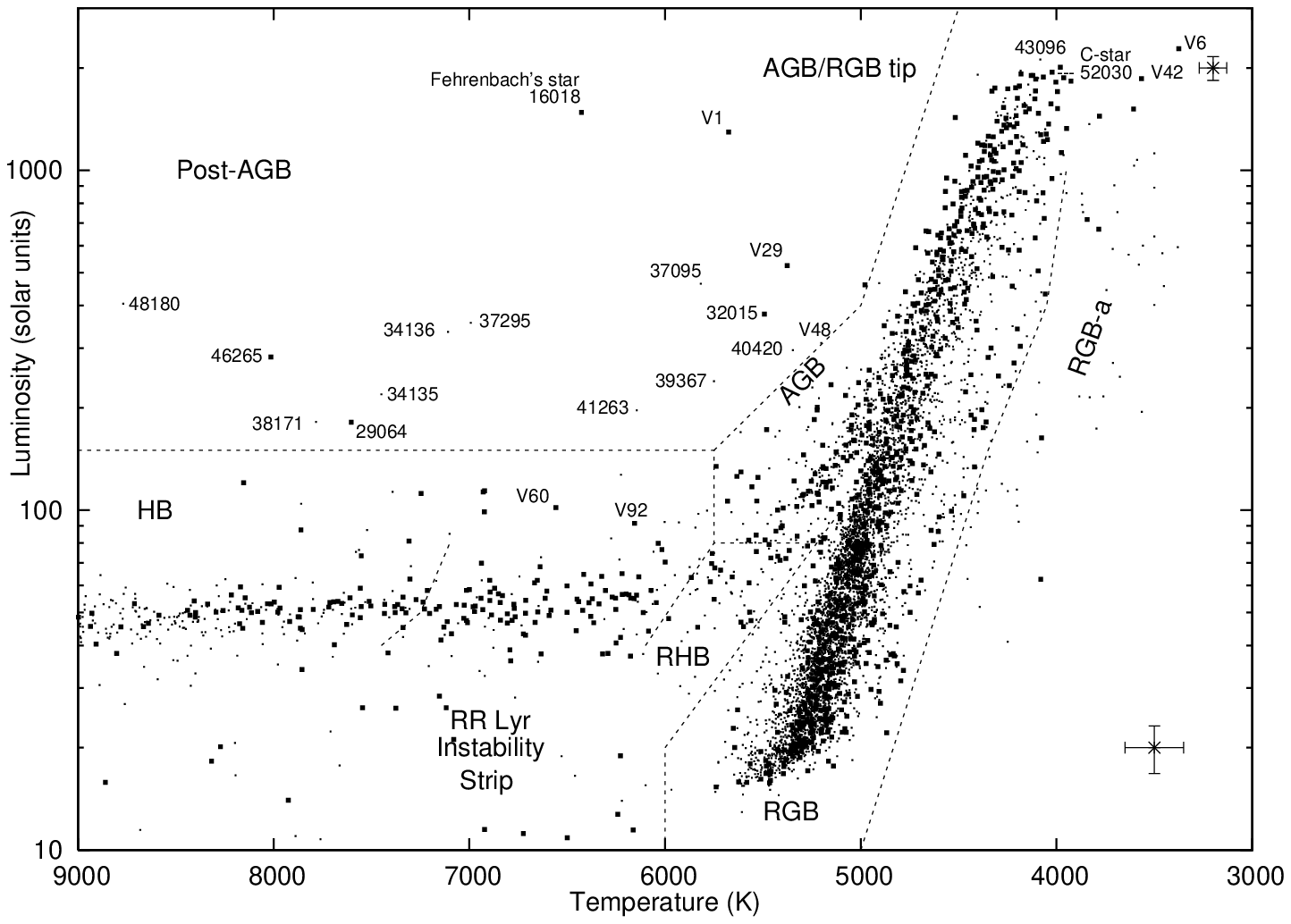}}
\caption[]{A physical HRD for the giant branch of $\omega$ Cen. Dots show proper motion members ($>$50\% probability); squares show objects that are (also) confirmed radial velocity members ($v_{\rm LSR}$ $>$ 100 km s$^{-1}$). Representative statistical errors for individual objects are shown in the lower- and upper-right corners. The limits to the RR Lyrae instability strip are taken from \cite{BCM95}.}
\label{HRDFig}
\end{figure*}

Literature distance estimates vary from 4.8 kpc (vL+00) to 5.52$\pm$0.13 kpc \citep{dPPS+06}, with 5.3 kpc being the median \citep{Peterson93,Harris96,TKP+01}. We adopted 5.0 kpc as an initial estimate, based on evolutionary isochrones (Section \ref{IsocSect}). Assuming this distance, a luminosity was calculated by integrating the final model spectrum. Due to the limits of integration ($>$130 nm) determined by the model spectra, we note that the luminosities of stars with considerable flux at $\lambda < 130$ nm (i.e. stars with $T \gg 8000$ K) will likely be more luminous than we have listed here, though this is not certain as our stellar models become progressively less reliable beyond 6500 K. The final stellar parameters are listed in Table \ref{ParamsTable}.

\begin{center}
\begin{table}
\caption[]{Average differences between modelled and observed fluxes, and standard deviations of those differences, for stars below 6500 K.}
\label{OffsetTable}
\begin{tabular}{lrrr}
    \hline \hline
Band	& Number   & Average    & Standard  \\
\ 	& of stars & difference & deviation \\
    \hline
B         & 5867 &  +5.1\% &  9.9\% \\
V         & 5861 & --6.9\% &  9.9\% \\
J         & 5645 &  +0.9\% &  7.2\% \\
H         & 5598 & --2.2\% &  6.7\% \\
K         & 5585 &  +4.2\% &  8.8\% \\
3.6$\mu$m & 5463 &  +2.6\% &  8.4\% \\
4.5$\mu$m & 5673 & --0.6\% &  8.3\% \\
5.8$\mu$m & 5319 &  +0.4\% & 14.5\% \\
8.0$\mu$m & 5759 & --1.3\% & 12.6\% \\
    \hline
\end{tabular}
\end{table}
\end{center}

\begin{center}
\begin{table}
\caption[]{Derived stellar parameters (full table on-line) ordered in decreasing luminosity, assuming $E(B-V) = 0.08$ mag and a distance of 5.0 kpc, including Leiden Identifier and proper motion (PM) percentage likelihood of membership from vL+00, and radial velocity from vL+07, where available.}
\label{ParamsTable}
\begin{tabular}{lll@{}c@{}c@{}c}
    \hline \hline
LEID & $T$ 			 & $L$  			& Gravity  & PM Mem & $v_{\rm LSR}$\\
\    & (K)         & (L$_\odot$) & log(cm s$^{-2}$) & \% & km s$^{-1}$\\
    \hline
33062 & 3375\,$\pm ^{257}_{204}$ & 2278\,$\pm ^{695}_{550}$ & 0.10 & 100 & 221 \\
43096 & 4082\,$\pm ^{ 90}_{108}$ & 2117\,$\pm ^{186}_{223}$ & 0.46 & 99  & \ \\
43099 & 3980\,$\pm ^{ 61}_{ 82}$ & 2012\,$\pm ^{123}_{167}$ & 0.44 & 100 & 210 \\
45232 & 4183\,$\pm ^{103}_{124}$ & 1964\,$\pm ^{193}_{233}$ & 0.54 & 100 & \ \\
52030 & 4022\,$\pm ^{ 89}_{121}$ & 1944\,$\pm ^{171}_{233}$ & 0.47 & 99  & 204 \\
48060 & 4181\,$\pm ^{ 65}_{ 76}$ & 1923\,$\pm ^{120}_{139}$ & 0.54 & 100 & 190 \\
47226 & 4402\,$\pm ^{106}_{123}$ & 1912\,$\pm ^{185}_{213}$ & 0.64 & 100 & \ \\
52017 & 4120\,$\pm ^{ 45}_{ 51}$ & 1903\,$\pm ^{ 83}_{ 95}$ & 0.52 & 100 & 206 \\
26025 & 4129\,$\pm ^{ 52}_{ 62}$ & 1900\,$\pm ^{ 96}_{113}$ & 0.53 & 100 & 229 \\
61015 & 4067\,$\pm ^{ 62}_{ 74}$ & 1881\,$\pm ^{114}_{138}$ & 0.51 & 99  & 218 \\
44277 & 3963\,$\pm ^{125}_{164}$ & 1873\,$\pm ^{236}_{310}$ & 0.46 & 100 & 211 \\
46062 & 4094\,$\pm ^{ 73}_{ 84}$ & 1869\,$\pm ^{133}_{154}$ & 0.52 & 100 & \ \\
44262 & 3565\,$\pm ^{385}_{281}$ & 1862\,$\pm ^{803}_{587}$ & 0.28 & 100 & 261 \\
...   & ...  & ...  &  ...  & ... & ... \\
    \hline
\end{tabular}
\end{table}
\end{center}

A physical Hertzsprung-Russell diagram (HRD) is shown in Fig.\ \ref{HRDFig}. The RGB and AGB can clearly be distinguished up to $\sim$150 L$_\odot$, with the Horizontal Branch (HB) extending towards high temperatures at $\sim$50 L$_\odot$. The expected errors suggest the majority of the spread in the diagram is real. The cooler giant branches should therefore represent the more metal-rich objects. Post-AGB stars are present at higher luminosities towards the warm side of the giant branch. The RGB-a stars are visible to the right of the main RGB. Interestingly, they do not appear to extend to the RGB tip, but as they are so few in number, this could merely be a stochastic effect.

\section{Comparisons with stellar isochrones}
\label{IsocSect}

With an HRD of this quality, a comparison with stellar evolution models can yield accurate determinations of the cluster's parameters, such as distance and reddening (we assume a fixed age of $\approx$12 Gyr as this is better estimated using the main-sequence turnoff). These parameters are important for both calibrating our luminosities and estimating mass-loss rates from individual stars. Evolutionary tracks and, in particular, the rate of stellar evolution, can already yield some constraint on mass loss on the RGB.

\subsection{Padova isochrones}

\begin{figure}
\resizebox{\hsize}{!}{\includegraphics[angle=270]{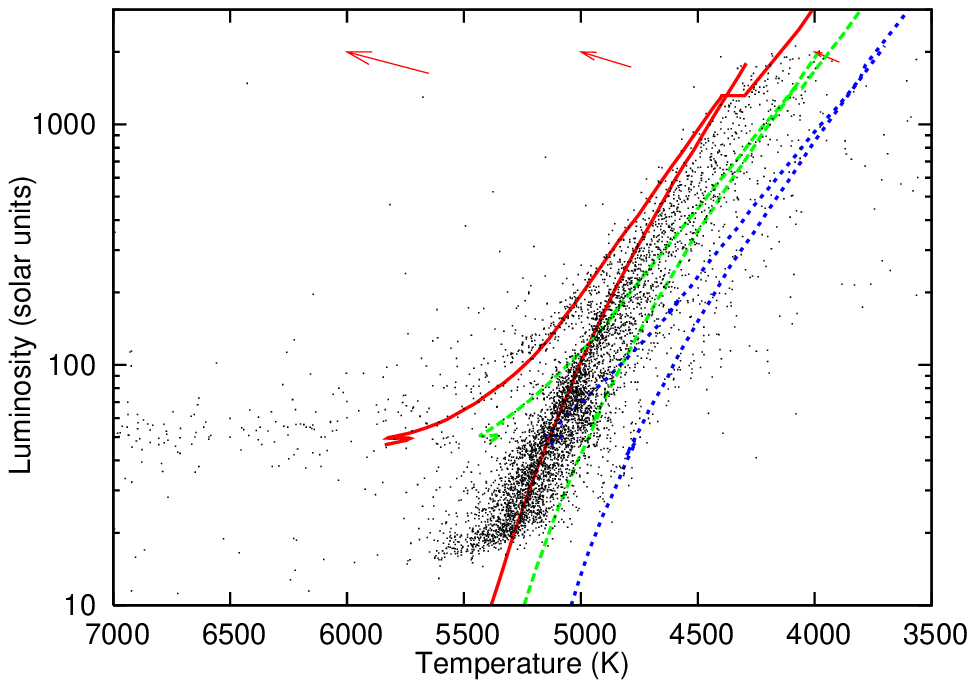}}
\resizebox{\hsize}{!}{\includegraphics[angle=270]{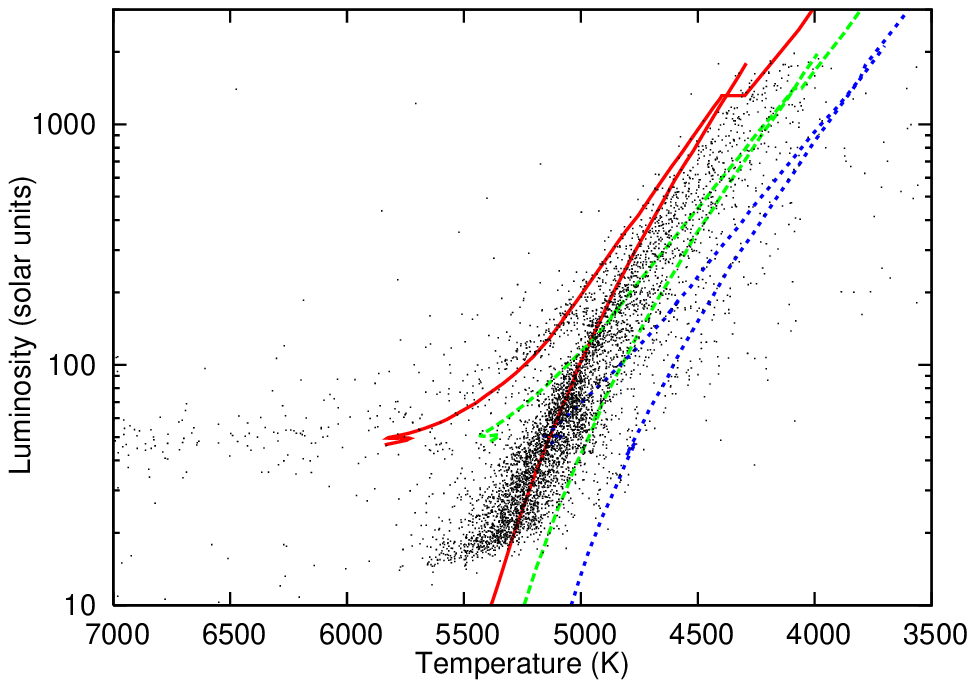}}
\caption[]{HRD of cluster members, including isochrones from \cite{MGB+08}. The top panel shows isochrones for metallicities of [Fe/H] = --1.6, --1.2 and --0.7 (from left to right), assuming a distance of 5000 pc. The arrows show the approximate reddening correction (taking $E(B-V) = 0.08$ mag) for that temperature. The bottom panel shows the same diagram, but with an \emph{approximate} correction to $E(B-V) = 0.09$ mag and a distance of 4800 pc. The isochrones are described in the text, with metallicity increasing from left to right.}
\label{IsoMFig}
\end{figure}

Fig.\ \ref{IsoMFig} shows Padova stellar isochrones from \cite{MGB+08}. The isochrones are for zero extinction, containing the dust solutions from \cite{BGS98}, with a log-normal initial mass function from \cite{Chabrier01}. They are shown for the metallicities and ages suggested by \cite{OFBP03} and \cite{HKRW04}, namely: 12.1 Gyr at [Fe/H] = --1.6 for the metal-rich population, 10.5 Gyr at [Fe/H] = --1.2 for the metal-intermediate population and 9.4 Gyr at [Fe/H] = --0.7 for the metal-rich population. Using these isochrones, we fitted the extinction and distance as $E(B-V) = 0.09 \pm 0.01$ mag and $4800 \pm 150$ pc, respectively.

We stress that these two parameters are somewhat correlated, and the errors given are based on single-parameter errors only. In this case, the distance is fitted primarily by the location of the zero-age horizontal branch (ZAHB), and the isochrones are then matched to the data by the correct reddening. While we have not undertaken a full quantitative analysis, comparing the two panels in Fig.\ \ref{IsoMFig}, we can see that a fit with a distance of 5000 pc and $E(B-V)$ of 0.08 mag also falls within visually plausible errors, however this does not reproduce the early-AGB and early-RGB so well.

Despite efforts to fit the distance and reddening, the Padova isochrones still do not provide a good fit to the slope of the RGB or AGB. We suggest that this is inherent in the models for three reasons: 
\begin{itemize}
\item on the basis of our above analysis, circumstellar dust starts reddening the stars at lower luminosities than their models predict, at around 1000 L$_\odot$ (c.f.~Fig.\ \ref{MdotFig}) -- this occurs both on the RGB and AGB;
\item RGB mass loss is not included in the Padova models (the accumulated mass loss is assumed to happen `instantaneously' at the RGB tip): stars which have lost mass will have larger radii and thus be cooler than the models predict;
\item the [$\alpha$/Fe] enhancement in the models is incorrect (Marigo et al.\ currently do not provide a mechanism with which to alter this) -- increased [$\alpha$/Fe] will make the stars cooler than the models.
\end{itemize}
It also appears that the mass-loss rates from \cite{MGB+08} for the RGB are not high enough for a large number of stars, as they do not reproduce the blueward extent of the HB that results from low mantle masses. We can therefore assume that, based on the Padova models, a star with an initial mass of 0.843 M$_\odot$, must typically lose $>$0.117 M$_\odot$ on the RGB.

\subsection{Dartmouth isochrones and ZAHB models}

\begin{figure}
\resizebox{\hsize}{!}{\includegraphics[angle=270]{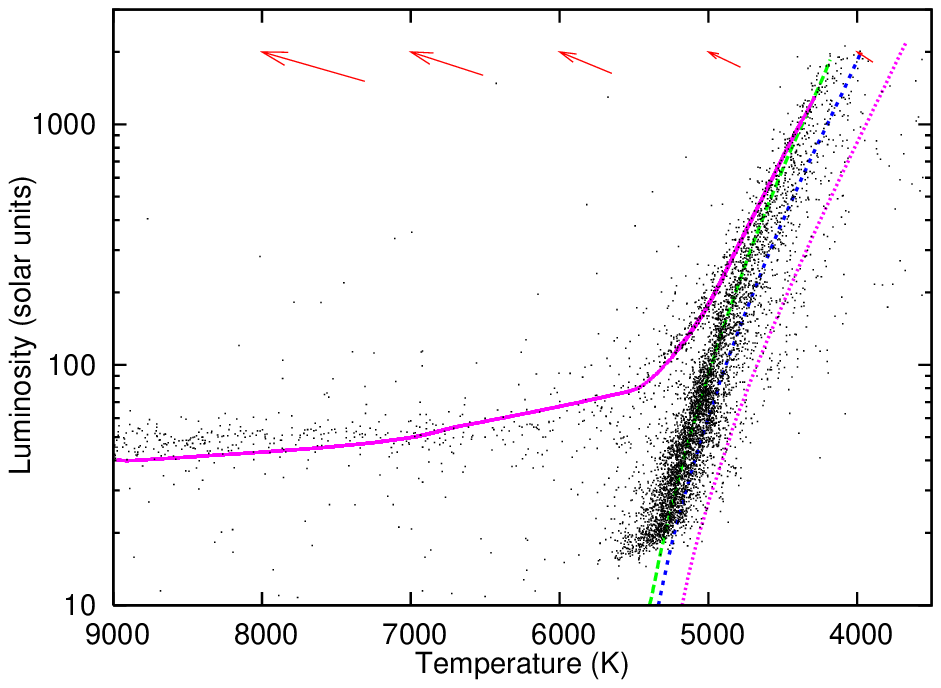}}
\resizebox{\hsize}{!}{\includegraphics[angle=270]{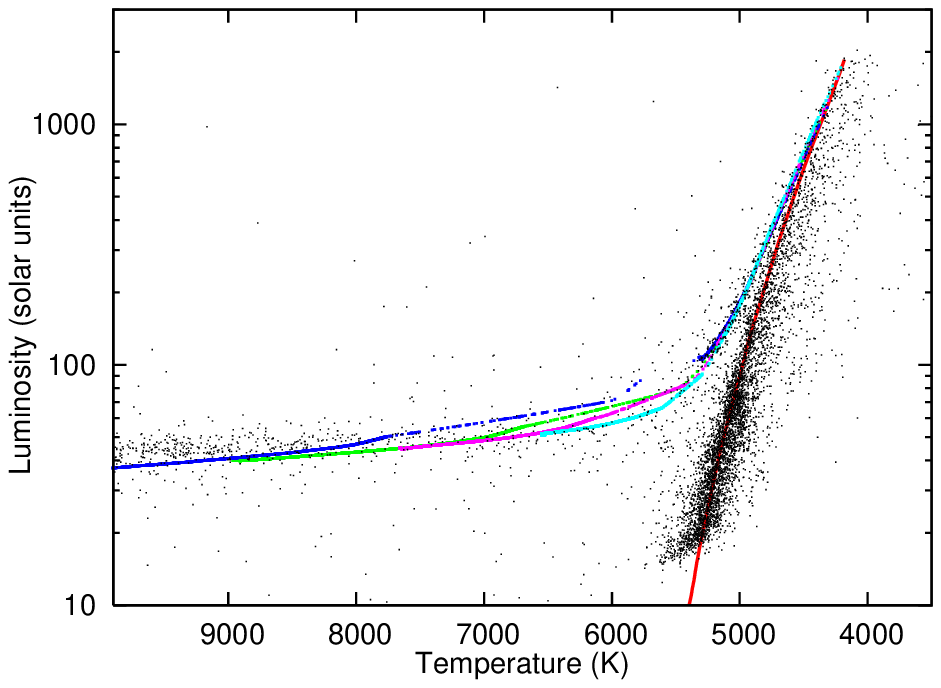}}
\resizebox{\hsize}{!}{\includegraphics[angle=270]{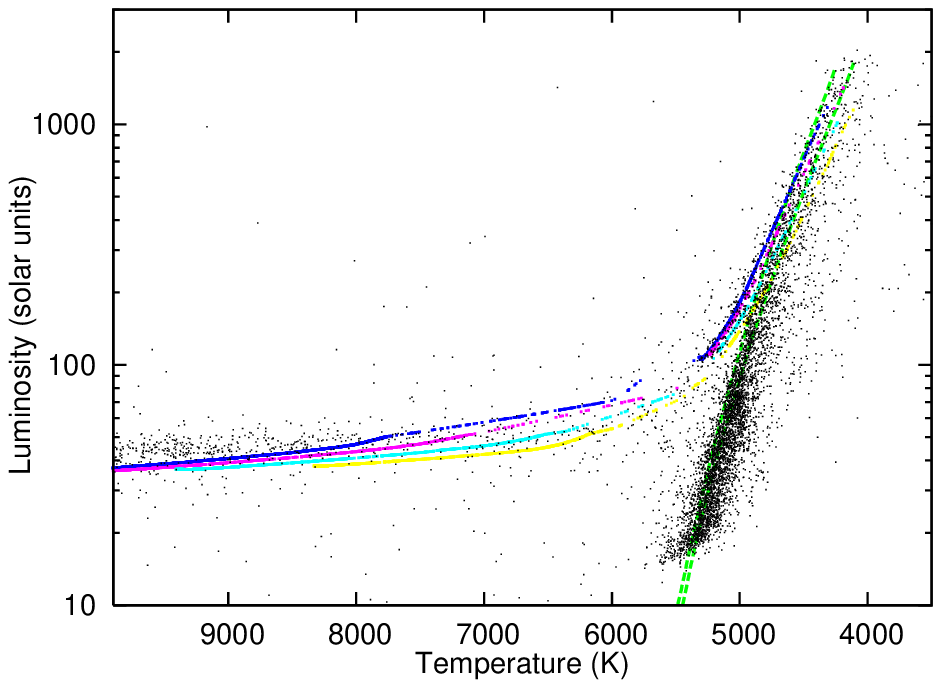}}
\caption[]{As Fig.\ \ref{IsoMFig}, using RGB isochrones and HB models generated using the Dartmouth database \citep{DCJ+08}. The top panel shows the same three RGB isochrones as Fig.\ \ref{IsoMFig}, with a solar-abundance-scaled HB track at 0.70 M$_\odot$. The middle panel shows the same diagram, using only the 12.1 Gyr isochrone, but with an \emph{approximate} correction to $E(B-V) = 0.08$ mag and a distance of 4900 pc, with HB models at (top to bottom) 0.65, 0.70, 0.75 and 0.81 M$_\odot$. The bottom panel is as the middle panel, but with [$\alpha$/Fe] enhancements of (bottom to top) 0.0, +0.2, +0.4 and +0.6; the RGB isochrones are shown for [$\alpha$/Fe] = 0.0 (left) and [$\alpha$/Fe] = +0.4 (right), both have helium abundances of $Y = 0.33$.}
\label{IsoDFig}
\end{figure}

Fig.\ \ref{IsoDFig} shows a similar plot, but for isochrones from the Dartmouth database \citep{DCJ+08}. In the top panel, we take a 12.1 Gyr model at [Fe/H] = --1.62, a standard helium abundance of $Y = 0.245 + 1.5 Z$ and an [$\alpha$/Fe] ratio scaled to solar abundances. We also include a HB model at $M = 0.70$ M$_\odot$, implying again that $\sim$0.11 M$_\odot$ is lost on the giant branch (the RGB tip corresponds to an initial stellar mass of 0.81 M$_\odot$ in this model).

By applying a reddening correction to $E(B-V) = 0.08 \pm 0.01$ mag and a distance of $4900 \pm 100$ pc, we can yield a good match to the isochrones: the RGB is matched nearly exactly, as is the AGB. The HB morphology is not well fit by solar-scaled helium abundance and [$\alpha$/Fe]. In the middle panel of Fig.\ \ref{IsoDFig}, we show the effect of varying mantle mass at solar [$\alpha$/Fe] with $Y = 0.33$. In the bottom panel, we show the effect of fixed stellar mass with varying [$\alpha$/Fe]: this affects the mass of the stellar core which, for fixed stellar mass, has the reverse effect on the mantle mass. We note that the errors given above are in that parameter only, however we can rule out a fit at $E(B-V) = 0.07$ mag and $d = 5100$ pc, or at $E(B-V) = 0.11$ mag and $d = 4500$ pc by examining the locations of the start of the AGB and the tip of the RGB.

It seems clear that variation of the mantle mass can account for the observed spread in the HB of $\omega$ Cen. This arises from a combination of varying core mass due to intrinsic metallicity and abundance differences. It can also arise from different efficiencies of mass loss on the RGB, which could well be due to the original metallicity and abundance differences themselves. At first sight, it is not clear whether varying mass loss or varying core mass is the primary factor. However, the spread of the early AGB is very narrow ($\sim$10--15\% in luminosity at a fixed temperature, 50--90 K in temperature at a fixed luminosity), most of which is due to temperature and luminosity errors of order 50--70 K, with small contributions from differential reddening across the cluster (implying, from Eq.\ (\ref{TCorrEqn}), $\Delta E(B-V) \ltsim 0.02$ mag) and varying distance within the cluster. It would therefore seem that we need a range of models that produce almost zero spread in the early AGB, a condition satisfied by changing the stellar mass, rather than [$\alpha$/Fe].

\begin{figure}
\resizebox{\hsize}{!}{\includegraphics[angle=270]{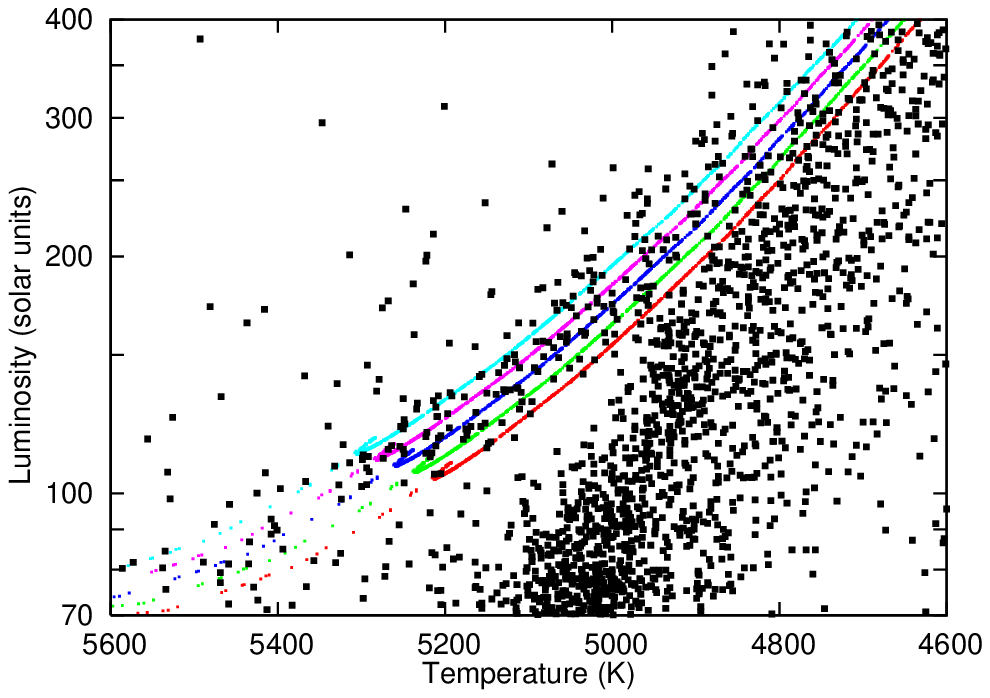}}
\caption[]{As Fig.\ \ref{IsoDFig}, middle panel, showing a Dartmouth model covering the early AGB for a stellar mass of 0.65 M$_\odot$, at [Fe/H] = --1.62 and [$\alpha$/Fe] = +0.2, with reddening (from left to right) of $E(B-V) =$ 0.11, 0.10, 0.09, 0.08 and 0.07 mag. The data are shown at $E(B-V) =$ 0.08 mag and $d = 4900$ pc.}
\label{IsoDEFig}
\end{figure}

We can further estimate the maximum value of $\Delta E(B-V)$ across the cluster. Fig.\ \ref{IsoDEFig} shows the effect of different values of $E(B-V)$ on the early AGB. In the absence of other variations leading to the dispersion of the AGB (which include photometric errors, and [$\alpha$/Fe], metallicity and distance variation), we can assume that the variation of $E(B-V)$ across the cluster, $\Delta E(B-V)$, is $< 0.04$ mag.

\begin{figure}
\resizebox{\hsize}{!}{\includegraphics[angle=270]{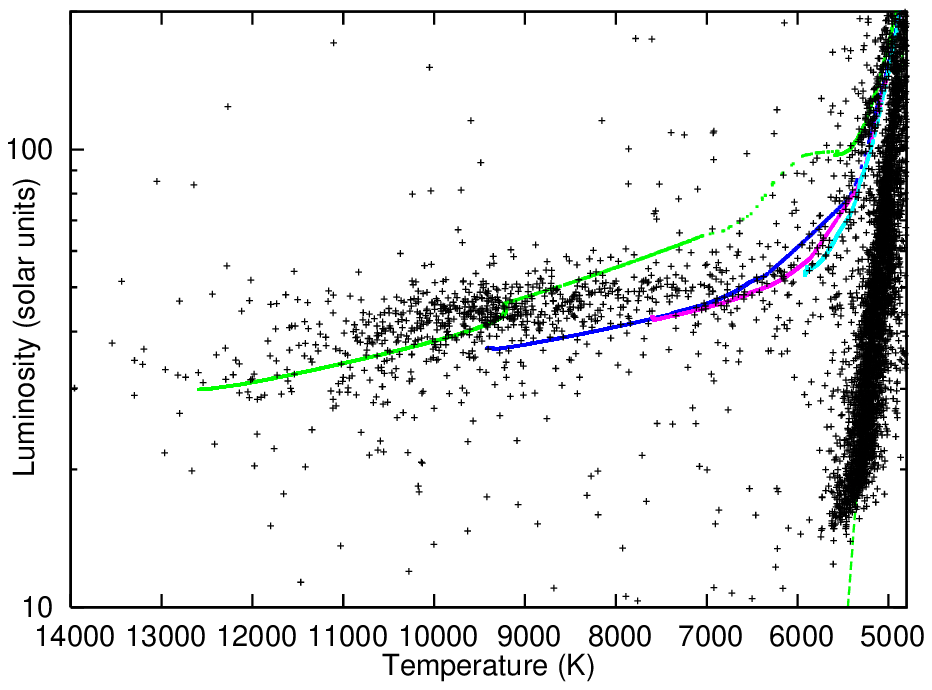}}
\caption[]{As Fig.\ \ref{IsoDFig}, middle panel, showing the HB for stellar masses of (top to bottom) 0.60, 0.65, 0.70 and 0.81 M$_\odot$, at [Fe/H] = --1.62 and [$\alpha$/Fe] = +0.4.}
\label{IsoDHBFig}
\end{figure}

By examining the HB (Fig.\ \ref{IsoDHBFig}), we can approximate the mass loss on the RGB. It would appear that the majority of the HB stars lie between the 0.60 and 0.65 M$_\odot$ models in both temperature and luminosity, but with some more closely following the higher-mass models. We must stress that our models do not cover the HB: while we expect temperatures to be broadly correct (within a few hundred to a thousand Kelvin), luminosities may be too low. We may also not cover the blue extent of the HB in its entirety, as bluer HB stars will be fainter at \emph{Spitzer}'s IR wavelengths than their redder counterparts.

A stellar mass of around 0.62--0.64 M$_\odot$ suggests that 0.18--0.20 M$_\odot$ is lost on the RGB, which is in keeping with \cite{Dotter08}, who suggests that the total mass lost on the RGB does not greatly depend on metallicity above [Fe/H] $\sim$ --2.

Globular cluster white dwarf masses (e.g.~\citealt{MKZ+04}) suggest that $\sim$0.3 M$_\odot$ must be lost on the RGB and AGB combined (the core mass in this case is 0.49 M$_\odot$), implying that around two thirds of the mass loss occurs on the RGB in $\omega$ Cen. Interestingly, we also expect about a third of the mass loss from the cluster to be in the form of chromospheric winds (see Section \ref{FateSect}), implying at least half the dusty mass loss occurs on the RGB.

Some concerns have been raised that the HB models from the Dartmouth models may be marginally too warm for low-metallicity stars (A.\ Dotter, private communication). Nevertheless, given the much more precise fit to the data, we would advocate the Dartmouth model values over those derived from \cite{MGB+08}.

\subsection{Victoria-Regina isochrones and ZAHB models}

\begin{figure}
\resizebox{\hsize}{!}{\includegraphics[angle=270]{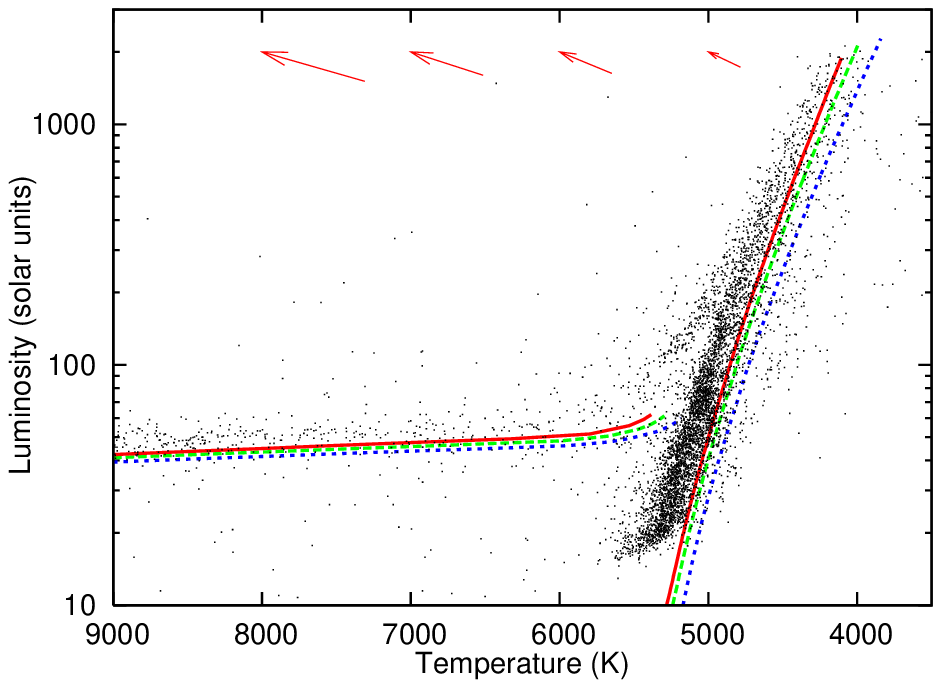}}
\resizebox{\hsize}{!}{\includegraphics[angle=270]{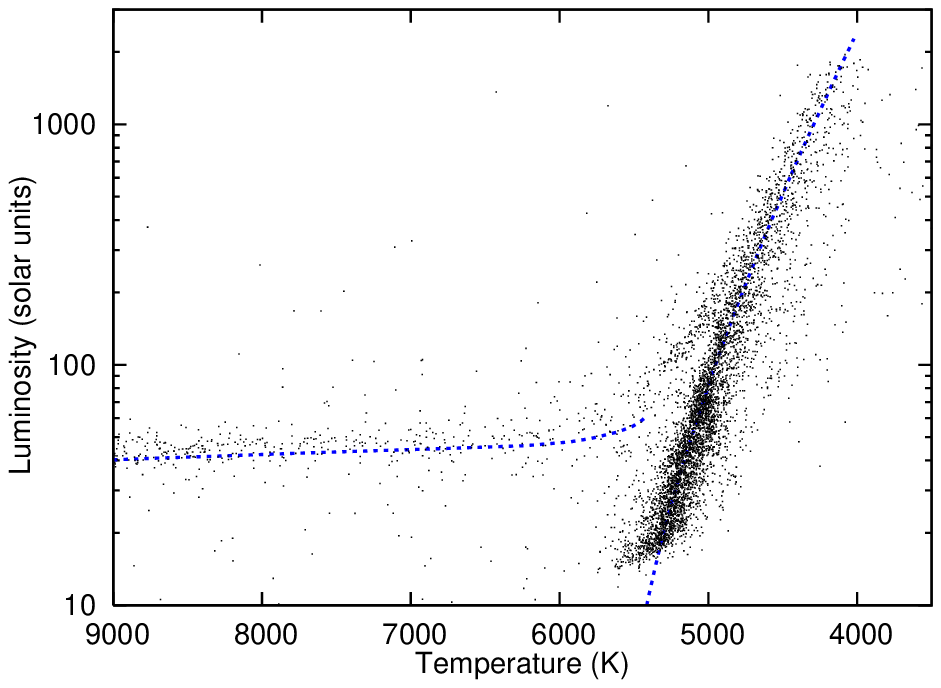}}
\caption[]{As Fig.\ \ref{IsoMFig}, using Victoria-Regina isochrones and ZAHB models from \cite{VBD06}. The top panel shows a 12 Gyr isochrone and ZAHB model for [Fe/H] = --1.61, with [$\alpha$/Fe] = +0.0 (solid line), +0.2 (dashed line) and +0.4 (dotted line). The bottom panel shows the fitted model at [$\alpha$/Fe] = +0.4, still at 12 Gyr, with E(B--V) = 0.09 mag and $d = 4800$ pc, with the model temperatures increased by 4.7\% (0.02 dex).}
\label{IsoVFig}
\end{figure}

We have performed a similar analysis using the Victoria-Regina (VR) isochrone and ZAHB models of \cite{VBD06}, which we show in Fig.\ \ref{IsoVFig}. Here, we take the models at [$\alpha$/Fe] = +0.0, +0.2, +0.4 for [Fe/H] = --1.61 at 12 Gyr. No values of reddening or distance, nor any sensible value of age can accurately match the VR isochrone on the RGB to our data for any value of [$\alpha$/Fe]. In order to achieve a fit, we must either increase the temperature of the VR model, or decrease the temperatures of our stellar data, by around 0.01 dex (2.3\%) in the case of [$\alpha$/Fe] $\sim 0$ or 0.02 dex (4.7\%) in the case of [$\alpha$/Fe] $\sim +0.4$. The value of [$\alpha$/Fe] is suggested to be around +0.3 for cluster stars with [Fe/H] $\ltsim -1$ \citep{Pancino03}. Under this assumption, we find a reddening of around $E(B-V) = 0.08$ mag and a distance of $4800 \pm 100$ pc.

\begin{figure}
\resizebox{\hsize}{!}{\includegraphics[angle=270]{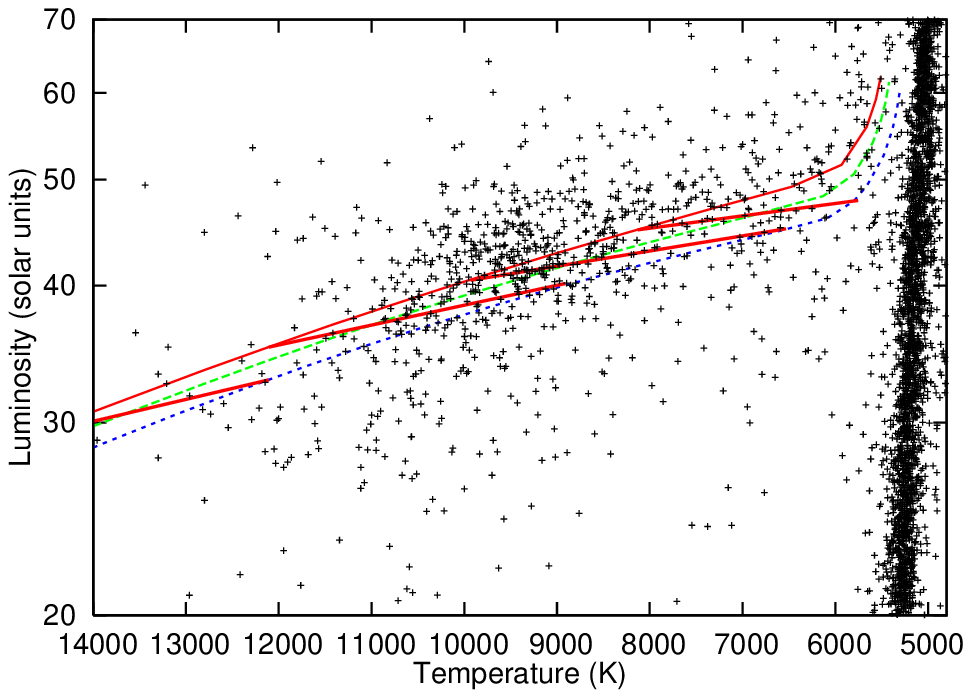}}
\caption[]{As Fig.\ \ref{IsoDHBFig}, with $d = 4750$ pc and $E(B-V) = 0.08$ mag, showing the ZAHB models from the Victoria-Regina models, at [Fe/H] = --1.61 and [$\alpha$/Fe] = +0.0, +0.2 and +0.4 (as in Fig.\ \ref{IsoVFig}). Thicker lines show the stellar masses on the ZAHB at 0.575, 0.600, 0.625 and 0.650 M$_\odot$.}
\label{IsoVRHBFig}
\end{figure}

On the assumption that the ZAHB location for the majority of metal-poor stars is well-represented by the clump around 8500--11\,000 K, and that [$\alpha$/Fe] $\approx +0.3$, we find that most stars have a ZAHB mass of around $0.61 \pm 0.02$ M$_\odot$ (Fig.\ \ref{IsoVRHBFig}). Obviously, this issue is clouded by HB evolution and lack of fully accurate measurements of temperatures and luminosities in this area, but the figure is broadly consistent with that from the Dartmouth models above, predicting only a slightly lower stellar mass. The 12 Gyr VR model predicts that the initial mass of RGB tip stars is 0.85 M$_\odot$ and that the core mass is 0.49 M$_\odot$ at $\alpha = +0.3$. This suggests that $\approx$0.24 M$_\odot$ is lost on the RGB and $\sim$0.12 M$_\odot$ on the AGB, if the entire envelope is ejected. This ratio of 2:1 is consistent with that derived from the Dartmouth models.

\subsection{BaSTI isochrones}

\begin{figure}
\resizebox{\hsize}{!}{\includegraphics[angle=270]{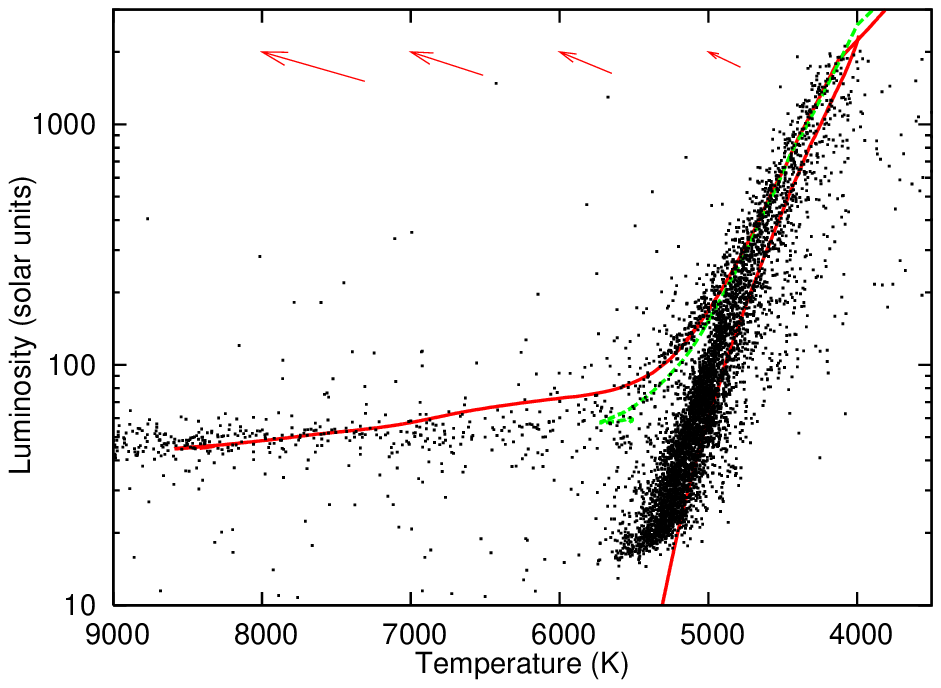}}
\resizebox{\hsize}{!}{\includegraphics[angle=270]{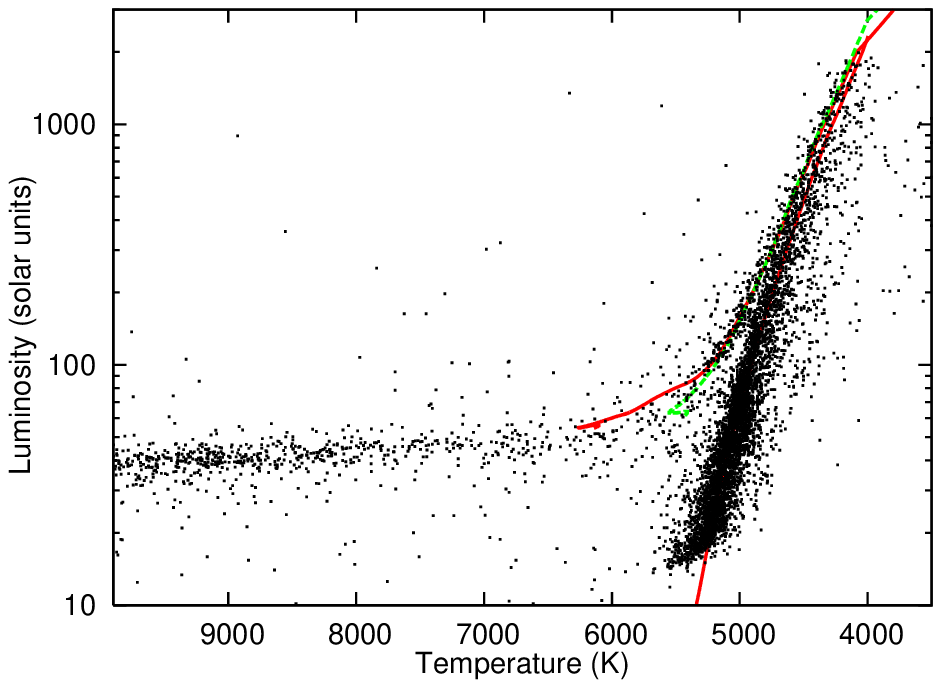}}
\caption[]{As Fig.\ \ref{IsoMFig}, using BaSTI isochrones. The top panel shows the same three RGB isochrones as Fig.\ \ref{IsoMFig}, with the $\alpha$-enhanced AGB-extended BaSTI model for $Z = 0.001$ ([Fe/H] = --1.62), at 12 Gyr. The bottom panel shows the fitted model, at 9.5 Gyr, with E(B--V) = 0.06 mag and $d = 4900$ pc. The solid lines assume Reimers' $\eta = 0.4$; the dotted lines show the AGB for $\eta = 0.2$.}
\label{IsoBFig}
\end{figure}

We show a fit using the BaSTI models (`A bag of stellar tracks and isochrones'; \citep{PCSC06}) in Fig.\ \ref{IsoBFig} (top panel). As before, we take a 12 Gyr model at [Fe/H] = --1.62 at a helium content of $Y = 0.246$. We use the $\alpha$-element enhanced models with $\eta = 0.2$ and $0.4$ \citep{Reimers75}, incorporating BaSTI's synthetic AGB evolutionary tracks from \cite{IT78}.

We note that, although the synthetic HB and AGB data fit the empirical cluster data well, the RGB is significantly cooler and under-luminous in the BaSTI model. The only way we can find to rectify the large gap between the early AGB and RGB is by assuming the cluster is much younger than has previously been calculated, at an age of $\ltsim 9.5$ Gyr. Taking models at this age, we found the reddening to only $E(B-V) = 0.06$ mag and $d = 4900$ pc.

The young age of the fitted model is at odds not only with the luminosities of the RGB and AGB tips, but with previous age estimations of typically 11--14 Gyr, derived using a variety of data and methods including isochrone fitting to colour-magnitude diagrams including stars down to $\sim 0.15$ M$_\odot$ \citep{TKP+01,KTK+02,CK02,Pancino03,PWH+03,HKRW04}. BaSTI models this old do not have the narrow separation between the RGB and AGB. Another implication of this is that mass-loss efficiency must range from $\eta \approx 0.2$ to $\eta \gg 0.4$ (probably $\approx 0.7$ to $0.8$ and perhaps higher for the Extreme HB, which we do not model). This level of mass-loss efficiency is not considered to be impossible, and very similar values have been suggested for $\omega$ Cen in the past \citep{DCDROC95}.

\subsection{Summary of isochrone fitting}

In summary, we find that the Dartmouth models provide the best fit to our data, although the models of \cite{MGB+08} may provide a similarly good fit if the mass-loss efficiency could be increased. The BaSTI models provide too small a temperature difference between the early RGB and AGB, which would appear unlikely to be due to a systematic error in our data, and can consequently only fit our data with a significantly decreased age. The Victoria-Regina models have an effective temperature which clearly differs from our data by up to 5\% in both the isochrones and the ZAHB models. 

Taken in combination, the models consistently suggest a distance to the cluster of around 4850$\pm$200 pc, and an average reddening to the cluster of $E(B-V) = 0.08 \pm 0.02$ mag. These values are both at the lower limit of those derived in the literature, which roughly span the ranges of 5160$\pm$360 pc and $E(B-V) = 0.12 \pm 0.03$ mag (see Section \ref{SEDSect} for references). The distance is very close to the 4740$\pm$160 pc in vL+00, whose data we use here. However, we note that this was derived using velocity dispersion data, not the photometry we use here, and is thus nearly independent. We cannot entirely rule out a distance of 5000 pc, though this fits the start of the early AGB less well, and requires a reddening of only $E(B-V) = 0.05$ mag. Conversely, a reddening of $E(B-V) = 0.12$ mag would suggest a distance of $\sim$4600 pc.

The potential for systematic error in our temperatures and luminosities (Section \ref{SEDSect}) is relatively small for stars within the range 4000--6500 K, and increases as one moves away from this region. We estimate that, excluding uncertainties in distance, and average and differential reddening, systematic differences on the RGB and AGB can be limited in temperature to $<$ 1\% and in luminosity to $\ltsim$ 5\%. Very red stars and HB stars have the potential for higher systematic errors due to the lack of {\sc marcs} models covering these temperature ranges. This has the potential to make a systematic offset to our distance of up to $\pm$120 pc and of up to $\pm$0.015 mag in $E(B-V)$. The 10--15\% discrepancy in our reddening correction found in Section \ref{SEDSect} and the assumption that the foreground ISM follows the approximation $A_{\rm V} / E(B-V) \approx 3.05$ probably inflates the systematic error in $E(B-V)$ to 0.020 mag. Treatment of convection by the stellar evolution models may also cause a systematic offset in the distance and reddening we determine to the cluster. However, all four models include physics describing convective core overshoot. The exact physics employed does not appear to have made a significiant difference to our results, as the distance we estimate to the cluster is roughly the same with all four models.

As interstellar extinction has a larger effect at bluer wavelengths, the alteration of $E(B-V)$ does not much affect the values we derive for our mass-losing stars, which are typically cooler than the giant branch, although a distance of 4850 pc would mean that their luminosities would still decrease by 6\% due to the distance correction. The net effect would be a decrease in our total mass-loss rates by 4.5\%, though this is included in the errors we list in Sections \ref{MdotV6V42Sect} \& \ref{TotalMdotSect}.

From arguments stemming from the location of evolutionary sequences on the HRD, we conclude that most of a cluster star's mass loss must occur near the tip of the RGB, rather than on the AGB. The HB morphology of the cluster requires that typically 0.20--0.25 M$_\odot$ of stellar atmosphere is lost from the (initially) 0.85 M$_\odot$ stars on the RGB. Stars then attempt to lose a further 0.05 M$_\odot$ or more on the AGB, but many may have insufficient atmospheric mass to lose and thus become post-early-AGB stars or even AGB-\emph{manqu\'e} stars (`failed' AGB stars -- see, e.g.~\citealt{OConnell99}, Section 6.2). That mass loss can work this efficiently at metallicities as low as [Fe/H] $\sim -1.6$ suggests that metallicity and mass loss have only a weak dependence in this metallicity regime.

\section{Deriving mass-loss rates}

\subsection{Mid-IR spectra and additional data of V6 and V42}

Spectra were taken of the two most extreme red giants in the cluster: V6 (LEID 33062, ROA 162) and V42 (LEID 44262, ROA 90). Taken with the T-ReCS spectrograph on the Gemini South Telescope \citep{DBF05}, the data are between 8 and 13 $\mu$m and have a resolving power of $R = 300-600$, along with $N^\prime$-band (11 $\mu$m) acquisition images, on each of the nights of 2007 August 16 \& 18, for V6; and on 2007 August 27 \& 28, for V42. Two spectra of 2.5 hours integration were taken each night, using the low resolution 10-$\mu$m grating, along with the 0.72'' slit. The full-width half-maximum image width of the standard star $\theta$ Cen in the $N^\prime$-band acquisition images was $\sim$0.44''.

The spectra were reduced using the {\sc iraf} package\footnote{{\sc iraf} is distributed by the National Optical Astronomy Observatories (NOAO), which is operated by the Association of Universities for Research in Astronomy, Inc., under co-operative agreement with the National Science Foundation.} designed for T-ReCS using the default settings. No wavelength offsets were observed, and differences among the spectra were small, so a flux-density-weighted average was taken to provide a single, high-signal-to-noise spectrum for each target. An absolute flux density for each target was estimated by comparing the acquisition images of the target stars and $\theta$ Cen using aperture photometry. Assuming an $N^\prime$-band flux density of 45.4 Jy for $\theta$ Cen\footnote{Extrapolated from http://www.gemini.edu/sciops/instruments/ \\ miri/T-ReCSBrightStan.txt}, a flux density of 203$\pm$12 mJy was calculated for V6 and 91$\pm$16 mJy for V42. The acquisition field for V42 also included the mid-M-type irregularly-variable red giant star V152 (LEID 44277), which is a confirmed cluster member (vL+00, vL+07). By comparison to V42, we estimate its $N^\prime$-band flux density is 36$\pm$8 mJy. Comparison with the 8- and 24-$\mu$m \emph{Spitzer} data shows this value to be consistent with a mid-infrared excess, but the error on the \emph{Gemini} photometry is too large to constrain whether a 10-$\mu$m silicate feature is present in V42.

\begin{figure}
\resizebox{\hsize}{!}{\includegraphics[angle=270]{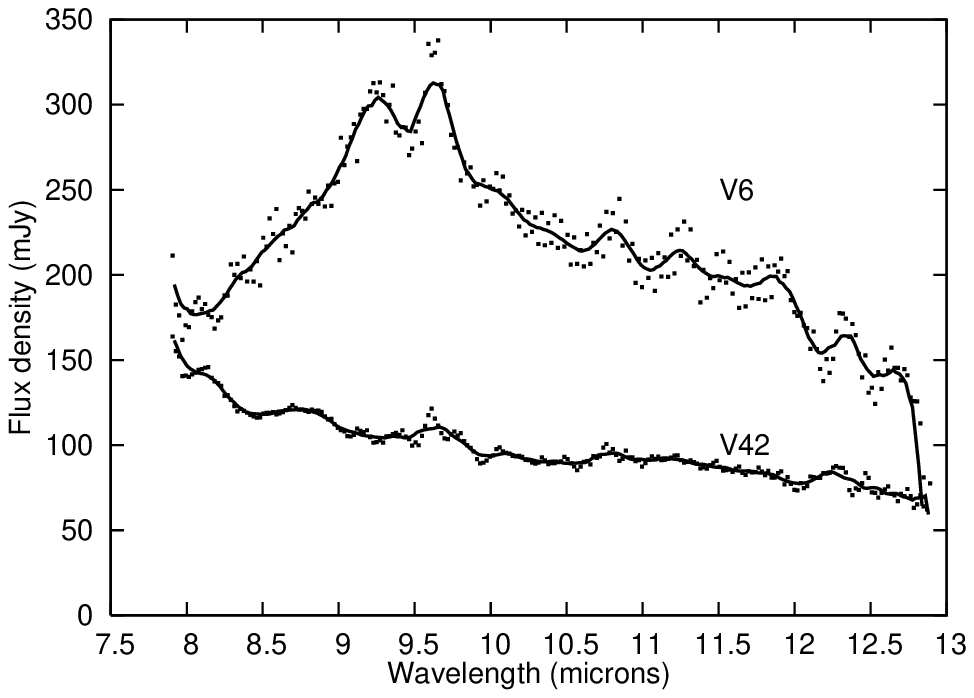}}
\caption[]{Flux-density-calibrated Gemini T-ReCS spectra of the two extreme red giant stars $\omega$ Cen V6 and V42, overlaid with a running boxcar average of 10 pixels.}
\label{SpectraFig}
\end{figure}

The flux-density-calibrated spectra for both stars are shown in Fig.\ \ref{SpectraFig}. Clear fringing in the spectrum of V6 can be seen. Attempts to correct for this by moving each spectrum by an integer number of pixels failed to improve this. We note that many of the `features' in the spectra are also the result of imperfect atmospheric correction, though the presence of 9.5-$\mu$m emission in V6 is certain.

In the case of V6, additional optical photometry was available from \cite{CS73}, \cite{LE83}, and \cite{Clement97}. Near-IR data from \cite{PCM+80} were also used. Of particular use were the works by Dickens et al.~(1972), and \citet{GF73,GF77}, which not only provide flux densities, but variability amplitudes, allowing us to see the temperature and luminosity changes in the star from $U$- to $L$-band. We have previously obtained a 2dF optical spectrum for this star (published in vL+07), where it was determined to have a temperature of $\sim$3500 K (the lowest-temperature model).

\begin{figure}
\vspace{3.8cm}
\resizebox{5.8cm}{!}{\includegraphics[angle=0]{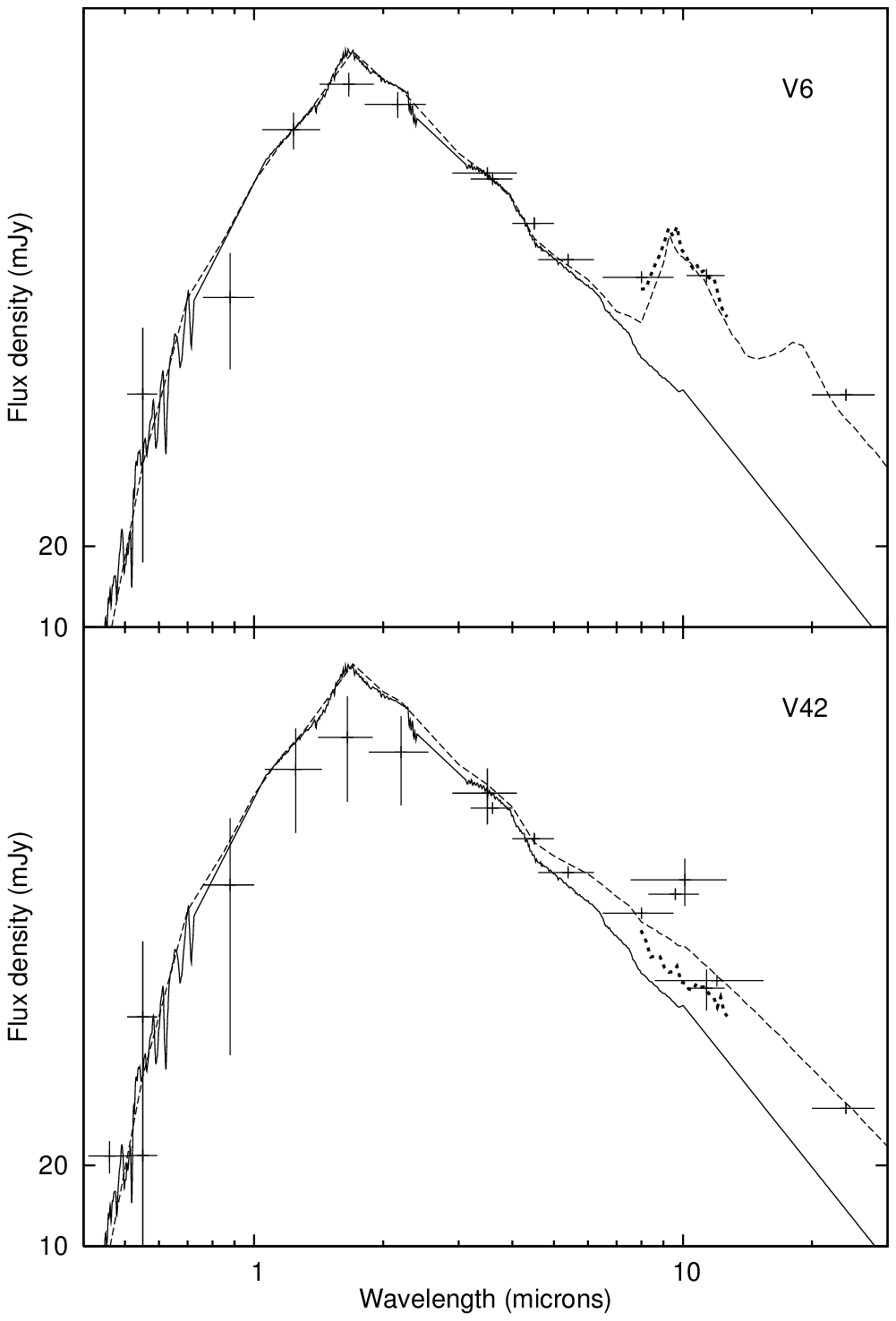}}
\caption[]{SED models of V6 (top) and V42 (bottom), showing {\sc marcs} models smoothed to $R = 200$ (solid lines), smoothed Gemini T-ReCS spectra (dotted lines) and literature photometry (crosses with error bars). The error bars in the latter reflect either the photometric variability or the measurement error (see text for details). The dashed curve shows the corresponding {\sc dusty} model, fitted to the Gemini spectrum and 24-$\mu$m \emph{Spitzer} datum in the case of V6 and the literature photometry, including \emph{Spitzer} data, in the case of V42.}
\label{SEDsFig}
\end{figure}

For V42, the same papers provided information on flux densities and variability. Additional data were available from the DENIS survey (Deep Near Infrared Survey of the Southern Sky; \citealt{FCC+00}); \cite{MW85}, which includes variability information; and 9.6-, 10- and 12-$\mu$m photometric data points (\citealt{OFP95}; \citealt{OFFPR02}, hereafter O+02). In addition to a 2dF spectrum, we also have an optical VLT/UVES spectrum for V42 (published in MvL07), though our previous temperature estimates have not been particularly reliable as a result of veiling by emission dissipated in the pulsation shock. The resulting SEDs for V6 and V42 are shown in Fig.\ \ref{SEDsFig}.

\subsection{Deriving a mass-loss rate for V6 and V42}
\label{MdotV6V42Sect}

\subsubsection{Mass loss of V6 from Gemini spectroscopy}
\label{MdotV6Sect}

Estimates of the mass-loss rate for the two stars V6 and V42 were determined using the {\sc dusty} modelling code \citep{NIE99}. Our Gemini T-ReCS spectrum of V6 (Fig.\ \ref{SpectraFig}) shows clear emission from silicate dust grains around 10 $\mu$m, typical of well-developed dusty winds seen around AGB Mira variables in the Solar Neighbourhood (e.g.~\citealt{SBSH00}). The spectrum is not well fit by the classic `astronomical' silicates alone \citep{DL84}, especially the 9.5-$\mu$m peak, regardless of the dust properties used. Our best fit was for 65\% (by number of grains) `astronomical' silicates \citep{DL84}, 15\% compact Al$_2$O$_3$ (optical constants from Jena database\footnote{http://www.astro.uni-jena.de/Users/database/entry.html}), and 10\% each of glassy and crystalline silicates \citep{JMB+94} (Fig.\ \ref{SEDsFig}). Although we have used these to produce the best fit, we do not claim that these are necessarily the particular species of silicates involved, nor that they are present in these proportions. We assume a radiatively-driven wind and a standard MRN grain size distribution \citep{MRN77}, where the number of grains of size $a$ is given by $n(a) = a^{-q}$, where we here take $q = 3.5$ over the range $a = 0.005 - 0.05$ $\mu$m. We derive an inner edge to the dust envelope at a temperature of 650 K, and a $V$-band optical depth of $\tau_{\rm V} = 0.08$ for V6. This dust temperature is relatively cool, and could potentially be related to mid-IR variability, which we discuss in Section \ref{VarySect}. An increase in the dust temperature to a more typical 1000 K would decrease the mass-loss rate by a factor of nearly two (Section \ref{DustTSect}).

The maximum grain size also has a weak effect on the dust mass-loss rate. Here, we have chosen 0.05 $\mu$m, which we believe is representative of this metal-poor environment, which may be subject to substantial UV flux from other cluster members. This also provides a marginally better fit to the region around 11 $\mu$m in our V6 spectrum (c.f.~\citealt{VH08}, Fig.\ 1). Our value is, however, less than the usual value of 0.1 $\mu$m, or higher, assumed in the literature (e.g.~\citealt{PP83}). Increasing the maximum grain size to 0.1 $\mu$m would decrease the mass-loss rate by $\sim$20\%, a factor that does not change much as one further increases the parameter. While theoretically possible to measure, it is very difficult to constrain the grain sizes involved purely from the SED alone \citep{CBM04}.

The dust mass-loss rate and total mass-loss rate from a star are related to the {\sc dusty} output by the following equations \citep{NIE99}:
\begin{equation}
	\dot{M}_{\rm dust} = \frac{\dot{M}_{\rm DUSTY}}{200} \left(\frac{L}{10^4}\right)^{3/4} \left( \frac{\psi}{200} \right) ^{-1/2} \left( \frac{\rho_{\rm s}}{3} \right) ^{1/2}	,
	\label{MdotDustDustyEqn}
\end{equation}
\begin{equation}
	\dot{M} = \dot{M}_{\rm dust} \psi ,
	\label{MdotTrueDustyEqn}
\end{equation}
where $L$ is the luminosity in solar units, $\psi$ is the gas-to-dust ratio and $\rho_{\rm s}$ is the bulk grain density in g cm$^{-3}$ (we assume it to be 3 g cm$^{-3}$ for silicates; e.g.~\citealt{Suh99}). 

Eq. (\ref{MdotTrueDustyEqn}) holds when condensation into grains occurs efficiently. We expect much of the condensible material to form silicate grains, as iron will typically form inclusions, rather than competing with silicon. We therefore expect silicate grain growth to scale inversely linearly with the metallicity, and specifically the abundance of silicon. Eq. (\ref{MdotTrueDustyEqn}) may not extend to stars where chromospheric mass-loss is important and dust production is not efficient, where the dust-to-gas ratio is expected to follow a square-law dependence on metallicity \citep{vL00,vL06}. We assume here that $\psi = 200$ for solar composition stars, which for V6 we take as [Fe/H] = $-$1.19 (\citealt{ZW84,NdC95b}; Vanture et al.~2002).

These mass-loss rates are uncertain due to:
\begin{itemize}
\item 30\% uncertainty in the dust mass-loss rate from internal errors in {\sc dusty};
\item $\ltsim$ 30\% uncertainty in the dust mass-loss rate due to uncertain dust chemistry, temperature and grain size;
\item 11\% and 6\% uncertainty in the photometric \emph{excess} above the SED at 8 and 24 $\mu$m, respectively;
\item $\sim$ 3\% uncertainty in the dust mass-loss rate due to inaccuracies in the stellar temperature;
\item $\ltsim$ 5\% uncertainty due to error in the integrated flux density from the SED;
\item 22\% uncertainty due to error in luminosity;
\item $\sim$ 5\% uncertainty due to error in the reddening;
\item $\sim$ 12\% uncertainty due to error in the distance.
\end{itemize}

When added in quadrature, this gives a total error of $\ltsim$ 52\% in the \emph{dust} mass-loss rate from V6, and is also based on the following assumptions, which we will discuss in Sections \ref{VarySect} \& \ref{MdotRGBSect}:
\begin{itemize}
\item that the mid-IR spectrum V6 has not varied substantially between the 8- and 24-$\mu$m exposures;
\item that $\psi \propto  10^{-{\rm [Fe/H]}}$;
\item that the wind velocity scales as prescribed in the {\sc dusty} code: $v \propto L^{1/4} \psi^{-1/2}$;
\item that the dust is coupled to the gas.
\end{itemize}
We thus find that for V6, under the above assumptions, $\dot{M}_{\rm dust} = 3.8 \pm ^{2.0}_{1.3} \times 10^{-10}$ M$_\odot$ yr$^{-1}$ and $\dot{M}_{\rm total} = 1.1 \pm ^{0.6}_{0.4} \times 10^{-6}$ M$_\odot$ yr$^{-1}$.

This total mass-loss rate for V6 is comparable to our earlier measurements in 47 Tuc \citep{vLMO+06}, though it is generally higher than more metal-rich, solar-mass Mira variables in the Solar Neighbourhood \citep{JK92}. It is also slightly inflated from those derived in B+08 purely on the basis of the \emph{Spitzer} data and the models of \cite{Groenewegen06}. We note that the above values and quoted errors depend on the invariance of V6 between the acquisition times of the \emph{Spitzer} IRAC and MIPS photometry, and of the Gemini T-ReCS spectrum.

\subsubsection{Mass loss of V42 from Gemini spectroscopy}
\label{VarySect}

The mass-loss rate of V42 is more complicated to derive, as there are considerable variations in the mid-IR spectrum, plus the metallicity of the star is unknown. The literature data on V42 (see Table \ref{VariabilityTable}) appear to show a clear silicate feature, which we have fitted using {\sc dusty}. We have set the dust envelope's inner edge to be at 1200 K, near the temperature limit for SiO condensation under normal circumstances \citep{NF06}. The fitted model has an optical depth of $\tau_{\rm V} = 0.25$, with the remaining input parameters staying the same as the model for V6. Assuming a metallicity of [Fe/H] = --1.62, this yields rates of $\dot{M}_{\rm dust} = 8.6 \times 10^{-11}$ M$_\odot$ yr$^{-1}$ and $\dot{M}_{\rm total} = 7.2 \times 10^{-7}$ M$_\odot$ yr$^{-1}$.

The Gemini spectrum shows excess above the Rayleigh-Jeans tail that could be attributable to dust. The rise towards 8 $\mu$m could represent the SiO fundamental band in emission, but this is near the edge of the atmospheric window and thus uncertain. The absence of a silicate feature in the spectrum is puzzling, given the clear feature apparently present in the literature broadband photometry (see Table \ref{VariabilityTable} \& Fig.\ \ref{SEDsFig}). These historical estimates for the mid-IR brightness of V42 have been high. Previous estimates of its mass loss have suggested a mass-loss rate of 7--10 $\times$ 10$^{-7}$ M$_\odot$ yr$^{-1}$, with a 200--350 K dust envelope, and it has been suggested that this is associated with a period of dust ejection around 1925, possibly linked to a thermal pulse event (Origlia et al.~1995; O+02). This is similar, though cooler than, the mass-loss rate estimated from the {\sc dusty} fit to our \emph{Spitzer} and Gemini data. However, what is obvious from Table \ref{VariabilityTable} is the factor of 2.4 range in the 10-$\mu$m flux received from V42 since 1994. The obvious solution would be that this is connected with a periodic phenomenon linked to the pulsation cycle: there is not an obvious link, however, with pulsation cycle phase, but neither is there sufficient evidence to refute one. The amplitude of variation is also somewhat larger than the factor of 1.6 change in the $L$-band flux observed by Dickens et al.~1972. Note that all flux values in Table \ref{VariabilityTable} show mid-IR excesses, strongly suggesting the presence of circumstellar dust.

\begin{center}
\begin{table}
\caption[]{Mid-IR photometry of $\omega$ Cen V42.}
\label{VariabilityTable}
\begin{tabular}{llllll}
    \hline \hline
Phase$^1$& Date       & Ref.$^2$	&	Instru-	& $\lambda^3$		& Flux$^4$ \\
\        & \          & \        &  ment    	& ($\mu$m)	& \  \\
    \hline
0.00	& 02 Jun 94		& a & TIMMI		& 10	& 3.1\\
\  	& Feb/Aug 97	& b & ISOCAM 	& 9.6	& 2.5\\
\  	& Feb/Aug 97	& b & ISOCAM 	& 12	& 1.8\\
29.87	& 02 Mar 06		& c & MIPS 		& 24	& 1.9\\
30.03	& 26 Mar 06		& c & IRAC 		& 8	& 1.6\\
30.87	& 29 Jul 07		& d & IRS 		& 5--37	& 2.3\\
31.07	& 27 Aug 07		& e & T-ReCS	& 8--13	& 1.3\\
    \hline
\multicolumn{6}{p{0.45\textwidth}}{\small Notes: (1) Phase in pulsation periods from $I$-band maximum, taken to be MJD 49506 after \cite{DFLE72}, with $P$ = 148.64 days (vL+00). (2) References: a -- \cite{OFP95}; b -- O+02; c -- B+08; d -- G.~Sloan, private communication; e -- this work. (3) Ranges denote spectra. (4) Flux as a multiple of the photospheric contribution, extrapolated from spectral models; flux calculated using a $F \propto \lambda^{-1.7}$ approximation for photometric observations that are not at 10 $\mu$m. \normalsize} \\
\end{tabular}
\end{table}
\end{center}

Surprisingly for this M-type star, the emission from the \emph{Spitzer} photometry and Gemini spectrum are well-fit by an \emph{amorphous carbon} wind (following \citealt{Hanner88}) with a constant temperature of 600 K and a mass-loss rate of between 1.7--4.3 $\times$ 10$^{-10}$ M$_\odot$ yr$^{-1}$ in dust and 1.4--3.6 $\times$ 10$^{-6}$ M$_\odot$ yr$^{-1}$ in total. The variance here comes from the uncertainty in the absolute flux calibration of the spectrum; and  the interpretation of carbon-rich dust also holds for the \emph{Spitzer} IRS spectrum. Note that this figure is some 2--4 times greater than that derived from fitting \emph{silicate} dust to the literature photometry. The carbonaceous nature of the wind is also puzzling, as this is not a \emph{bone fide} carbon star (vL+07).

It is clear that our Gemini spectrum, and a similar \emph{Spitzer} IRS spectrum taken four weeks prior (G. Sloan, private communication) show no silicate emission. What is interesting is that, although the spectrum does not change appreciably between the two epochs, the \emph{flux} received from V42 decreases considerably, mirroring the wide range in the historic photometry measurements. We discuss possible reasons for this in Section \ref{V42Sect}.

The key consequence of the mid-IR variability observed in V42 is that it now becomes virtually impossible to accurately pin down a mass-loss rate to better than a factor of $\sim$2 without simultaneous observations of the flux at $\ltsim$ 4 $\mu$m, where the dust component does not dominate, to define the stellar temperature and luminosity.

With respect to the rest of our data, it would appear that the accuracy of the mass-loss rates of other stars may depend on their optical variability, plus the optical properties of their circumstellar dust. This is a potential source of error in our analysis, but given that V6 and V42 are by far the most extreme pulsators, it would appear that V6 is the only other likely candidate for significant mid-IR variability. Our only comparison here is between the \emph{Spitzer} photometric data and the Gemini spectrum, but would appear to limit the effect to a level much less than the known errors present in our mass-loss analysis.

\subsection{Derivation of mass-loss rates for other stars}

\subsubsection{Correction of systematic differences between Spitzer photometry and model atmospheres}
\label{SystematicSection}

\begin{figure}
\resizebox{\hsize}{!}{\includegraphics[angle=270]{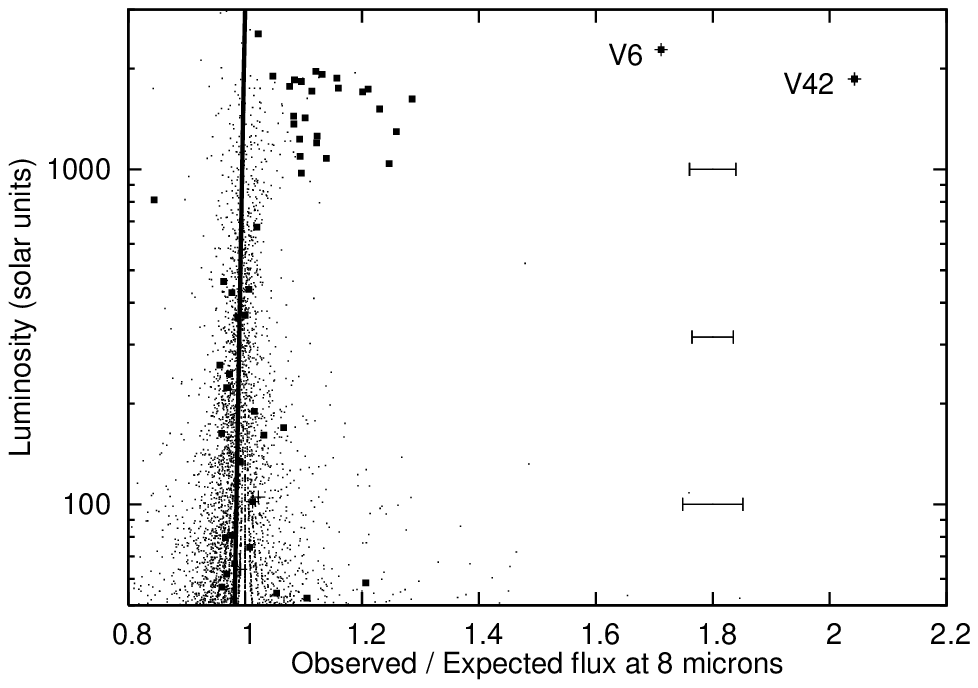}}
\resizebox{\hsize}{!}{\includegraphics[angle=270]{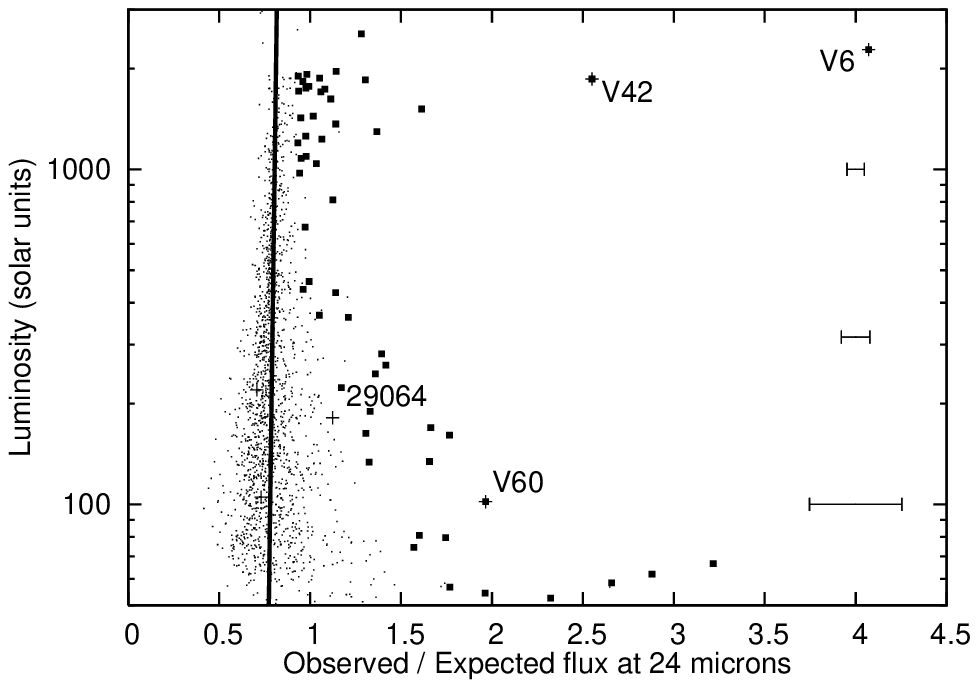}}
\caption[]{Observed mid-IR excess (attributable to dust) at 8 and 24 $\mu$m, derived from fitting to model atmospheres. Over-plotted is the line we have taken marking `zero' dust mass loss from Eqs.~(\ref{CorrEqun1}) and (\ref{CorrEqun2}). Representative 1$\sigma$ error bars are shown to the right of the plots -- stars which have mid-IR excess above the 3$\sigma$ level at 24 $\mu$m are marked by large squares.}
\label{OEFig}
\end{figure}

Fig.\ \ref{OEFig} shows the distributions of excess above the model at 8 and 24 $\mu$m, respectively. Clearly, V6 and V42 are the most excessive stars, although there are a few others exhibiting excesses at both 8 and 24 $\mu$m. Those objects with flux excesses at low luminosities are likely background galaxies or mis-identifications. 

There is an observable offset between the model atmospheres provided by {\sc marcs} and the 8- and 24-$\mu$m photometry data for the warmer, less luminous stars (the giant branch is offset from unity towards the bottom of Fig. \ref{OEFig}). This effect is only $\sim$1\% for the brightest stars, but increases to $\sim$4\% and $\sim$10\% for the fainter stars at 8 and 24 $\mu$m, respectively. The reason behind this is unclear, as from a modelling perspective one would assume that the Rayleigh-Jeans tail that has been used in extending the model atmospheres out to the limit of the 24-$\mu$m filter would be more accurate in a warmer atmosphere. A silicate absorption feature caused by interstellar material (c.f.~\citealt{EGS+02}) can also be ruled out, as it is only expected to affect our 8- and 24-$\mu$m photometry by about 0.5\% in each case, based on a 100 K interstellar medium at $A_{\rm V} = 0.4$ mag. A combination of SiO and H$_2$O absorption in the region of the 24-$\mu$m filter bandpass may affect the cooler giants. It may be that this under-luminosity in the fainter stars is a result of imperfect data reduction of the \emph{Spitzer} MIPS data near the sensitivity limit of the instrument.

While this correction is not significant on the level of individual stars, it is important when considering the cluster as a whole: without a correction for this effect, the total mass-loss rate for the cluster would be grossly incorrect. In order to minimise the effects of statistical scatter in the photometry, we have assumed that stars with less-than-expected 8- and 24-$\mu$m fluxes have negative mass-loss rates. This is unphysical, but by making this assumption we can examine stars with low mass-loss rates in a statistical manner, with a minimum bias from random uncertainty in the photometry.

To calculate the correction, we have assumed that the average star produces negligible dust, therefore will have no flux excess. We then split the data into luminosity bins of 0.1 dex, and for each, calculate the average offset. These form a tight correlation, except at the lower and upper limits, where poor-quality data and significantly dusty atmospheres (respectively) deviate the median. Taking the intermediate points (representing 21 bins, containing 1624 stars at 8 $\mu$m and 1207 stars at 24 $\mu$m), we have computed a best-fit line and removed this from the data. These corrections are:
\begin{equation}
	\left(\frac{F_{\rm obs}}{F_{\rm exp}}\right)^\prime_{\rm {\rm }8 \mu m} = \left(\frac{F_{\rm obs}}{F_{\rm exp}}\right)_{\rm {\rm }8 \mu m}
		+ 0.01 \log(L) - 0.035,
	\label{CorrEqun1}
\end{equation}
\begin{equation}
	\left(\frac{F_{\rm obs}}{F_{\rm exp}}\right)^\prime_{\rm {\rm }24\mu m} = \left(\frac{F_{\rm obs}}{F_{\rm exp}}\right)_{\rm {\rm }24\mu m}
		+ 0.025 \log(L) - 0.27.
	\label{CorrEqun2}
\end{equation}

\subsubsection{Dust Temperatures}
\label{DustTSect}

For the remainder of the stars we examine, we have only the 8- and 24-$\mu$m data from which we derive a mass-loss rate. At this point, we are forced to make some assumptions about the mineralogy of the dust. We assume here for simplicity that all stars show `astronomical' silicate dust, following \cite{DL84}. This thus means that our dust mass-loss rates will tend to under-estimate the total mass-loss rate of the star, as our wavelength coverage misses the 12-$\mu$m Al$_2$O$_3$ and 13-$\mu$m MgAl$_2$O$_4$ bands seen at the start of the proposed dust condensation sequence \citep{LPH+06}. Carbon-rich dust is also not considered for the majority of stars: if it is present, the assumption of silicate dust alone will again lead to an under-estimation of total mass-loss rate, as in the case of V42, by a factor of two to four.

Our first priority was to derive an estimated dust temperature for the stars as, for a given optical depth, mass loss can vary significantly with dust temperature. In order to do this, we constructed a grid of {\sc dusty} models at 4000 K, with similar dust properties as those used for V6, but with a pure `astronomical' silicate dust component and $\tau_{\rm V} = 0.02$ and a standard MRN distribution. The temperature of the inner edge of the dust envelope was varied from 200 to 1500 K, in steps of 100 K.

The dependence of the calculated mass-loss rate on the assumed photospheric temperature of the star is relatively weak: for a constant IR excess and stellar luminosity, a comparison of {\sc dusty} models shows that $\dot{M}$ scales roughly as $T^{1.25}$ in the region of 4000--4500 K, providing a reasonable (within 10\%) fit to photospheric temperatures up to 6000 K. This factor takes into account the increase in wind velocity (from increased radiation pressure) that {\sc dusty} calculates for higher-temperature stars. A velocity scaling relation of $v_\infty \propto T^{1/3}$ can be shown to match the output from {\sc dusty} to within 2\% over temperatures from 3500 to 6000 K. We make these corrections in our final rates, but they do not add a considerable effect over the range of temperatures at which stars appear to be losing mass by dust-driven winds.

Errors in the stellar temperature will have an impact here, though their primary influence is on the inferred luminosity of the star, since our dust mass-loss rates use photometry taken from the Rayleigh-Jeans tail of the stellar spectra. This will add a 10\% error in the mass-loss rates of the most-luminous non-variable stars, and up to 30\% in the mass-loss rates of the most-luminous variable stars, due to their higher temperature uncertainties. Due to the stochastic nature of these errors, the effect on the sample as a whole will be greatly reduced from this.

We can then derive the temperature of the dust, under the assumption that the underlying photospheric spectrum follows a Rayleigh-Jeans tail. The [8]--[24] colour \emph{of the excess emission from the dust} will then be directly indicative of the dust temperature, provided the chemical make-up of the dust does not change. For each model, we derive the [8]--[24] colour we would observe by convolving the {\sc dusty} model spectrum with the \emph{Spitzer} filter responses and taking the ratio of the resulting fluxes. We then compare our observed [8]--[24] colour with the models and linearly interpolate between them to find the dust temperature.

Unfortunately, the excess flux from the dust component is typically of order of the errors within the photometry or less: only 30 stars -- 27 cluster members and three non-members -- show excess flux above the model SED at 24 $\mu$m with a significance of over three standard deviations (for comparison, only four stars show a similar flux deficit, all of which are under five standard deviations). The dust temperature is also \emph{very} sensitive to the zero points of the photometry: the additional factor of 1.015 in IRAC 4 (8 $\mu$m) from \cite{RBD+08} raises the temperature of the warmest dust surrounding V6 by 110 K and V42 by 340 K and even more dramatically alters the dust temperatures of all other stars. We were therefore only able to obtain only dust temperature estimates for these two stars, plus a handful of others, listed in Table \ref{DustTempsTable}. LEID 40220 is omitted as, despite showing clear mid-IR excess, its photometric errors are still too large for a temperature to be determined.

For many of these 30 stars, the [8]--[24] colour of the excess emission is sufficiently high or low that it is outside the bounds of our test range (100--1500 K). We would not expect significant amounts of circumstellar dust hotter than 1500 K, and we would expect the inner edge of the dust envelope to exceed 100 K unless the star had recently completed a period of mass loss and was now in (mass-losing) quiescence. The stars with very cool dust are all low-luminosity objects, suggesting they are blends with background galaxies (see Section \ref{LowLumSect}). Stars with warm dust may have emission in the IRAC 8-$\mu$m band. Several of the stars with evidence for warm dust are known variables. It is quite possible that, in the cases of V152, V138, V148 and V161, changes in the stellar luminosity may impact our derivation of mid-IR excess and thus dust temperature. It therefore appears that most of the cluster's mass-losing stars have warm circumstellar dust, with the most evolved, most IR-excessive stars (V6, V42 and possibly V17) showing probably slightly colder dust.

It is noteworthy that we find much colder dust around V6 using only the \emph{Spitzer} data points than we do when we also consider the Gemini T-ReCS spectrum. This is due primarily to the marginally different dust chemistry, which has a substantial effect on the derived dust temperature, but may be related to low-level variability of the star between the 24- and 8-$\mu$m observations, or the presence of an additional, lower-temperature dust component.

For the remainder of this analysis, we have assumed that the temperature of the inner edge of the dust envelope is the typical 1000 K \citep{Habing96} in all stars other than V1, V6, V17, V42, and LEIDs 39105 and 45232, where we take the temperatures or lower limits as given in Table \ref{DustTempsTable}. In terms of mass-loss rates, this represents a roughly average value out of the possible range, with a possible variation of $\pm$80\% between the coldest dust likely based on Table \ref{DustTempsTable} and the hottest dust likely to form ($\ltsim$ 1500 K, based on interferometric observations and condensation temperatures \citep{Salpeter77,Draine81,KHS84,DBD+94,TDH+00,GP04}).

\begin{center}
\begin{table}
\caption[]{Dust temperatures (based on pure `astronomical' silicate dust) for stars with the highest computed dust mass-loss rates. Errors are on the basis of \emph{Spitzer} photometry only and do not include the comparatively small uncertainties in estimating stellar luminosity. Variable numbers are from \cite{Clement97}.}
\label{DustTempsTable}
\begin{tabular}{l@{\ \ }l@{\ }c@{\ }c@{\ }c@{\ }c}
    \hline \hline
LEID  & Variable    & Max.\ Dust 	& \multicolumn{2}{c}{Inner radius}	& Notes \\
\     & number      & temperature 	& \multicolumn{2}{c}{of dust envelope}	& \ \\
\     & \           & (K) 		& (R$_\ast$)	& (AU)		& \ \\
    \hline
55114 & \    & $>$1400             & \ & \ & 4 \\
44277 & V152 & $>$1400             & \ & \ & 4 \\
49123 & V138 & $>$1400             & \ & \ & 4 \\
41455 & V148 & $>$1400             & \ & \ & 4 \\
39105 & \    & $>$801                & $<$23                 &  $<$8.8                      & \ \\
32029 & V1   & $>$685                & $<$50                 &  $<$8.8                      & 1 \\
47153 & V161 & $>$595                & $<$26                 &  $<$8.4                      & \ \\
45232 & \    & $>$556                & $<$27                 & $<$10.7                     & \ \\
44262 & V42  & 937 $^{+207} _{-143}$ &    15 $^{+ 6} _{- 5}$ &     7.6 $^{+ 3.0} _{- 2.5}$ & 6 \\
34041 & V2   & 715 $^{+109} _{-88}$  &    22 $^{+ 6} _{- 6}$ & \                           & 2 \\
35250 & V17  & 509 $^{+177} _{-138}$ &    50 $^{+44} _{-24}$ &    23.2 $^{+20.5} _{-10.7}$ & \ \\
33062 & V6   & 320 $^{+48} _{-48}$   &   111 $^{+43} _{-27}$ &    72.6 $^{+27.9} _{-17.7}$ & \ \\
37098 & \    & $<$706                & \ & \ & 3 \\
45169 & \    & $<$371                & \ & \ & 5 \\
47135 & \    & $<$328                & \ & \ & 5 \\
40015 & \    & $<$240                & \ & \ & 5 \\
31159 & \    & $<$186                & \ & \ & 5 \\
44457 & \    & $<$100              & \ & \ & 5 \\
54135 & \    & $<$100              & \ & \ & 5 \\
51019 & \    & $<$100              & \ & \ & 5 \\
    \hline
\multicolumn{6}{p{0.45\textwidth}}{\small Notes: (1) Post-AGB, proper motion membership probability 3\%, radial velocity 188 km s$^{-1}$; (2) proper motion and radial velocity non-member V825 Cen; (3) proper motion membership probability 0\%, no radial velocity data, variable identified in vL+00 with $P$ = 1.0253 days; (4) suspected SiO or other emission near 8 $\mu$m; (5) faint stars, suspected blends with background galaxies; (6) for silicate dust. \normalsize} \\
\end{tabular}
\end{table}
\end{center}

\subsubsection{Calculating the mass-loss rates along the RGB/AGB}

Using these assumed temperatures, we can derive the expected excess flux at 8 $\mu$m by taking the ratio of the 8-$\mu$m flux from each {\sc dusty} model and dividing it by the flux from a similar {\sc dusty} model with no dust. This is repeated for the 24-$\mu$m excess. Our mass-loss rate then scales similarly to Eqs.~(\ref{MdotDustDustyEqn}) and (\ref{MdotTrueDustyEqn}), with:
\begin{equation}
	\dot{M} _{\rm dust} = \frac{\dot{M}_{\rm DUSTY}}{200}\left(\frac{\psi}{200}\right)^{-1/2} \left(\frac{L}{10^4}\right)^{3/4} R_{\rm E} ,
\label{MdustEqun}
\end{equation}
where:
\begin{equation}
	R_{\rm E} = \frac{F_{\rm e}^{\rm o}(8) + F_{\rm e}^{\rm o}(24) - 2}{F_{\rm e}^{\rm d}(8) + F_{\rm e}^{\rm d}(24) - 2} ,
\end{equation}
$F_{\rm e}^{\rm o}$ is the flux observed above the Rayleigh-Jeans tail in our SED fit at 8 and 24 $\mu$m, and $F_{\rm e}^{\rm d}$ is a similar value, calculated for our {\sc dusty} model at the relevant dust envelope temperature and scaled to the star's luminosity. The factor of 200 in Eq.~(\ref{MdustEqun}) comes from the conversion from the {\sc dusty} \emph{total} mass-loss rate, which assumes a gas-to-dust ratio of 200, to a mass-loss rate purely for the dust component.

\begin{figure}
\resizebox{\hsize}{!}{\includegraphics[angle=270]{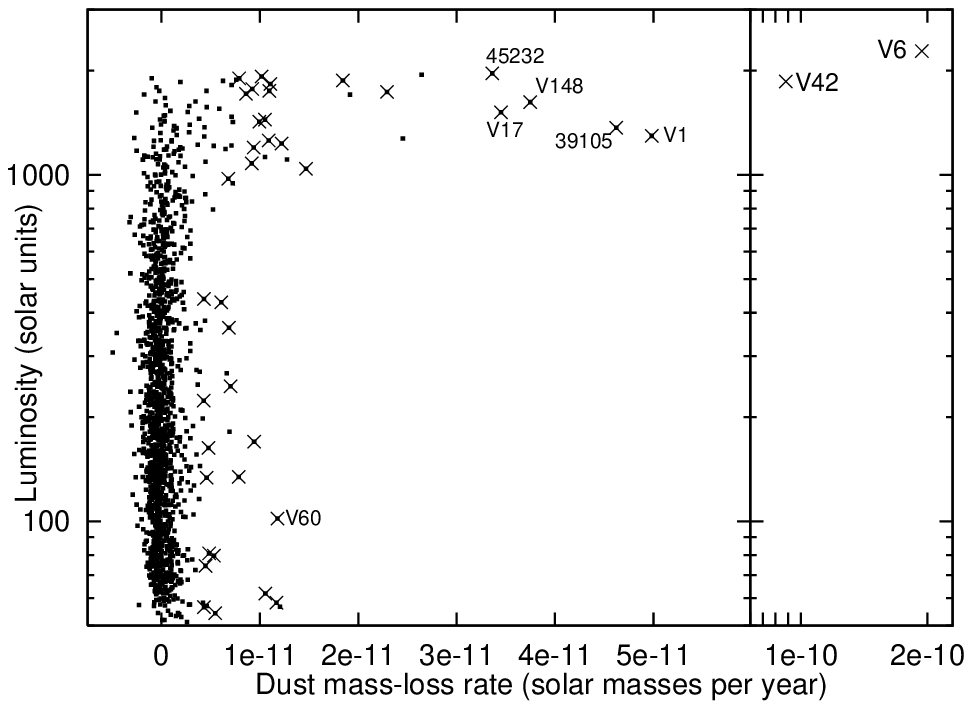}}
\caption[]{Correlation between mass-loss rate and luminosity for confirmed proper-motion cluster members. Objects with 24-$\mu$m excess over 3$\sigma$ are marked by crosses.}
\label{MdotLFig}
\end{figure}

Our computed mass-loss rates are listed in Table \ref{MdotTable}, which excludes those stars identified as blends at 24 $\mu$m in B+08. The correlation between mass-loss rate and luminosity is shown in Fig.\ \ref{MdotLFig}, which confirms that it is the most luminous objects that are losing observable quantities of dust, and that most of the dust mass loss is indeed coming from V6 and V42. We also note that V6 and V42 have very similar mass-loss rates in our analysis here, which only contains the \emph{Spitzer} data, as in our analysis in Section \ref{DustTSect}, which also includes our Gemini spectra and other data. This is despite the assumed dust temperatures being considerably different in both cases.

Also noteworthy are the stars with apparent significant dust mass-loss rates at \emph{low} luminosities ($L \ltsim 500 $L$_\odot$, $\dot{M}_{\rm dust} \sim 10^{-11} $M$_\odot $yr$^{-1}$) in Fig.\ \ref{MdotLFig}. Close inspection of these targets show that they are mostly warm stars (some on the HB) with very little or no 8-$\mu$m excess, but considerable 24-$\mu$m excess. It appears most likely that these are cluster stars blended with background galaxies (see Section \ref{LowLumSect}).

\begin{center}
\begin{table}
\caption[]{List of mass-loss rates from corrected IR excesses, in order of decreasing total mass loss, excluding stars with known blending issues from B+08, stars with 24-$\mu$m excesses of $> 3 \sigma$ and stars dimmer than 500 L$_\odot$. Variable numbers are from Clement (1997).}
\label{MdotTable}
\begin{tabular}{ll@{\,}l@{\,}ll}
    \hline \hline
LEID  & $\dot{M}_{\rm dust}$ & $\dot{M}_{\rm total}$ & [Fe/H] & Notes\\
\     & ($10^{-10}$\,M$_{\odot}$\,yr$^{-1}$) & ($10^{-7}$\,M$_{\odot}$\,yr$^{-1}$) & \  & \ \\
    \hline
44262 & 0.88 & 7.34 & --1.62 & 1, 3, 6, V42\\
32029 & 0.50 & 5.87 & --1.77 & 2, 3, V1\\
33062 & 1.96 & 6.02 & --1.19 & 1, 3, V6\\
35250 & 0.35 & 2.88 & --1.62 & 3, 6, V17\\
45232 & 0.34 & 2.80 & --1.62 & 6, \\
49123 & 0.23 & 1.91 & --1.62 & 3, 6, V138\\
48060 & 0.10 & 1.90 & --1.97 & 3\\
56087 & 0.11 & 1.82 & --1.92 & 3\\
55114 & 0.19 & 1.67 & --1.64 & 3\\
44277 & 0.18 & 1.54 & --1.62 & 3, 6, V152\\
41455 & 0.37 & 1.46 & --1.29 & 3, V148\\
32138 & 0.09 & 1.38 & --1.87 & \\
37110 & 0.15 & 1.22 & --1.62 & 3, 6\\
25062 & 0.09 & 1.16 & --1.83 & \\
47153 & 0.12 & 1.02 & --1.62 & 6, V161\\
48150 & 0.11 & 0.92 & --1.62 & 6, \\
43351 & 0.11 & 0.91 & --1.62 & 6, \\
42044 & 0.11 & 0.88 & --1.62 & 6, V184\\
42302 & 0.10 & 0.83 & --1.62 & 6, \\
39165 & 0.09 & 0.78 & --1.62 & 6, \\
36036 & 0.03 & 0.78 & --2.05 & 4\\
42205 & 0.09 & 0.77 & --1.62 & 6, \\
26025 & 0.08 & 0.75 & --1.68 & \\
52030 & 0.26 & 0.67 & --1.10 & 4, 5\\
39105 & 0.46 & 0.65 & --0.85 & \\
52111 & 0.07 & 0.57 & --1.62 & 6, \\
    \hline
\multicolumn{5}{p{0.45\textwidth}}{\small Notes: (1) Values based solely on $Spitzer$ photometry; (2) post-AGB star;  (3) also shows significant ($>3\sigma$) excess at 8 $\mu$m; (4) shows 24-$\mu$m excess \emph{below} 3$\sigma$, but shows 8-$\mu$m excess above 3$\sigma$; (5) carbon star, see Section \ref{CStars}; (6) no metallicity known, we have used the cluster average, [Fe/H] = --1.62. \normalsize} \\
\end{tabular}
\end{table}
\end{center}

\section{Discussion}

\subsection{Notable objects besides V6 and V42}

\subsubsection{Post-AGB objects}
\label{PostAGBSect}

The cluster contains two confirmed post-AGB objects: Fehrenbach's Star (LEID 16018) and V1 (LEID 32029). It is notable that, while Fehrenbach's Star does not appear to show a dusty circumstellar envelope, V1 does. From the spectra of vL+07, Fehrenbach's Star also has very weak emission lines, but those of V1 are strong.

It is difficult to reconcile the similarities between the dust shells in V1 and the mass-losing AGB/RGB-tip stars (see Table \ref{DustTempsTable}), given that V1 has evolved off the AGB tip. It is worth noting that V1 is almost four times more metal poor, so we may not necessarily expect the historic wind from V1 to be the same as in V6. The proximity of V1's dust shell to the star does raise the interesting possibility that the dust shell in V1 is relatively stationary, supporting the low velocity rates given by {\sc dusty}'s scaling of $v \propto \psi^{-1/2}$.

For Fehrenbach's Star, we can surmise that the absence of dust is due to either the dust having dispersed and become too cold to detect with our \emph{Spitzer} data, or the dust having been destroyed. It does not appear to be present at any significant level in the longer-wavelength \emph{AKARI} data from \cite{MMN+08}.

Several other potential post-AGB stars have been highlighted in our HRD (Fig.\ \ref{HRDFig}). These are listed in Table \ref{PostAGBTable}. Of these, several show bad cross-identifications or poor quality 2MASS data. As we expect such objects to scatter excessively in the HRD, this should not be surprising, and we anticipate these are the exceptions, rather than the standard in our dataset. In particular, LEID 34135 is identified as an RGB-a star, LEID 40240 is a known blend in our 24-$\mu$m data. The exact status of LEID 38171 remains unresolved: it shows a mid-IR deficit in our \emph{Spitzer} 5.8- and 8-$\mu$m data, though it may be an early-AGB star with poor 2MASS photometry. The remainder have SEDs visually consistent with their ascribed temperature and luminosities, although we note that, of these, only LEIDs 29064, 30020, 32015, 43105, 46054 and 46265 have radial velocity measurements that confirm their cluster membership. These could potentially be post-early-AGB stars, which have left the AGB sequence before reaching the thermally-pulsating stages. None show a mid-IR excess at 24 $\mu$m above the photometric errors.

\begin{center}
\begin{table}
\caption[]{List of post-AGB candidates identified from the HRD (Fig.\ \ref{HRDFig}), showing luminosities, temperatures, proper motion membership percentage probabilities (PM Mem.) from vL+00, radial velocities from vL+07 and individual notes.}
\label{PostAGBTable}
\begin{tabular}{l@{\ \ \ \ \ \ }c@{\ \ \ }c@{\ \ \ }c@{\ \ \ }c@{\ \ \ }l}

  \hline\hline
LEID  & $T$    & $L$   & PM & RV  & Notes \\
\     & (K)    &(L$_\odot$)&\% Mem. & (km\,s$^{-1}$)  & \  \\
  \hline
46162 &   5201 &   310 &   98 & \    & V148, 1, 4\\
40420 &   5347 &   296 &   72 & \    & 5 \\
43105 &   5376 &   524 &  100 &  212 & V29\\
32015 &   5493 &   378 &  100 &  205 &  \\
32029 &   5675 &  1297 &    3 &  188 & V1, 6 \\
39367 &   5750 &   239 &   99 & \    & 2, 9 \\
37095 &   5817 &   464 &   72 & \    &  \\
41263 &   6145 &   197 &  100 & \    &  \\
16018 &   6427 &  1480 &   18 &  225 & 6, 7 \\
37295 &   6993 &   356 &  100 & \    &  \\
34136 &   7109 &   335 &   76 & \    &  \\
34135 &   7450 &   219 &   99 & \    & 2, 9 \\
29064 &   7604 &   181 &  100 &  206 &  \\
38171 &   7784 &   182 &  100 & \    & 8 \\
46265 &   8014 &   282 &  100 &  224 &  \\
48180 &   8769 &   405 &  100 & \    & 2, 9 \\
43188 &   9166 &  1017 &   97 & \    & 2, RGB \\
50182 &  10\,051 &   157 &  100 & \    & 3, HB? \\
46054 &  11\,106 &   178 &  100 &  226 &  \\
30020 &  11\,438 &   335 &  100 &  207 &  \\
30120 &\llap{$>$}18\,953&   488 &  100 & \    &  \\
32163 &\llap{$>$}18\,953&  1728 &  100 & \    & 3, HB? \\
41457 &\llap{$>$}18\,953& 10625 &  100 & \    & 1, HB? \\
56071 &\llap{$>$}18\,953&   626 &  100 & \    & 1, HB? \\
  \hline
\multicolumn{6}{p{0.45\textwidth}}{\small Notes: (1) Poor or no 2MASS data; (2) blend in 2MASS; (3) bad cross-identification in 2MASS; (4) $L \sim 190$ L$_\odot$, probable early-AGB star with low mantle mass; (5) young post-early-AGB or very bright early-AGB star; (6) confirmed post-AGB star; (7) Fehrenbach's Star; (8) isolated object with very steep spectrum, nature unresolved; (9) likely early-RGB/AGB object. \normalsize} \\
\end{tabular}
\end{table}
\end{center}

\subsubsection{Carbon stars}
\label{CStars}

Of the known carbon stars in the cluster (LEIDs 52030, 41071, 32059, 14043, and 53019), only the most luminous of these, LEID 52030, shows significant dust mass loss, and even this is only a 4$\sigma$ detection. The values we quote here are for silicate dust. Carbonaceous dust reproduces the observed 8-$\mu$m excess marginally better than silicate dust. However, any analysis of this star is uncertain as our photospheric models are for \emph{oxygen}-rich stars.

We estimate that the mass-loss rate for amorphous carbon dust with the same grain parameters as our silicate dust model is roughly 50\% higher, at $\sim 4 \times 10^{-11}$ M$_\odot$ yr$^{-1}$, though the error in the mass-loss rate itself is over 50\%. It would appear at first sight that these stars are no more effective at producing dust than their M-type counterparts.

\subsubsection{Low-luminosity giant stars}
\label{LowLumSect}

\begin{figure}
\resizebox{\hsize}{!}{\includegraphics[angle=270]{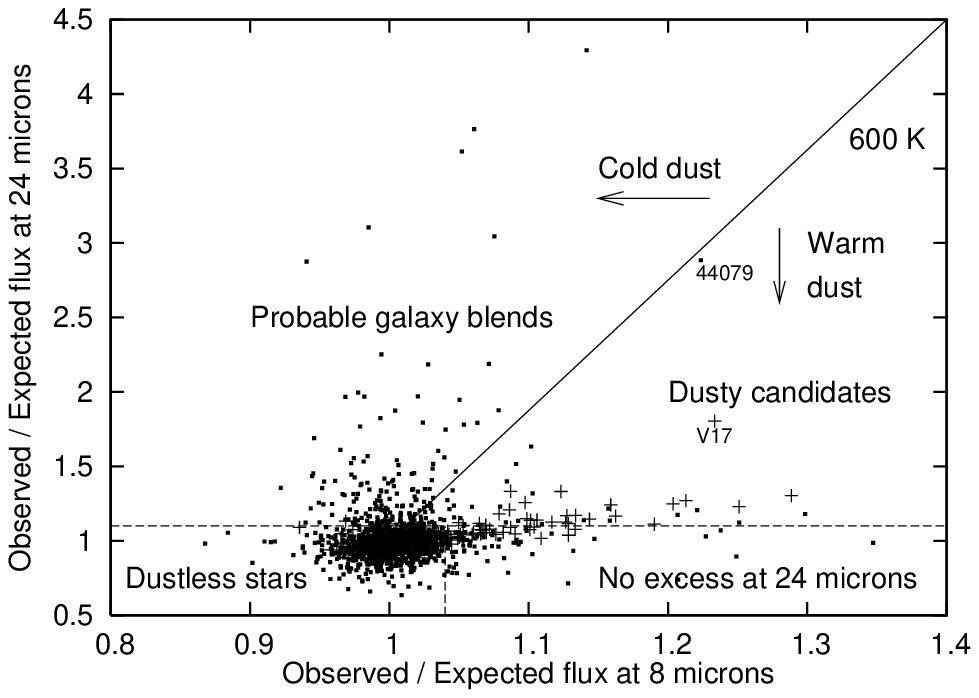}}
\caption[]{Observed mid-IR excess (attributable to dust) at 8 versus 24 $\mu$m, which corresponds (for given dust properties) to the inner temperature of the dust envelope. Potentially dusty stars and probable blends with background galaxies can be split by a line corresponding to a 600 K silicate dust envelope. Stars above 500 L$_\odot$ are shown as crosses, stars below 500 L$_\odot$ as dots.}
\label{DustColsFig}
\end{figure}

We find a number of stars below 500 L$_\odot$, both on the giant branches and the HB, which have significant IR excess. As they appear to occur with roughly equal prevalence across the HRD, we suggest that these are merely blends with background galaxies (c.f.~Section \ref{DustTSect}). Fig.\ \ref{DustColsFig} shows that the mid-IR-excessive stars nicely bifurcate into these two colour groups: we expect large 24-$\mu$m excesses but little 8-$\mu$m excess (`cold dust') to most likely be blends with background galaxies (c.f.~2MASX J13272621--4746042 in B+08), while we expect fairly comparable 8- and 24-$\mu$m excesses for stars with circumstellar dust. Note that V6 and V42 lie off the diagram to the right, as does the low-luminosity RGB star LEID 31159, which lies just below the 600 K line. The low-luminosity stars with mid-IR excess also spatially follow the distribution of \emph{all} stars, not just cluster members, and show no concentration towards the centre of the cluster, ruling out crowding in the cluster core as a primary factor for generating these detections.

In total, 37 stars with luminosities below 500 L$_\odot$ show significant ($> 3 \sigma$) excess at 24 $\mu$m. Of these, however, 34 have little or no excess at 8 $\mu$m. While we have made every effort to remove 24-$\mu$m blends from our sample, the glare from nearby stars up to 150 times brighter may prevent an accurate 24-$\mu$m flux measurement being made in these cases. The remaining three stars are: LEID 31159, an 18 L$_\odot$ RGB star; LEID 44457, a 49 L$_\odot$ RGB star; and LEID 47135, an early-AGB star. We have no reason, \emph{a priori}, to assume that any of these is likely to harbour circumstellar dust.

\subsubsection{Non-members}

Our SED modelling also includes a number of stars that have been confirmed as non-members, either by proper motion or radial velocity measurement. These include V2 (V825 Cen, LEID 34041), an emission-line M-giant with a fitted temperature of 3346 K and a strong IR excess. It was identified as a radial velocity non-member by \cite{Feast65} and has a period of 236 days \citep{KKS+96}.

\begin{figure}
\resizebox{\hsize}{!}{\includegraphics[angle=270]{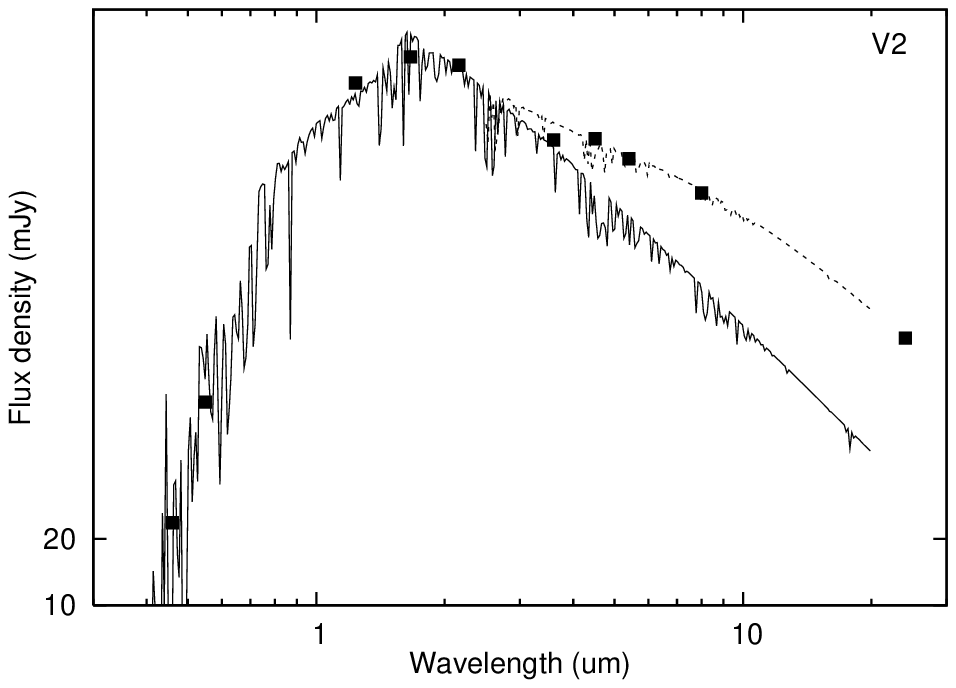}}
\caption[]{The SED for the radial velocity non-member V2 (filled squares). Over-plotted is a 3500 K spectral model (solid line) and the same model with a 850 K blackbody (dashed line).}
\label{V2Fig}
\end{figure}

A fit to the IR excess in V2 is shown in Fig.\ \ref{V2Fig}, with a 850 K blackbody. The temperature of this blackbody is comparable to the 715$\pm ^{109}_{88}$ K we find in Section \ref{DustTSect}. This blackbody has a flux density scaling constant ($L/R^2$) that is ten times that of the underlying photosphere; though without a good distance measure, we cannot estimate its mass-loss rate. We cannot attain a good fit to the spectrum using oxygen-rich dust, due to the high flux in the IRAC 4.5- and 5.8-$\mu$m bands and the absence of any silicate-like feature (which would show up at 8 and 24 $\mu$m). Carbon-rich dust could provide a fit, though without a mid-IR spectrum of the star, we have no reason to believe it harbours such dust.

The star LEID 37098 (VLR O-123) was identified in vL+00 as a variable star with a period of around a day. It exhibits little excess at 8 $\mu$m, but considerable excess at 24 $\mu$m. We estimate the mass-loss rate at $1.4 \times 10^{-8} \ (d / {\rm kpc})^{-1.5}$ M$_\odot$ yr$^{-1}$, though the absence of 8-$\mu$m emission suggests that the star is most likely blended with a background galaxy.

\subsection{Total mass-loss rate of $\omega$ Centauri}
\label{TotalMdotSect}

Our calculation of the total mass-loss rate for $\omega$ Cen is subject to broadly similar errors as discussed in Section \ref{MdotV6V42Sect}:
\begin{itemize}
\item 30\% uncertainty in the dust mass-loss rate from internal errors in {\sc dusty};
\item $\ltsim$ 30\% uncertainty in the dust mass-loss rate due to uncertain dust chemistry and grain size;
\item negligible uncertainty in the cluster's metallicity, which would affect the gas-to-dust ratio;
\item $\ltsim$ 10\% uncertainty due to errors in scaling the mass-loss rate of the stars to the photospheric temperature (see Section \ref{DustTSect});
\item $\ltsim$ 5\% uncertainty due to error in the integrated flux density of the stars from fitting the SED, including errors in stellar temperature;
\item $\sim$ 10\% uncertainty due to error in the interstellar reddening;
\item $\sim$ 20\% uncertainty due to error in the distance;
\item errors due to the corrections we have made to the photometric excesses above. Though difficult to discern analytically, we have altered the parameters of the constants in Eqs.\ (\ref{CorrEqun1}) and (\ref{CorrEqun2}), and obtained a visual estimate of the error involved, finding it to be at most 30\%;
\item uncertainty in the dust envelope temperature, although we do not expect the associated error in mass-loss rate to be as high as the 80\% error in individual stars (Section \ref{DustTSect}), as this arises from (random) photon noise. However, since we do not have any sensible upper limits on dust temperature, bar the maximum dust condensation temperature around 1500 K, the error involved may still be as high as $\sim$30\%.
\end{itemize}
Added in quadrature, this gives a total error of 65\%.

For the remainder of this analysis, we have removed any objects thought to be 24-$\mu$m blends on the basis of B+08. We include V1, which is listed as a proper-motion non-member in vL+00 (at a membership probability of 3\%), but a radial velocity member in vL+07. We do not include mass-loss rates from stars with $L < 500$ L$_\odot$, as these appear not to be losing mass (Section \ref{LowLumSect}). We also include the mass-loss rates derived from our Gemini spectra for V6 and V42, in preference to those found from Spitzer data alone. Integrating the mass-loss rates of the remaining cluster members, we find the total stellar dust mass-loss rate within $\omega$ Cen, under the same assumptions listed in Section \ref{MdotV6V42Sect}, to be:

\begin{equation}
	\dot{M}_{\rm dust} = 1.3 \pm ^{0.8}_{0.5} \times 10^{-9} \  {\rm M}_\odot \  {\rm yr}^{-1} .
\end{equation}
Due to the incompleteness of our knowledge on precise dust composition, detailed in Section \ref{DustTSect}, we suggest that, in reality, this is a lower limit. It is notable that, of this value, V6 and V42 therefore presently produce around half of the dust produced by the entire cluster, with V6 being the marginally larger contributor due to its higher metallicity. Also, the cluster's entire dust mass loss appears to come entirely from stars within a magnitude of the giant branch tip. This is in contrast to the case of 47 Tuc, where O+07 find that most of the dust mass loss occurs in the more numerous stars much further down the giant branch. Even the shallow exposure of M15 \citep{BWvL+06} revealed dusty giants at least 1.5 magnitudes below the tip of the RGB. It is possible that O+07's result was affected by blending at longer wavelengths in this visually more compact cluster, which would produce apparent excesses at these wavelengths. Observations of other clusters are needed to confirm the extent of the dust mass loss.

In our analysis, we have not considered gas mass loss by methods other than a dust-driven wind. The most important of these is chromospheric mass loss. There is weak evidence (\citealt{SD88}; Mauas et al.~(2006); MvL07) that the dust-driven wind begins following the extension of the stellar chromosphere by pulsation, although the exact sequence of events is yet to be fully explained. For this reason, as well as those described in Section \ref{DustTSect}, we here derive only a lower limit to the total mass-loss rate from the cluster. Multiplying the dust mass-loss rate for each star by its gas-to-dust ratio (see Section \ref{MdotV6Sect}), we can again integrate the total mass-loss rate of all the dusty cluster stars, finding it to be: $\dot{M}_{\rm total} \gtsim 8 \pm ^{5}_{3} \times 10^{-6}$ M$_\odot$ yr$^{-1}$. If we include the contribution from dustless (chromospherically-driven) winds as prescribed by \cite{SC05}, this lower limit rises to around $\dot{M}_{\rm total} \gtsim 1.2 \pm ^{0.6}_{0.5} \times$ 10$^{-5}$ M$_\odot$ yr$^{-1}$, depending on the point at which chromospheric mass loss is replaced by dust-driven and/or pulsation-induced mass loss.

It is interesting to note here that the fraction of the \emph{total} mass loss provided by V6 and V42 is somewhat less, due to the higher metallicity (hence presumed gas-to-dust ratio) of V6 and the inclusion of dustless winds. V42 now produces around 20\% of the gas produced by dust-producing stars (not including mass loss from giants with chromospheres), but V6 is reduced to around 9\% of this total.

\subsection{Mass loss along the giant branch}
\label{MdotRGBSect}

\subsubsection{The onset and evolution of dust formation}

\begin{figure}
\resizebox{\hsize}{!}{\includegraphics[angle=270]{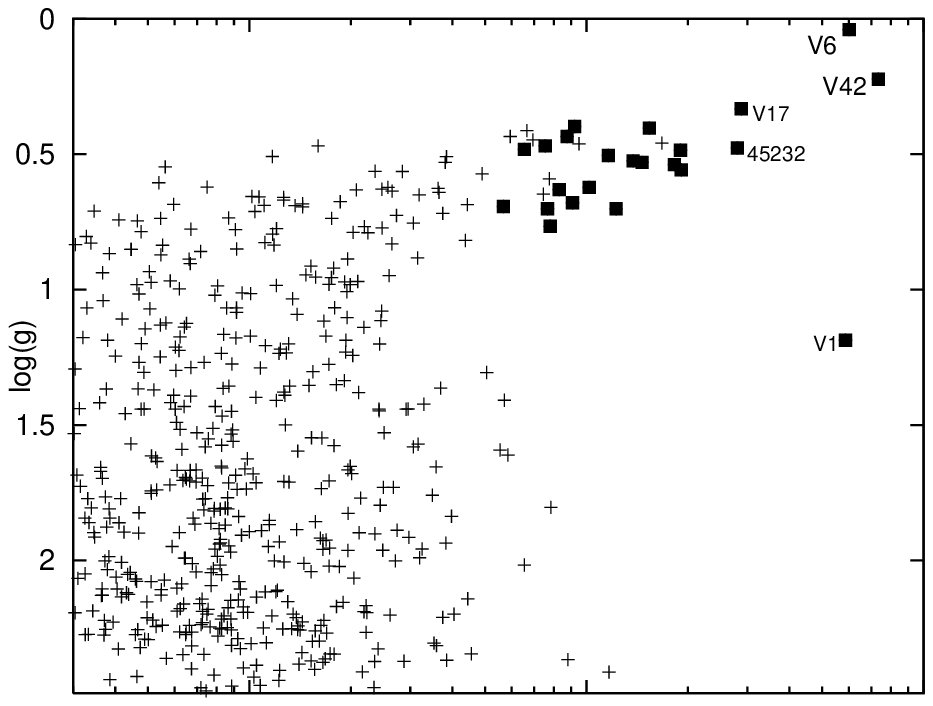}}
\resizebox{\hsize}{!}{\includegraphics[angle=270]{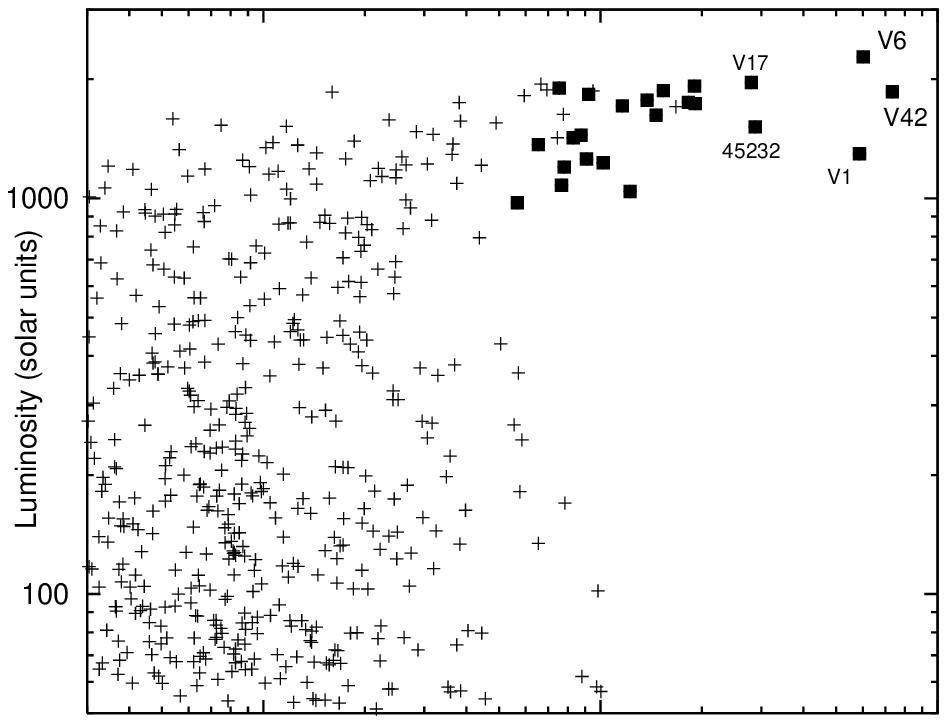}}
\resizebox{\hsize}{!}{\includegraphics[angle=270]{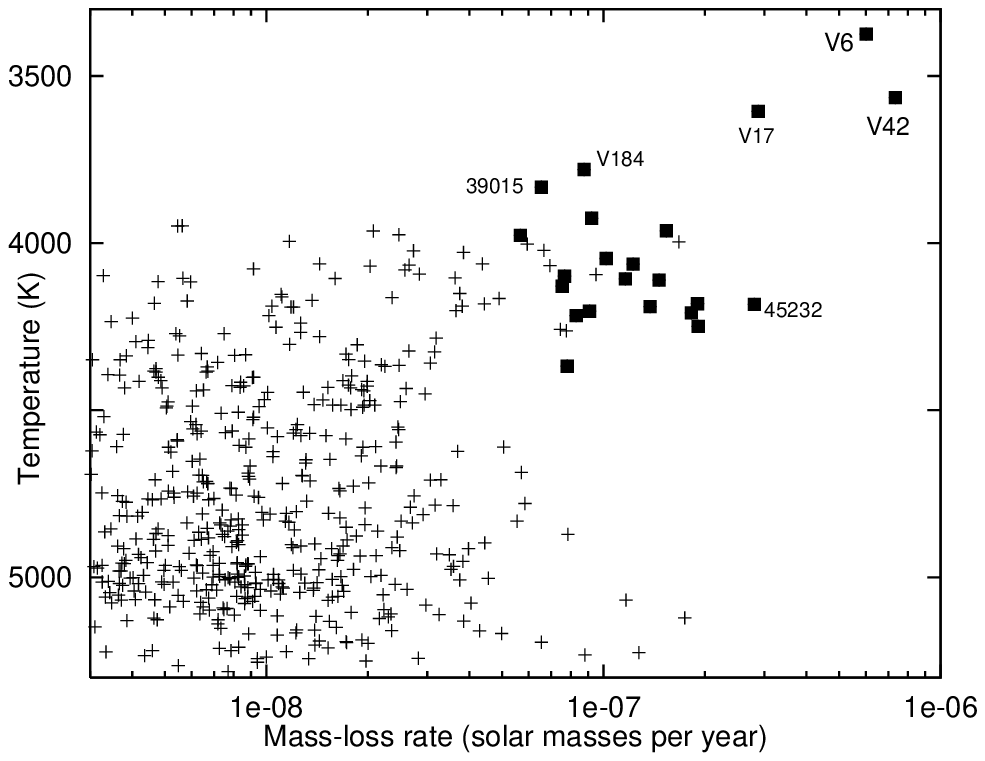}}
\caption[]{Computed total mass-loss rates and their dependence on log($g$) (top panel), luminosity (centre panel) and stellar effective temperature (lower panel). Large squares denote stars above $L > 500$ L$_\odot$ with individually statistically-significant mass loss, based on a 3$\sigma$ cut on 24-$\mu$m excess.}
\label{MdotFig}
\end{figure}

While we have no clear measure in the data presented here of mass loss by means other than a dusty wind, we can place constraints on when dust production starts to become efficient and explore the parameters upon which it depends. Fig.\ \ref{MdotFig} shows the variation of mass-loss rate with gravity, luminosity and stellar effective temperature. It is difficult to say what happens in denser, fainter, warmer stars due to lack of sensitivity in the photometry. Despite this, the fact that extreme dust mass loss occurs towards minimum gravity and maximum luminosity is obvious, although it is not clear which parameter is \emph{physically} responsible. There are arguments for mass loss to be linked to any or all of these parameters.

Surface gravity is a measure of escape velocity and of the ease with which dust grains can be accelerated away and lost from the star. It has previously been suggested that this is the primary determining parameter of mass loss \citep{JS91}, and we do indeed observe a smooth trend in Fig.\ \ref{MdotFig}.

Radiation pressure on the dust grains increases with increasing stellar luminosity (and likely temperature, depending on the grain opacities). Stars evolving along the giant branch increase monotonically in luminosity, but at the same time their photospheres cool as their radii increase, also in a monotonic fashion. As the stars in GCs have very similar masses, the differences in surface gravity amongst stars along the giant branch essentially depend on the radius. Therefore, it is not surprising that if a correlation is found with any of the three parameters luminosity, temperature and surface gravity, that a similar correlation is found with the remaining parameters. We do observe an increasing trend in mass-loss rate with luminosity. The post-AGB star V1 fits the luminosity relation better, possibly since its mass-loss rate was determined while it was still on the AGB at a similar luminosity, but very different surface gravity.

Dust formation requires a sufficiently low temperature at still a sufficiently high pressure -- a condition most easily met in the atmospheres of the coolest stars. At warmer temperatures, we see that no stars with $T_{\rm eff} \gtsim 4500$ K are producing dust significantly. Following the trend to cooler temperatures, however, we find that there is a clear transition to statistically-significant dust mass-loss rates for around half the stars below $\sim$4300 K. By 3900 K, all stars seem to be producing dust at a substantial rate. The onset of dusty winds in $\omega$ Cen therefore seems to occur at a somewhat cooler and more luminous threshold than found in the more metal-poor M15 (M\'{e}sz\'{a}ros et al.~2008).

The sudden transition at $\sim$4400 K could correspond to a sudden transition to allow significant dust production to occur in dense environments, as \cite{WWSS08} suggest occurs in carbon-rich stars. This may be a particularly important factor if the temperature gradient is small and pressure (i.e.~density) gradient is high in these metal-poor stars, as suggested by \cite{MWHE08}. The fact that there is a regime where some stars show significant mass-loss rates, while others show insignificant mass-loss rates could be due to episodic mass loss, where stars produce dust for increasing periods of time until they reach $\sim$4000 K, at which point the continuous dust production may take place. The fraction of dust-producing stars at a given temperature would then yield the relative duration of the dust-producing phase. It may also be possible that the AGB stars (which typically have a lower surface gravity at a given surface temperature) can maintain a significant dusty wind already at higher surface temperatures than their RGB counterparts. The gas-to-dust ratio may also play a r\^ole: it is possible to have higher mass-loss rates at higher temperatures, but in winds containing fractionally less dust.

\subsubsection{Comparison with literature relations}
\label{PsiSect}

\begin{figure}
\resizebox{\hsize}{!}{\includegraphics[angle=270]{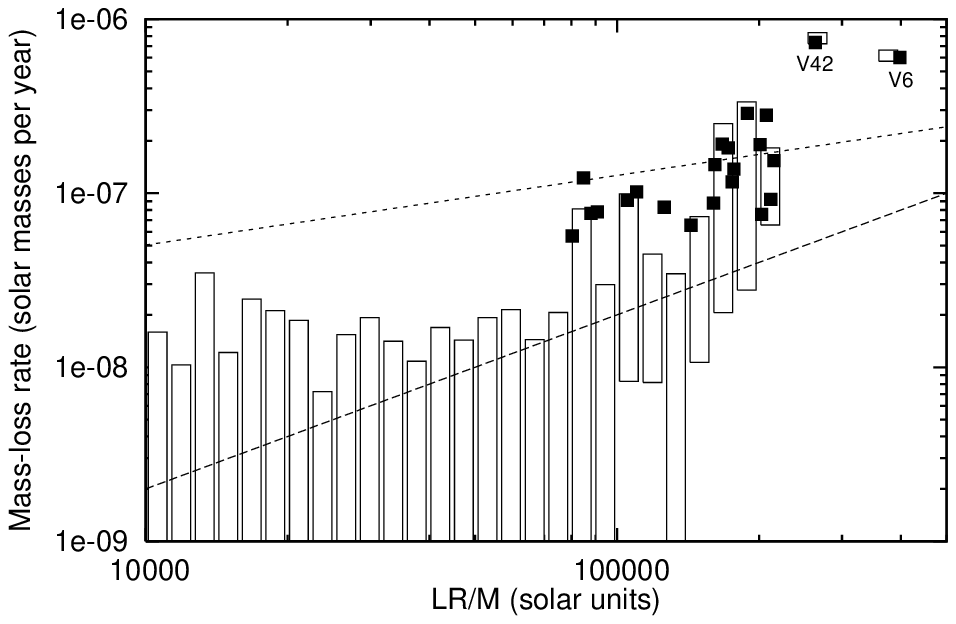}}
\resizebox{\hsize}{!}{\includegraphics[angle=270]{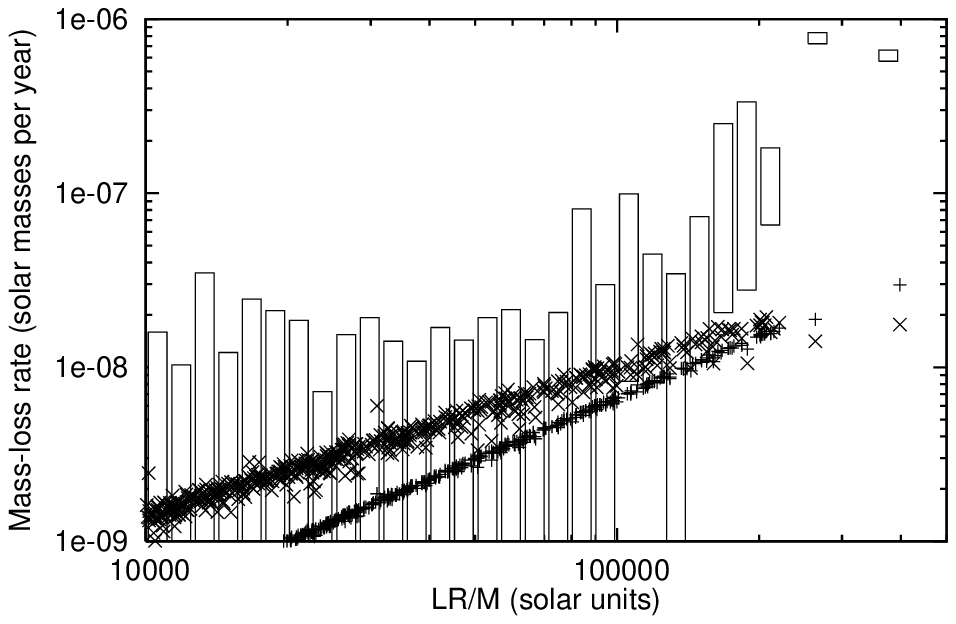}}
\resizebox{\hsize}{!}{\includegraphics[angle=270]{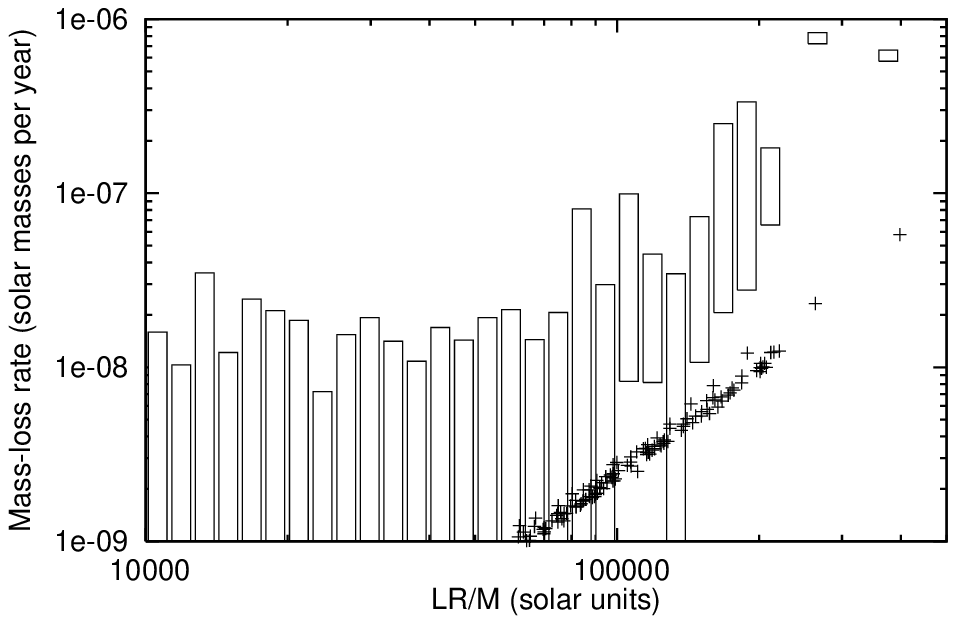}}
\caption[]{Comparison of our estimated total mass-loss rates to literature relations. Our data have been binned in units of $0.05 \log(LR/M)$ and are shown as boxes with heights of $\pm 1 \sigma$. Individual stars with statistically-significant mass-loss rates have been shown in the top panel as filled squares. Comparisons: \cite{Reimers75} -- top panel, dashed line; \cite{ORFF+07} -- top panel, dotted line; \cite{SC05} -- middle panel, crosses; \cite{NdJ90} -- middle panel, plus signs; \cite{WWSS08} -- bottom panel, crosses.}
\label{MdotCompsFig}
\end{figure}

The real drivers behind mass loss are likely to be determined by a combination of these parameters. Such has been the supposition of previous empirical determinations of mass-loss laws, such as those of \cite{Reimers75}, \cite{NdJ90}, \cite{SC05}, and O+07.

Reimers' law is based on the following simple parameterisation for chromospheric mass-loss:
\begin{equation}
	\dot{M} = 10^{-13} \eta_{\rm R} \frac{L_\ast R_\ast}{M_\ast} \ {\rm M}_\odot \ {\rm yr}^{-1},
\end{equation}
where $L_\ast$, $R_\ast$ and $M_\ast$ are the luminosity, radius and mass of the star in solar units, and $\eta_{\rm R}$ is a constant, generally assumed to be in the range 0.4 to 3 \citep{SJ07}, but most commonly 0.5 \citep{SC05}, which we assume here. The modified relation given by \cite{SC05} is based on the energetics of momentum transfer in a chromospheric outflow, namely:
\begin{equation}
	\dot{M} = \eta_{\rm S} \frac{L_\ast R_\ast}{M_\ast} \left(\frac{T_{\rm eff}}{4000 {\rm K}}\right)^{3.5} \left(1 + \frac{g_\odot}{4300 g_\ast}\right) ,
\end{equation}
where $\eta_{\rm S}$ is suggested to be $(0.8 \pm 0.1) \times 10^{-13}$ M$_\odot$ yr$^{-1}$. \cite{NdJ90} give a parameterisation similar to that of Reimers, but fitted to the entire HRD:
\begin{equation}
	\dot{M} = \eta_{\rm N} L_\ast^{1.42} R_\ast^{0.81} M_\ast^{0.16} ,
\end{equation}
where $\eta_{\rm N} = 9.63 \times 10^{-15}$ M$_\odot$ yr$^{-1}$. O+07 base their relationship on a similar method to ours, using ($K$--[8]) colour excess to attribute dust mass-loss rates, and fit the following parameterisation:
\begin{equation}
	\dot{M} = 4 \times 10^{-10} C \left(\frac{L_\ast}{g_\ast R_\ast}\right)_\odot^{0.4} \ {\rm M}_\odot \ {\rm yr}^{-1},
\end{equation}
where:
\begin{equation}
	C = (\psi/200)^{0.5}(v_\infty/10)(\rho_{\rm s}/3),
\end{equation}
with gravity ($g_\ast$) and radius ($R_\ast$) in solar units. As $\psi$ and $v_\infty$ vary among our stars, we here take $C = 1$.

We compare the expected relations from these laws to our data in Fig.\ \ref{MdotCompsFig}. In this figure, we have assumed $M = 0.8$ M$_\odot$. For the O+07 relation, we used  $\psi = 2000$, on the basis that $\omega$ Cen is an order of magnitude less metal-rich than 47 Tuc, where O+07 assumed $\psi = 200$. The other relations have no dependence on $\psi$. We also compare the carbon-star relation of \cite{WWSS08} for the SMC (at [Fe/H] $\sim -0.7$), namely:
\begin{equation}
	\dot{M} = C^\prime \left(\frac{L_\ast}{L}\right)^{2.84} \left(\frac{T_\ast}{2600 {\rm K}}\right)^{-6.81} M_\ast^{-3.01}.
\end{equation}
Though Wachter et al.\ use $C^\prime = 2.34 \times 10^{-5} \ {\rm M}_\odot \ {\rm yr}^{-1}$ as their constant, based on their modelled scaling with metallicity, we would expect $C^\prime$ to be roughly $1 \times 10^{-5} \ {\rm M}_\odot \ {\rm yr}^{-1}$ for $\omega$ Cen, which we use in the following comparison of the aforementioned relations.

We can scale our mass-loss rates to match the above relations given our assumption that $\dot{M} \propto \sqrt{\psi}$, with the solar-metallicity value being 200. This places V6 at $\psi = 2600$ and the majority of the cluster (probably including V42) at $\psi = 8300$. We do not expect the gas-to-dust ratio to attain solar-metallicity values (i.e.~$\psi \sim 200$) in these stars, simply due to the lack of condensable material which can form dust \citep{vanLoon08}.

Fig.\ \ref{MdotCompsFig} shows the relation from O+07 successfully reproduces the distribution of our dusty stars in the range $7.5 \times 10^5 \ltsim LR/M \ltsim 2.5 \times 10^6$ times the solar value, suggesting that $\dot{M} \propto \sqrt{\psi}$ does hold between 47 Tuc and $\omega$ Cen. However, the quality of the fit decreases dramatically once we include the dust-free stars and stars where we have a marginal detection of dust. The gas-to-dust ratios we require for V6 and V42 to be consistent with O+07 are $\psi \approx 800$ and 500, respectively. This may not be unreasonable, if V42 is part of the cluster's most metal-rich sub-population, though Origlia et al.\ themselves find several stars showing dusty emission above their own model, suggesting the relation does not describe these more unusual objects very accurately. We also note that our data do not reproduce the high mass-loss rates seen in O+07 below $LR/M \approx 7.5 \times 10^5$ times solar.

The classic Reimers' relation is meant to describe mass loss in solar-metallicity environments where a chromosphere, \emph{not} dust, is the primary wind driver. We should therefore not be especially surprised that it does not fit the upper giant branch of this metal-poor cluster well, where we expect the chromospheres to be disrupted and a dust-driven wind to be present. That it over-predicts the dusty mass-loss rate below $LR/M \approx 7.5 \times 10^5$ times solar suggests that chromospheric mass loss may dominate here, while its under-estimate of mass-loss for greater values of $LR/M$ confirms that a dusty wind is the primary mass-losing mechanism here. The formula from \cite{SC05} is essentially a downward revision of Reimers' law, derived for similar environments. This provides a similar conclusion, although the over-prediction around $LR/M \approx 2$ to $7.5 \times 10^5$ times solar is less severe.

The \cite{NdJ90} relation provides a slightly better match to the slope of the relation than the other laws, but grossly under-estimates the mass loss, requiring $\psi \sim 100-500$ for the dustiest stars but $\psi < 100$ for V6 and V42. It is, however, in agreement with the upper limits further down the giant branch. The values of $\psi$ required near the giant branch tip are unphysical, but perhaps this should not be surprising, given that Nieuwenhuijzen \& de Jager themselves state that the relation does not work well for dusty stars.

\begin{figure}
\resizebox{\hsize}{!}{\includegraphics[angle=270]{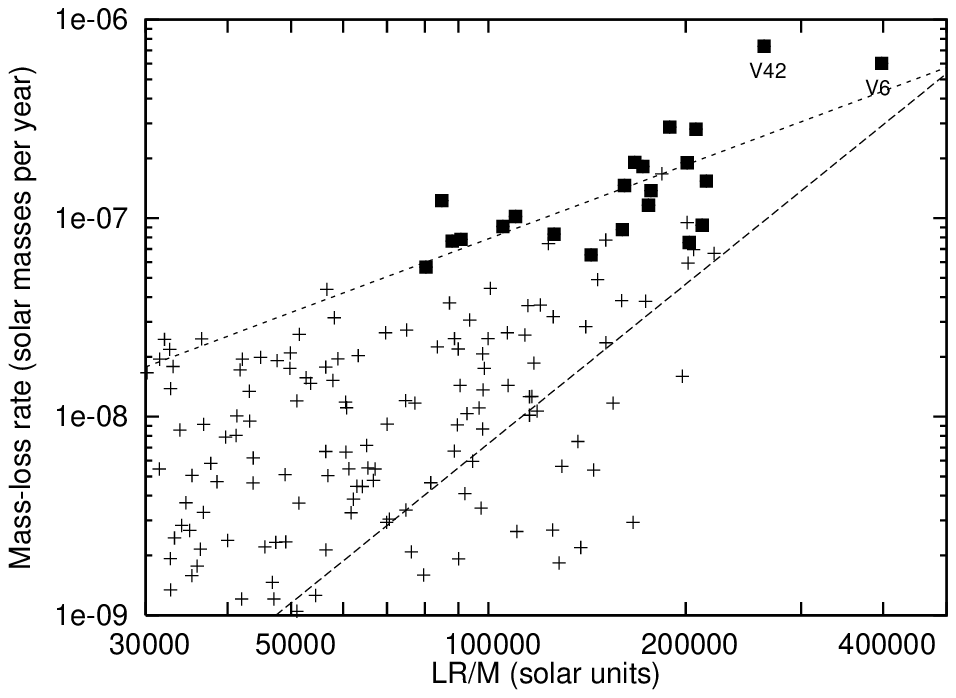}}
\caption[]{Correlation between total mass-loss rate and physical stellar parameters. The dashed line denotes the fit to the entire data above $LR/M = 80\,000$, described in Eq.~(\ref{LRMFitEqun}); the dotted line fits only the stars with significant mass loss, described in Eq.~(\ref{LRMBadFitEqun}). Individual stars with statistically-significant mass-loss rates have been shown as filled squares.}
\label{LRMFig}
\end{figure}

\emph{Empirical} literature relations therefore do not seem to be able to reproduce the mass-loss rates we find. Without data on the gas-to-dust ratios in the wind, we cannot tell whether it is truly the mass-loss rate which changes, or merely the gas-to-dust ratio, which in turn influences our derived mass-loss rates. In Fig.\ \ref{LRMFig}, we show two fits to the data, given by:
\begin{equation}
	\dot{M} = 3.3 \times 10^{-22} \left(\frac{L_\ast}{g_\ast R_\ast}\right)_\odot^{8/3} \ {\rm M}_\odot \ {\rm yr}^{-1} ,
\label{LRMFitEqun}
\end{equation}
and:
\begin{equation}
	\dot{M} = 4.5 \times 10^{-14} \left(\frac{L_\ast}{g_\ast R_\ast}\right)_\odot^{5/4} \ {\rm M}_\odot \ {\rm yr}^{-1} .
\label{LRMBadFitEqun}
\end{equation}

These two fits show, respectively, the fit to all stars with $LR/M > 80\,000$ and the fit to only those stars with significant mass-loss rates (defined, as above, by a $>3\sigma$ flux excess at 24 $\mu$m). These two fits are markedly different, and highlight the necessity to treat all stars, not just those which are observed to lose mass, when creating relationships between mass loss and other parameters.

We stress that Eq.\ (\ref{LRMFitEqun}) is not a replacement for the theoretical relations examined above, but simply a parametric fit to our data. Taking $T \propto R^{-2}$, it is in relatively good agreement with the model of \cite{WWSS08}, aside from a factor of $\sim$10 in the constant, despite the fact that their relation is modelled on carbon-rich dust producers and not the oxygen-rich dust we have in the majority of $\omega$ Cen's stars. If we assume their almost-linear scaling of mass-loss rate with metallicity, our models have a product of $\psi v_\infty$ that is about ten times too high. If we assume the scaling from {\sc dusty}, the same is true for the product $\sqrt{\psi} v_\infty$. A decrease in outflow velocity would be surprising, given how low the observed outflow velocities we calculate are (see Section \ref{VSect}). A decrease to the value of $\psi$ by a factor of 10 would be more promising, though a factor of 100 would again bring $\psi$ to sub-solar values.

\subsubsection{Wind velocity}
\label{VSect}

{\sc Dusty} prescribes by simple argument that the wind velocity (if the dust and gas are coupled) must scale as $L^{1/4} (\psi \rho_{\rm s})^{-1/2}$, and also comes with the proviso that the mass-loss rates provided are only valid if the calculated wind velocity exceeds 5 km s$^{-1}$. The main reason for this is that the turbulence in the wind is expected to be of order 1--2 km s$^{-1}$ (e.g.~\citealt{SOW+04}), suggesting a wind that is highly fractal in nature, as large-scale turbulence causes some parts of the wind to sink back towards the star (c.f.~\citealt{IFH92}).

From our {\sc dusty} models, we find the calculated velocities for V6 and V42 are only 1.16 and 0.76 km s$^{-1}$, respectively (assuming [Fe/H]$_{\rm V42}$ = --1.62). A typical RGB-tip star, with $L \sim 2000$ L$_\odot$ and [Fe/H] = --1.62, then has a wind velocity of $\sim$1 km s$^{-1}$. Clearly this is much less than {\sc dusty}'s minimum valid wind velocity, and gives us some cause for concern. Observationally, very similar very-low-velocity winds have been suggested in Galactic halo carbon stars \citep{Mauron08}, as well as the likely more metal-rich Galactic M5--M6 IIIe star L$_2$ Pup and the M8 IIIv star EP Aqr, which show hints of some asphericity or inhomogeneity in their outflows \citep{JCP02,WlBPN08}.

Rotation and magnetic activity could also play a r\^ole in modifying mass loss. Perhaps the largest effect, however, is present in the stars near the AGB tip. Here, pulsation-induced shocks could increase the wind velocity considerably beyond that which {\sc dusty} predicts (c.f.~\citealt{Bowen88b}). Depending on the strength of the shocks, controlled by the piston amplitude of the pulsation, a wind speed of 10 km s$^{-1}$ could theoretically be obtained. This is similar to the situations in \cite{MWHE08}, where the kinetic energy input (along with the amount of condensable carbon) controls the wind outflow: this is corroborated by observational evidence \citep{vLCO+08}. Shocks could facilitate dust formation, which would only occur when the pulsation becomes strong enough to cause shocks capable of accelerating the wind. Given the highest mass-loss rates we find are in LPVs, this is perhaps not surprising. A side-effect of this is that, since mass-loss rate depends linearly on velocity, the mass-loss rates observed for these stars would be significantly higher than the values we publish here.

\subsubsection{Evolutionary status of dusty stars}

A comparison of the relative abundance of stars on the RGB and AGB can provide evidence of the evolutionary stage of our dusty stars. To compute this ratio, we have used the relative evolutionary rates (d$t/$d$L$, proportional to the number of stars in a given luminosity interval -- d$N/$d$L$) along the Padova and BaSTI RGB and AGB isochrones:
\begin{eqnarray}
	F_{\rm AGB}(L) &=& \frac{({\rm d}N/{\rm d}L)_{\rm AGB}}{({\rm d}N/{\rm d}L)_{\rm AGB}+({\rm d}N/{\rm d}L)_{\rm RGB}} , \\
	R_{\rm A/R}(L) &=& \frac{({\rm d}N/{\rm d}L)_{\rm AGB}}{({\rm d}N/{\rm d}L)_{\rm RGB}} .
\end{eqnarray}
These ratios define, respectively, the fraction of stars on the AGB, and the ratio between AGB and RGB stars, at a given luminosity on the giant branch. In practice, these values are upper limits, signifying the fact that some stars may evolve off the AGB as a post-early-AGB or AGB-\emph{manqu\'{e}} star before reaching the model's canonical AGB-tip (defined here by the last thermal pulse). The Padova isochrones suggest $F_{\rm AGB}$ declines from 30\% to 20\% between 100 and 2000 L$_\odot$, while the BaSTI isochrones suggest the decline is from $\sim$19\% to $\sim$13\%. We note that this percentage depends on the physics of convection used in the models, but that empirical constraints on convection mean that the physics used is unlikely to lead to significant alteration of our conclusions (Vandenberg et al.\ 2006).

\begin{figure}
\resizebox{\hsize}{!}{\includegraphics[angle=270]{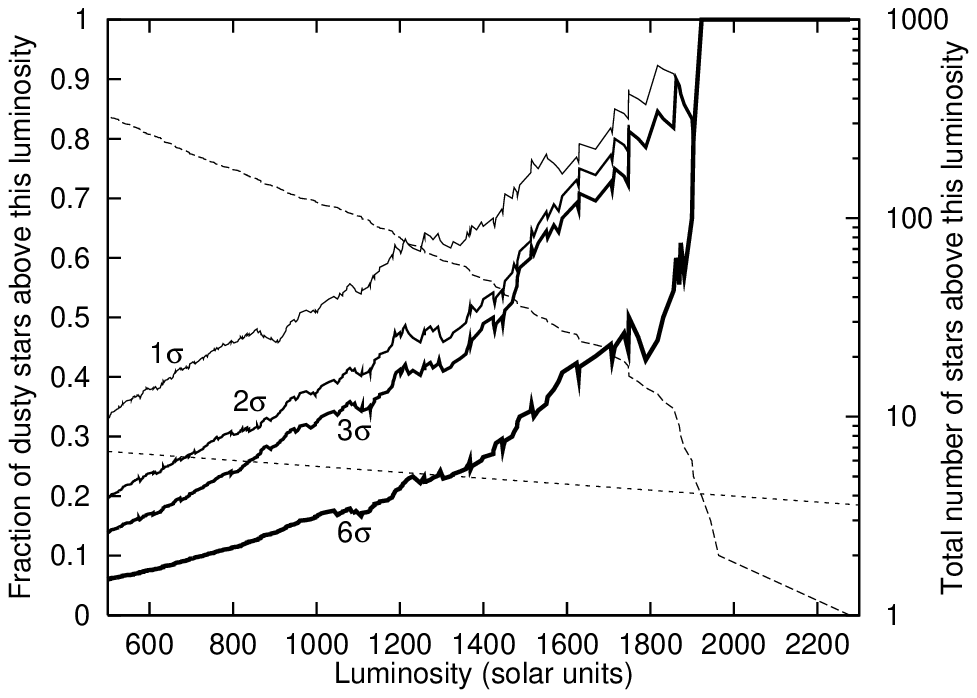}}
\caption[]{The fraction of dusty stars in the most evolved regions of the RGB+AGB, as a function of luminosity. The solid lines show the fraction of objects, above a given luminosity, with combined 8 + 24-$\mu$m excesses of at least one, two, three and six times the photometric errors, in order of increasing thickness. The dashed line (right axis) shows the total number of stars above that luminosity. The dotted line (left axis) demarcates the maximum likely fraction on the AGB: $\int_L^\infty F_{AGB} $d$L$. Dusty stars above this line therefore imply dust production on the RGB.}
\label{FDustyFig}
\end{figure}

We have used the Padova isochrones as an upper limit for $F_{\rm AGB}$ in Fig.~\ref{FDustyFig}, where we present the fraction of dusty stars as a function of statistical significance. Here, we can afford to relax our criteria for significance from our previous value of $3 \sigma$, as we are dealing with a statistical sample, rather than individual stars, thus we expect errant data due to random noise will average out.

This figure suggests that even from the start of measurable dust mass loss ($\sim 1300$ L$_\odot$), at least some stars on the RGB must be producing dust, at a confidence level of $\sim 6 \sigma$. It also suggests that all stars which reach the RGB tip become dust producers.

Both the Padova and BaSTI isochrones imply a reduction in d$N/$d$L$ by a factor of 2--7 as for luminosities above the RGB tip, suggesting we might see 9--30 stars in the 1000 L$_\odot$ above the RGB tip. In fact, we see none, with the possible exception of the metal-rich V6. This includes all post-AGB objects, both confirmed and unconfirmed, which further suggests that the majority of AGB stars do not regain the luminosity of the RGB tip in $\omega$ Cen.

We can see how this may occur if we consider the implied mass-loss efficiency on the RGB. Our isochrone fitting (particularly using the Dartmouth models) suggests that RGB/AGB stars initially have masses of 0.85 M$_\odot$, that \emph{typically} 0.20--0.25 M$_\odot$ of that is lost on the RGB and around half that value on the AGB, leaving a remnant of around 0.5 M$_\odot$. A significantly larger spread of RGB mass loss than this is required to explain the entire HB distribution. If mass loss on the AGB occurs at the same rate as on the RGB for a given luminosity, we can estimate that the AGB mass loss is given simply by the RGB mass loss, divided by the comparative rate of evolution (and hence d$N/$d$L$) between the RGB and AGB, thus:
\begin{equation}
	\dot{M}_{\rm AGB}(L) = \frac{\dot{M}_{\rm RGB}(L)}{R_{\rm A/R}(L)} .
\end{equation}
The value of $R_{\rm A/R}(L)$ is 0.20--0.25 near the tip, where most mass loss takes place, implying a mass loss of 0.04--0.06 M$_\odot$ on the AGB. In practise, the mass loss on the AGB is likely to be more than this, as it is easier to levitate mass off AGB stars as their mass (hence gravity) is less. This would suggest that many (perhaps most) stars either do not reach the AGB (AGB-\emph{manqu\'e} stars) or `peel off' the AGB before they reach the tip. This further corroborates with the large number of low-luminosity post-early-AGB candidates we find in Section \ref{PostAGBSect}, though these may also be very-low-mantle-mass horizontal-branch stars on the transition to the AGB. This in turn again suggests that most dusty cluster members are actually \emph{RGB}-tip rather than \emph{AGB}-tip stars.

\subsection{Comparisons between stellar groups}

\begin{figure}
\resizebox{\hsize}{!}{\includegraphics[angle=270]{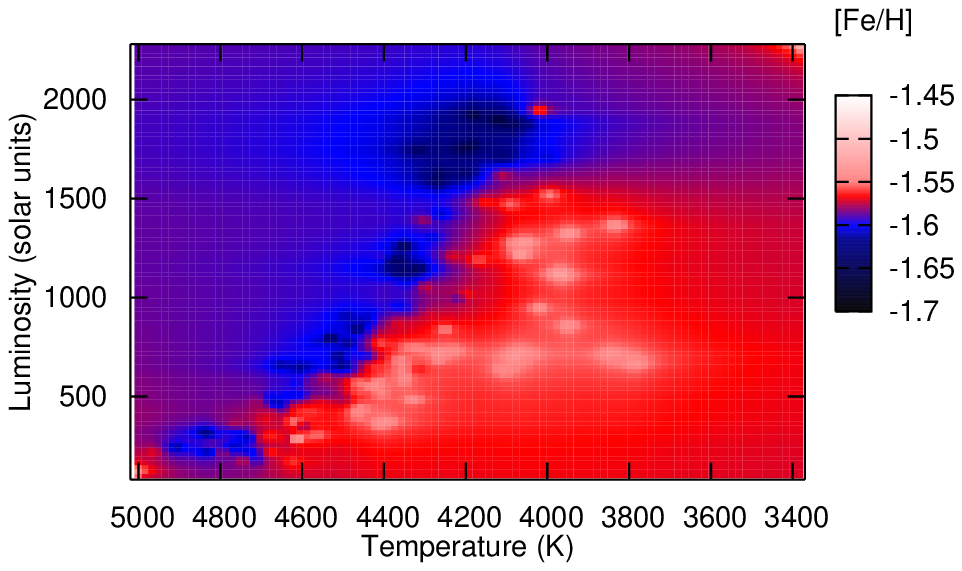}}
\caption[]{HRD, using our luminosity and temperature values, but showing interpolated literature spectroscopic metallicities from \cite{CdSSR01} and \cite{JPSS08}.}
\label{HRDMetalFig}
\end{figure}

\begin{figure}
\resizebox{\hsize}{!}{\includegraphics[angle=270]{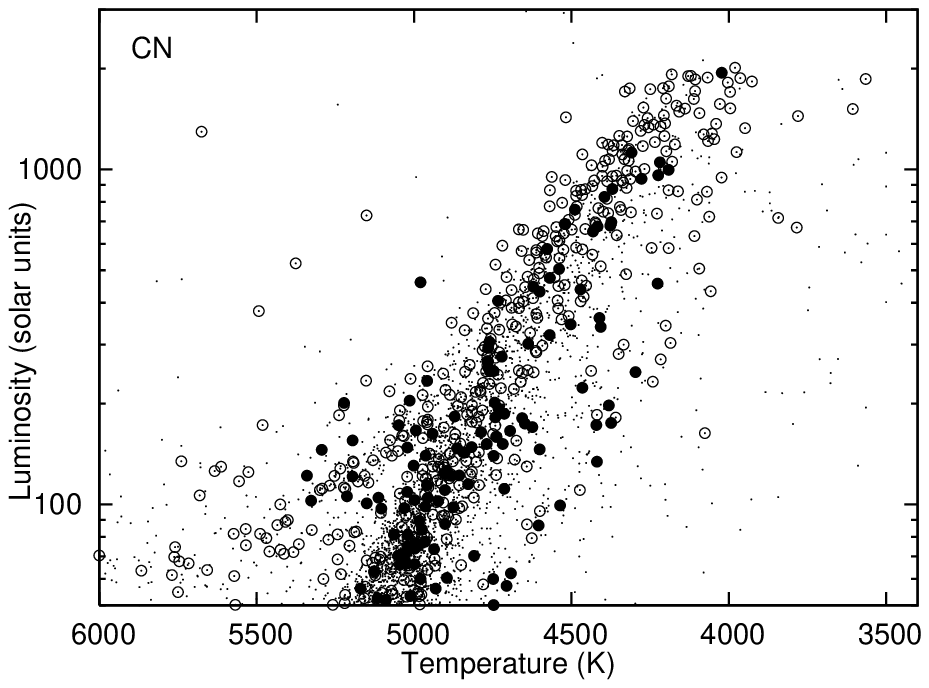}}
\resizebox{\hsize}{!}{\includegraphics[angle=270]{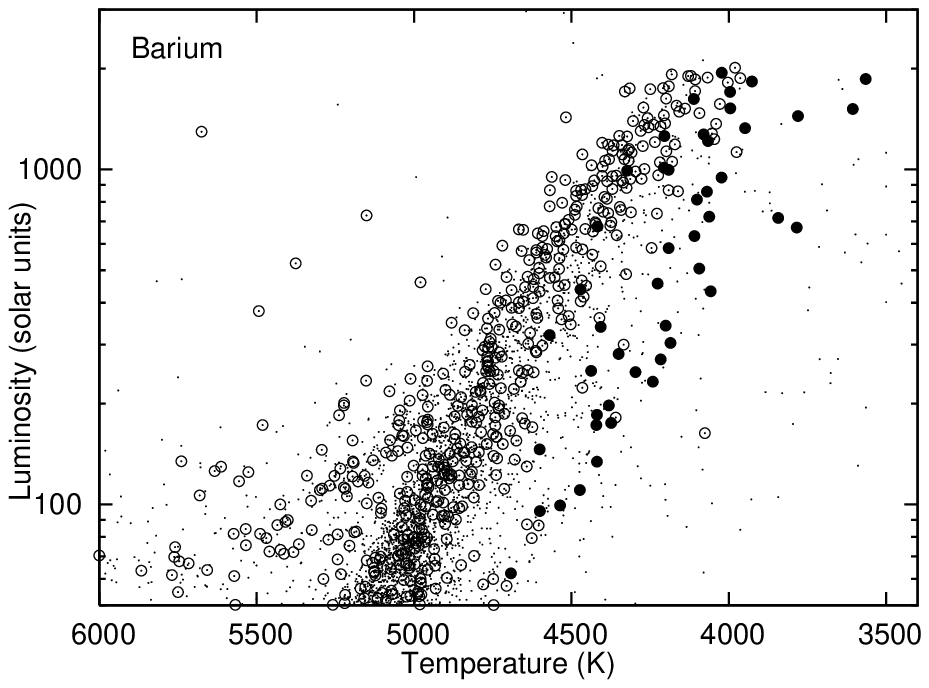}}
\caption[]{HRD, showing stars with measured CN (top) and barium (bottom) line strengths from vL+07. Small points show stars with no measurements, circles show stars with measured line strengths, CN- and Ba-rich stars are shown as filled circles.}
\label{HRDCHBaFig}
\end{figure}

We have taken the metallicity data available on $\omega$ Cen giants from \cite{CdSSR01} and \cite{JPSS08} and super-imposed them onto a HRD of our data. These are displayed in Fig.\ \ref{HRDMetalFig}. In Fig.\ \ref{HRDCHBaFig}, we show a similar diagram, identifying those stars measured as CN- or Ba-rich in vL+07.

As expected, the stars determined to be metal-rich occupy a region much further to the cooler side of the giant branch. Of the stars with measured line strengths, the CN-rich stars do not appear to cluster to any one region more so than the CN-poor stars. This confirms that CN enrichment is primordial and not a result of the stars' evolution themselves. The Ba-rich stars are more likely to be found towards the cool side of the sequence, down to luminosities below the HB, suggesting that these stars are probably metal-rich, and not associated with the AGB (c.f.~vL+07). The strong temperature-dependence of the Ba $\lambda$4554 line used to determine barium richness also suggests that the strong lines in the Ba-rich stars are due to the stars' low temperatures and higher metallicities, and that the relative abundances do not differ as much as is apparent in vL+07.

\begin{figure}
\resizebox{\hsize}{!}{\includegraphics[angle=270]{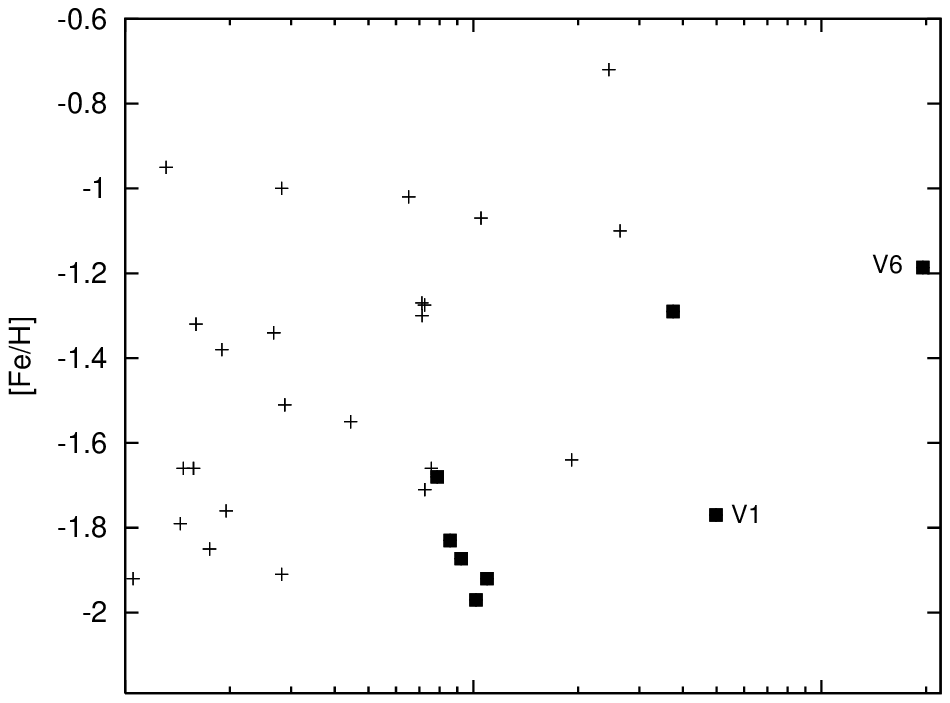}}
\resizebox{\hsize}{!}{\includegraphics[angle=270]{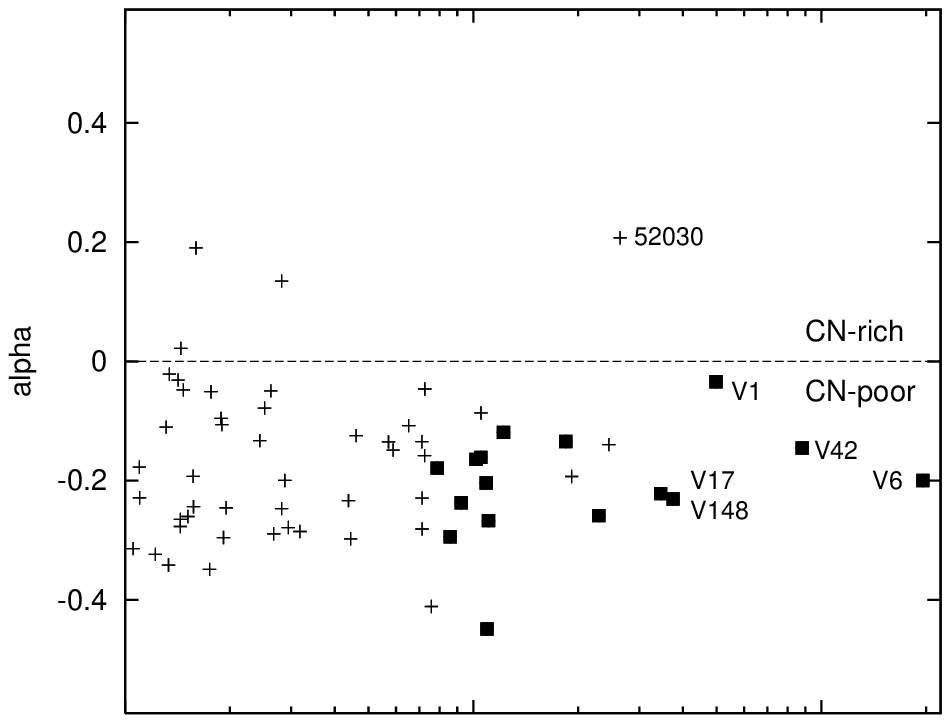}}
\resizebox{\hsize}{!}{\includegraphics[angle=270]{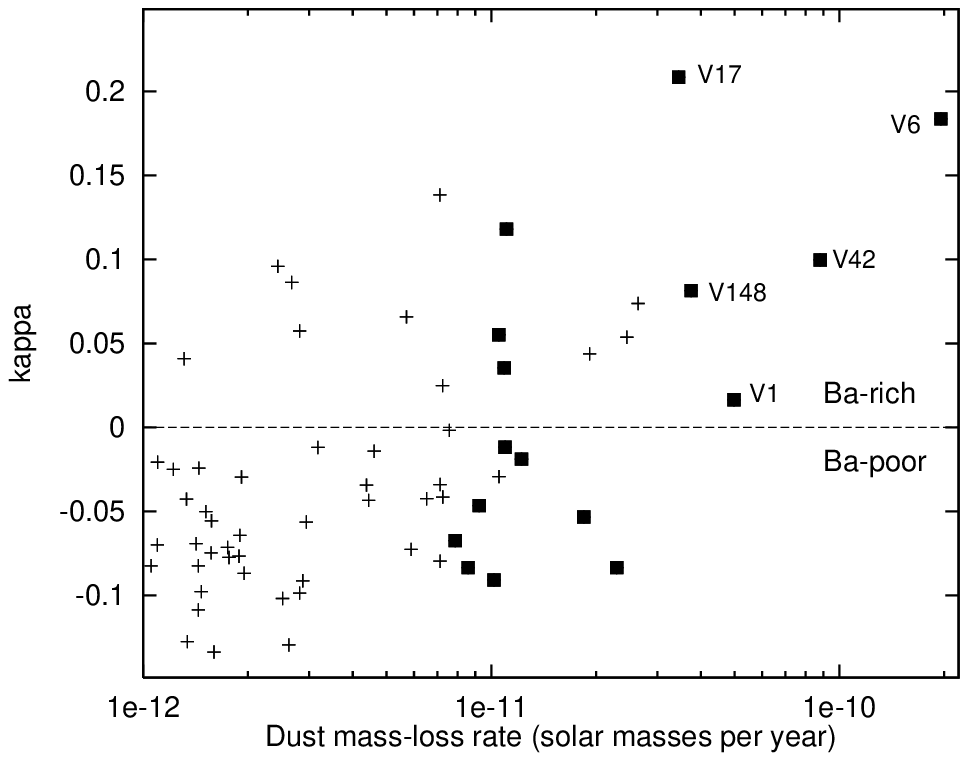}}
\caption[]{Dust mass-loss rates for stars with $L > 500$ L$_\odot$ and their dependence on [Fe/H] (top panel), and the CN (centre panel) and barium (lower panel) richness indicators $\alpha$ and $\kappa$ (see text). Large squares show stars with individually statistically-sigificant ($>3\sigma$) mid-IR excesses.}
\label{MdotChemFig}
\end{figure}

Fig.\ \ref{MdotChemFig} shows the variation of the same indices with dust mass-loss rate. Here we define two indices to measure the richness of CN and barium. These are taken relative to the demarcations for richness marked in Figs.\ 13 and 20(f) of vL+07, and are described thus:
\begin{eqnarray}
	\alpha &=& {\rm S3839} - 2.2({\rm CH4300} - 0.1) ,\\
	\kappa &=& {\rm Ba4554} - 0.127({\rm S3839} + 1.26) ,
\end{eqnarray}
where S3839, CH4300 and Ba4554 are the line indices as described in vL+07. The values of $\alpha$ and $\kappa$ are positive for CN- and Ba-rich stars, respectively. Interestingly, there does not appear to be any significant variation of mass-loss rate with measured [Fe/H] values, unless one takes only those stars with highly significant excesses. However, there does appear to be a strong relation with barium-richness. This correlation is strengthened if we remove the special case of V1 -- the dusty post-AGB star. We also note that then, the four most-mass-losing stars are metal-rich, suggesting a weak correlation may still be present. The barium may simply trace the coolest and metal-richest subpopulations in $\omega$ Cen, which one would expect to be the most prolific dust producers.

Using the above criteria, and setting a cut at 500 L$_\odot$, there is no obvious dependence of dust production on CN-richness: of the cluster members with observable mass loss ($\gtsim 10^{-12}$ M$_\odot$ yr$^{-1}$), four are CN-rich, and 53 CN-poor. This is comparable to the figures of 12 and 95 in the non-dusty sample. The only CN-rich star losing large amounts of mass is, perhaps unsurprisingly, the carbon star LEID 52030. The lack of CN-rich M-type giants has been noted in vL+07.

\subsection{Implications and further work}

\subsubsection{Mass-loss rate and variability of metal-poor giants}

There are several implications that result from our analysis.
\begin{itemize}
\item Mid-infrared excess, thus the inferred amount of dust, appears highly variable in pulsating stars on timescales of a month or less, much shorter than thought possible. Actual mass-loss rates may thus be considerably different from literature values. The true mass-loss rate may not be attainable without mid-IR photometry and spectroscopy throughout the pulsation cycle.
\item An accurate measure of dust and gas mass loss requires knowing: how the gas-to-dust ratio varies both with evolution and metallicity, whether gas is coupled to dust, and how low the outflow velocity can reasonably be to support strong mass loss.
\item Lack of correlation between our results and literature relations and studies (Sections \ref{PsiSect} and \ref{VSect}) highlights the difficulty in obtaining empirical relations between mass loss and other stellar parameters in an inhomogeneous sample of stars. From our analysis of V42 (Section \ref{VarySect}) and other stars (MvL07), it seems more likely that other non-LTE effects, such as shocks and chromospheres, are significantly modifying the winds.
\item Our mid-IR spectrum and critical evaluation of other data of V42 suggest that the analysis of \cite{HA07} may be at least partially correct in assuming that some C-rich mass loss can occur in conditions that depart from LTE. This may provide the opacity source required to drive O-rich dust-driven winds \cite{Woitke06b}.
\end{itemize}

\subsubsection{The fate of $\omega$ Cen's lost mass}
\label{FateSect}

The fate of the lost mass is also interesting. The mass loss from the cluster must be stochastic at some level if it is being fed by only a very small number of stars, as these stars do not maintain a constant dust production throughout their evolution. Based on our observations, the current mass loss from the cluster should be around 1.2 $\times$ 10$^{-5}$ M$_\odot$ yr$^{-1}$ or more, depending on the contribution from chromospherically-driven winds, the gas-to-dust ratio in winds, and the gas-to-dust coupling efficiency.

With this total mass-loss rate and assuming 48 Myr since the last Galactic plane crossing (when tidal friction is expected to strip away intra-cluster material) based on the proper motions from vL+00, we might expect $\gtsim$ 290 M$_\odot$ of gas and $\sim$0.05 M$_\odot$ of dust to have built up within the cluster. Our \emph{Spitzer} observations (B+08) suggest the amount of dust in the cluster is $< 10^{-4}$ M$_\odot$ (c.f.~\cite{BWvL+06}, who detect $9 \times 10^{-4}$ M$_\odot$ of dust in M15), and data in the literature imply that $< 2.8$ M$_\odot$ of neutral gas can be present \citep{SWFW90}. The possibility remains that the intra-cluster medium (ICM) in $\omega$ Cen is in a hot, ionised state -- the ICM in 47 Tuc contains at least 5\% ionised gas \citep{SWFW90,FKL+01}. The fact that cold ICM in both gas and dust phase is observed in M15 suggests that neutral ICM is likely an existing phase in many GCs.

On this basis, it would seem likely that \emph{neutral} gas is being processed or removed from the cluster environment on timescales of $\ltsim 3 \times 10^5$ years, and dust within $1 \times 10^5$ years. With a tidal radius of around 85 pc \citep{Harris96}, it would take a wind 8 Myr to escape the cluster if it were escaping at 10 km s$^{-1}$, or 1.2 Myr if swept clean as a result of ram pressure by the hot Galactic Halo ($\omega$ Cen has a space velocity of 69.1$\pm$5.6 km s$^{-1}$ with respect to the Halo; \citealt{OKYM07}).

Another possibility is that the dust mass-loss rates (and thus also the gas mass-loss rates) we derive are too high due to their intrinsic variability, however we would expect this to average out through addition of errors in quadrature so that our value is only affected by this at the $\sim$15\% level, which mostly comes from the error in the mass loss of V42. Without knowing the long-term trend of this mass loss it is difficult to tell, but we can argue based on the sustainability of the mass-loss rate of V6, given the total mass loss we expect over its RGB or AGB evolution. Based on the total mass-loss rate we derive in Section \ref{MdotV6V42Sect}, we find that, at the current rate, the star would lose the $\sim$0.1 M$_\odot$ required by evolutionary theory in 90\,000 years. Given its position at the top of the giant branch, this does not appear too unreasonable, so we can assume our global rates are relatively sound.

We can also consider the dynamical effect of lost mass on the cluster. If a mass-loss rate of $1.2 \times 10^{-5}$ M$_\odot$ yr$^{-1}$ is indeed representative, this equates to some $\sim$0.46\% of the cluster's mass per gigayear \citep{MMM97,dSR05,vdVvdBVdZ06}. If assumed constant in the past, the cluster will thus have lost 5--7\% of its initial mass, with only small dynamical consequences. However, more massive stars lose a much larger fraction of their mass, and thus the cluster mass loss will have been more significant in the past -- even when the cluster was already a Gyr old and intermediate-mass AGB stars were the main contributors to this mass loss. Cluster mass loss is therefore not a phenomenon limited exclusively to the first 10 Myr or so when exploding massive stars assured significant mass loss from the cluster's gravitational potential.

\subsubsection{Are globular clusters only producing dust episodically?}

Our analysis appears to show that the total mass loss from GCs is highly episodic. In $\omega$ Cen, V6 and V42 appear to be producing most of the dust mass loss in the cluster, and the total mass loss from the cluster appears dominated by a small number of stars.

Mass loss in $\omega$ Cen would seem to be stochastically variable. The high number of low-luminosity post-(early)-AGB stars and candidates suggests many stars may leave the AGB early, and that there may be a considerable number of AGB-\emph{manqu\'e} stars, leaving a significantly depleted fraction of stars actually reaching the AGB tip. By implication, the mass loss from other GCs of similar metallicity would have even more stochastically-variable mass-loss rates, as they are less populous and therefore have (numerically) even fewer stars exhibiting high mass-loss rates.

Mass loss from a cluster may increase dramatically in the event that a single star undergoes the ejection of a planetary nebula. Several of these have now been observed in GCs: notably Pease 1 (K648) in M15 \citep{Pease28}, but also IRAS 18333--2357 (GJJC1) in M22 \citep{GJJ+89}, JaFu1 in Pal6, and JaFu2 in NGC 6441 \citep{JMF+97}. The fact that, out of all the clusters observed to date, only M15 shows convincing evidence of hosting an observable intra-cluster medium (\citealt{BWvL+06}; Barmby et al.~2008), could be due to a random effect, rather than due to M15's particularly low metallicity. This implies that the diffuse matter seen in M15 could merely be a diffusing planetary nebula that was ejected $\gtsim 10^4$ years ago, however its large mass ($\gtsim$ 0.3 M$_\odot$) would suggest that a single such event is unlikely to be the sole producer of this much material. Potentially this free gas and dust could also be the results of a stellar collision, which by their nature are also episodic. Simulations in GCs fail to reproduce this level of mass loss in the case of colliding main-sequence stars \citep{LRS96}, however a collision between a red giant and a neutron star may have the potential to eject the entire $\sim$ 0.3 M$_\odot$ stellar mantle \citep{RS91}. Such collisions have been suggested as a method of clearing GCs of ISM \citep{UCR08}. The true fate of the lost mass, and the precise origin of the cloud in M15, remains at this time unknown.

\subsubsection{The nature of V42's mass loss}
\label{V42Sect}

It is clear from Section \ref{VarySect} that V42, with its mass-loss rate of 1.4--3.6 $\times 10^{-6}$ M$_\odot$ yr$^{-1}$, stands out from the rest of the group. One possibility for this is that this represents a `superwind' phase akin to that seen in more-massive long-period variables (e.g.~\citealt{vLGdK+99, SWS99}). Here, the mass loss increases dramatically shortly before the star makes the transition from AGB-tip star to post-AGB star. If this were the case, we would expect V42 to denote the AGB-tip in Fig.\ \ref{HRDFig}. As we are unable to separate the RGB from the AGB, we are unable to tell whether this is the case, as the RGB tip may be more luminous. V42 is also clearly less luminous than V6 both in the data we have used to determine its luminosity, and the variability studies of Dickens et al.~(1972). We should also note that the luminosity of V42 is likely variable on timescales significantly longer than the pulsation timescale -- it is possible that V42 has recently experienced a thermal pulse, as suggested by Origlia et al.~(1995), and that it is currently in an associated luminosity minimum.

The position of V42 on the HRD (Fig.\ \ref{HRDFig}) also suggests a second possibility -- that the star is part of one of the metal-rich sub-populations, like V6. If it is 0.5 dex more metal-rich, this would decrease the calculated outflow velocity of its wind and hence its dust mass-loss rate by 0.25 dex and decrease its gas mass-loss rate by 0.75 dex to better match V6 in Fig.\ \ref{MdotCompsFig}.

The possible presence of carbon-rich dust highlighted in Section \ref{VarySect} is surprising, given the oxygen-rich nature of the star. We put forward three suggestions to explain this behaviour: firstly, that a large variation in dust production is producing significant variations in the amount of dust in the wind; secondly, that luminosity changes in the star are altering the amount of light reprocessed by the dusty wind; or finally, that the grain size varies sufficiently that the silicate feature disappears periodically. In the first case, it can be argued that the criteria for dust formation to occur are only met during part of the pulsation cycle, leading to episodic dust formation under non-LTE conditions (e.g.\ \citealt{HA07}). This may lead to carbon-rich dust at some epochs, with oxygen-rich dust forming at others, which we propose as a possible solution to the disparity between the mid-IR spectra (suggesting carbon-rich dust) and accumulated literature photometry (suggesting oxygen-rich dust).

Under the second scenario, the rapid, large-scale changes seen in the mid-IR amplitude of V42 are due to the variation of light from the pulsation cycle: during periods when the star is warm, there is more optical emission to be re-processed by the dusty wind, which is much more opaque at these wavelengths. During periods when the star is cool, the optical emission is suppressed, the re-processing of optical radiation is decreased, and the mid-IR emission drops accordingly.

It would, at face value, appear unlikely that the grain size in V42 could become sufficient (i.e.~a few tens of microns) to completely mask the silicate feature, due to the low metallicity and thus the infrequency of particle collisions that would lead to the aggregation of large grains. However, we note that it is not presently possible to observationally distinguish between these three scenarios.

Episodic and/or variable mass loss near the RGB/AGB tip is not a new concept: previous studies, including O+02 and \cite{MDS08}, have argued on observationally-based grounds that episodic mass loss is occuring. Applying this principle to other stars, the consequences of this could be quite far ranging: historical mass-loss rates have been assumed to be accurate probes of the mass lost over many years, yet many observed stars have very strong optical variations. It is thus important that, when calculating dusty mass-loss rates from such stars, that investigators have a good estimation of the stellar luminosity and temperature \emph{at the time of measurement}.

\section{Conclusions}

We have here presented stellar parameters derived from spectral energy distribution fitting to stars over two orders of magnitudes in luminosity down the RGB and AGB of $\omega$ Centauri, creating a physical HRD of luminosity versus temperature, outlining the RGB, HB and AGB in great detail and accuracy, and identifying several new post-AGB star candidates. From isochrone and spectral energy distribution fitting, we estimate the following parameters for the cluster:
\begin{itemize}
\item Distance: $d = 4850 \pm 200$ (statistical) $\pm 120$ (systematic) pc;
\item Reddening: $E(B-V) = 0.08 \pm 0.02$ mag (statistical) $\pm 0.02$ mag (systematic) $\pm 0.02$ mag (differential);
\item Total dust mass loss: $\dot{M}_{\rm dust} = 9 \pm ^{6} _{4} \times 10^{-10}$ M$_\odot$ yr$^{-1}$;
\item Total gas mass loss: $\dot{M}_{\rm gas} \sim 1.2 \times 10^{-5}$ M$_\odot$ yr$^{-1}$ ($\sim 2/3$ from dusty winds, $\sim 1/3$ from chromospheric mass-loss);
\item Timescale to clear the intra-cluster medium from cluster: $\ltsim 10^5$ yr;
\end{itemize}
under the following assumptions:
\begin{itemize}
\item gas-to-dust ratio, $\psi \propto 10^{-{\rm [Fe/H]}}$;
\item wind velocity, $v \propto L^{1/4} \psi^{-1/2}$;
\item absence of substantial mid-IR variability;
\item constant dust chemistry along the giant branch;
\item inner dust envelope temperature of 1000 K;
\item our low wind velocities ($\sim 1$ km s$^{-1}$) are accurate and still produce an accurate mass-loss rate.
\end{itemize}
We derive inner dust envelope temperatures for a handful of stars (including V6 and V42), suggesting they are typically $\gtsim$ 600 K.

We show that V6 and V42 contribute $\sim$25\% of the cluster's dust production. V42 may be variable at mid-infared wavelengths on timescales of as little as a few weeks, suggesting equally variable dust production, meaning a single estimate of mass-loss rate and wind conditions may not accurately reflect the long-term conditions in this and similar stars. V42, an oxygen-rich star, may also be producing carbon-rich dust.


Dusty mass loss appears to start suddenly at a threshold of $\sim$4400 K, or $\sim$ 1000 L$_\odot$, for a significant fraction of stars. There appear to be additional star-to-star differences, suggesting that other factors also influence the mass-loss rate. Empirical literature relations do not accurately reflect the dust production rates we measure in $\omega$ Cen.

By comparing our stellar parameters with optical line strengths, we have deduced that the Ba-rich stars in $\omega$ Cen belong to the more metal-rich populations and not to the AGB. These stars are characterised by dustier winds.

Mass loss along the RGB appears quite efficient with typically 0.20--0.25 M$_\odot$ lost on the RGB and $\gtsim$0.05 M$_\odot$ on the AGB, possibly leading to large numbers of AGB-\emph{manqu\'e} and post-early-AGB stars. Dust production in $\omega$ Cen appears not confined to the AGB, with most dust producing stars in the cluster near the RGB-tip. 

\ \\

\noindent
{\bf Acknowledgements}

\noindent
We thank Aaron Dotter, Noriyuki Matsunaga, Greg Sloan, and the many others who have given help in this project for their words of wisdom. I.M.\ acknowledges a PPARC/STFC studentship award. L.D. acknowledges financial support from the Fund for Scientific Research -- Flanders (Belgium). M.L.B., C.E.W., and R.D.G.\ are supported in part by NASA through \emph{Spitzer} contracts 1276760, 1256406, and 1215746 issued by JPL/Caltech to the University of Minnesota. A.K.D.\ acknowledges research support from \emph{Spitzer} contract 1279224. Finally, we thank the referee for his very helpful advice. This publication makes use of data products from the Two Micron All Sky Survey, which is a joint project of the University of Massachusetts and the Infrared Processing and Analysis Center/California Institute of Technology, funded by the National Aeronautics and Space Administration and the National Science Foundation. This manuscript is also based on observations obtained at the Gemini Observatory, which is operated by the Association of Universities for Research in Astronomy, Inc., under cooperative agreement with the NSF on behalf of the Gemini partnership: the National Science Foundation (United States), the Science and Technologies Facilities Council (United Kingdom), the National Research Council (Canada), CONICYT (Chile), the Australian Research Council (Australia), CNPq (Brazil), and CONICET (Argentina).



\bsp

\label{lastpage}

\end{document}